\documentclass[pra,11pt,letterpaper,draft,bibnotes,notitlepage,final,balancelastpage,nofootinbib]{revtex4}
\usepackage{times,bm,bbm,bbold,amssymb,amsbsy,amsthm,amsmath,amsfonts,mathrsfs,graphics,graphicx,color,xcolor,multirow}

\definecolor{myurlcolor}{rgb}{0,0,0.7}
\definecolor{myrefcolor}{rgb}{0.9,0.4,0}
\usepackage{hyperref}
\hypersetup{colorlinks,linkcolor=myrefcolor,citecolor=myurlcolor,urlcolor=myurlcolor}

\begin{document}

\begin{titlepage}
\vspace*{10mm}
\begin{center}
{\Large \bf Space Isotropy and Homogeneity Principles Determine the Maximum Nonlocality of Nature}


\vspace*{10mm}

\large{Akbar Fahmi} \footnote{fahmi@theory.ipm.ac.ir}

\vspace*{.1cm}

\small{(\today)}

\vspace*{.5cm}





{\bf Abstract}
\end{center}
One of the fundamental questions in physics concerns the relation between spacetime and quantum entanglement. The spacetime is usually considered as a fixed background physical space, and the quantum entanglement is usually manifested as a  ``spooky action at a distance" or the existence of ``nonlocality" in nature. Here, we propose the flat-space isotropy and homogeneity principles as the fundamental criteria for determining the maximum degree of nonlocality of nature. More specifically, we consider abstract and deterministic nonlocal-box models which have stronger correlations than in quantum mechanics, whereas therein instantaneous communication remains impossible. We impose space-symmetry group structures on these models and derive a measure for the degree of space symmetries. Surprisingly, there is a tradeoff or inconsistency between the degree of space symmetries and the degree of nonlocality, where this inconsistency is exactly lifted at the Tsirelson bound, as predicted by quantum physics and also predicted in the experiments. Moreover, we prove this result in the general framework of deterministic nonlocal models and conclude that the probabilistic interpretation of the nonlocal box models is an emergent property of the flat-space symmetries.



\end{titlepage}

\section*{}

The development of quantum mechanics in the early twentieth century forced physicists to radically change some of their concepts and even introduce new concepts to describe the world. For example, quantum entanglement was first viewed as a source of some paradoxes, most noticeably the Einstein-Podolsky-Rosen paradox (EPR) \cite{EPR}, which states that any physical theory must satisfy both local and realistic conditions. These conditions manifest themselves in the Bell-type inequalities \cite{Bell,Bell1,CHSH}. These inequalities are violated by quantum predictions and have passed a wide variety of experimental tests \cite{Han,Sha, Ant,Win,Pan,Sup}, which demonstrated strong indications against local hidden variable theories. These phenomena are often interpreted as nonlocality in quantum mechanics. Although quantum mechanics is nonlocal, it is not as nonlocal as allowed by the no-signaling principle, the minimum necessity condition which any physical theory shall satisfy in order to be consistent with special relativity \cite{PR1}. Specifically, there are wide range of abstract and deterministic ``postquantum" models which have higher degree of nonlocality than the upper bound of nonlocality in quantum mechanics framework while instantaneous communications still remain impossible \cite{PR1,Sv,Non1,BCP,JM,BP1,BP2,Mas}. A famous subset of postquantum models is nonlocal (NL) box models which enable lead to the violation the Clauser-Horne-Shimony-Holt (CHSH) inequality \cite{CHSH} (an experiment friendlier version of the Bell inequality) by the algebraic maximum value of $4$ \cite{BCP,JM,BP1,BP2,Mas}, while quantum correlations achieve at most $2\sqrt{2}$ (Tsirelson bound) \cite{Tsi}.


Now, this natural question arises: ``Is there a deep principle of nature that limits the amount of nonlocality?" \cite{Pop}.
Intensive efforts have been devoted to search for principles characterizing quantum correlations. From an information-processing point of view,
the communication complexity \cite{Dam,Bra2}, information causality \cite{IC}, macroscopic locality \cite{ML,ML1}, many-box locality (a refined version of the macroscopic locality) \cite{MBL},
and the local orthogonality \cite{LO} have supplied fundamental limitations on the amount of nonlocality in NL-box models. These principles have been suggested as principles to distinguish physical theories from nonphysical ones. However, all the mentioned approaches provide only partial answers which single out quantum mechanics with a possible exception, the information causality principle \cite{IC}. In the information causality scenario \cite{IC}, the mutual information between two observers (conventionally called ``Alice" and ``Bob") is bounded by classical bits communicated from Alice to Bob. In this approach, the maximum value of the violation the CHSH inequality \textit{asymptotically} approaches the quantum critical value of $2\sqrt{2}$ when the number of NL-boxes are increased \cite{ML1}.


In this article, we revisit the above question from another independent and fundamental perspective. We aim to investigate whether one can make a \textit{conclusive} answer by using some other principles of nature which are more elementary in that they apply even to simplest systems and at the same time hold universally for all theories of nature. We show that the \textit{isotropy and homogeneity of the real space} play the role of such elementary principles. All theories have experimental predictions, and all experiments are performed in a fixed background real space. This is an immediate question whether basic properties of this background space and its symmetries, in particular its isotropy/homogeneity, can put fundamental constrains on a physical theory. These two principles are believed to be respected by all observable and experimental evidence of all theories of nature, independent of their contents or mathematical models describing these theories \cite{MM,LoM1,LoMM,LoMM9}.


The isotropy/homogeneity symmetry principle is understood in the form that all orientations/locations of an experiment are physically equivalent. That is, if all elements of an experiment (systems, sources, channels, or other measurement settings of the lab) are rotated/translated equally by the same amount, such rotations/translations do not have any impact on the value of any observable quantities. In quantum mechanics as a mathematical model, the isotropy/homogeneity symmetry is represented by applying rotation/translation unitary transformations on state vectors and measurement operators such that the probability distributions (correlation functions) all remain invariant. As a hint supporting our proposal, in another work \cite{Almsy} we have imposed rotation/translation symmetry transformations on one set of postquantum models (called almost-quantum correlations \cite{Alm}) and have shown they would reduced to standard quantum mechanics.

Here, we aim to discuss the impact of the isotropy/homogeneity symmetry transformations on the abstract and deterministic NL-box models. Although the isotropy/homogeneity principles have unambiguous and straightforward geometrical pictures in the real space, their implications on purely mathematical models (e.g., with Boolean algebra) is unclear. To realize all of mathematical models in the experiment, we always imply a correspondence relation between the inputs/outputs of experiment and the inputs/outputs of the mathematical model. This argument interestingly explained in one of the EPR \cite{EPR} axioms to recognize a complete physical theory: ``every element of the physical reality must have a counterpart in the physical theory." Similarly, we employ a correspondence between the experimental settings of the CHSH inequality and an NL-box model. This correspondence naturally indicates that the isotropy/homogeneity of the flat space have immediate implications on the NL-box model.






To study this implication, we translate the isotropy/homogeneity symmetry in the abstract NL-box framework (employing our correspondence) and obtain pertinent symmetry conditions in the NL-box model. We prove that these symmetry conditions induce a symmetry parameter which is inconsistent with the violation of the CHSH inequality for the NL-box model \cite{CHSH} at the maximum nonlocality of $4$. In addition, in the imperfect version of the NL-box models, we show that there is trade-off between the probability of violation of the CHSH inequality at the maximum value of $4$ and the probability of the symmetry parameter holds. Interestingly, we prove that the Tsirelson bound \cite{Tsi} is \textit{exactly} the threshold for which the internal consistency of the nonlocal models remains valid. Moreover, without the need for \textit{imperfect} NL-box models, we derive the Tsirelson bound as the natural consequence of the invariance of observable quantities under the action of symmetry group elements. This result indicates that the Tsirelson bound and imperfection of NL-box models (probabilistic interpretation of measurement outcomes) are both \textit{emergent} consequences of the symmetries of the flat space.




The typical Bell-like scenario is defined with two spacelike parties (Alice and Bob) which share a bipartite system and perform local measurements on it. Alice and Bob measurement inputs are represented by three-dimensional unit vectors $\mathbf{a}$ and $\mathbf{b}$, and their outputs by $A\in\{+1,-1\}$ and $B\in\{+1,-1\}$, respectively. The isotropy and homogeneity principles in these experimental setups are illuminated by following assumptions/``facts":


\textbf{Assumption $1$}: The spin-perfect correlation (anti-correlation) conservation law. 
In any experiment, including the Bell-like experiments, the probability of obtaining output $A$ ($B$) for the measurement setting $\mathbf{a}$ ($\mathbf{b}$) is equal to the probability of obtaining output $-A$ ($-B$) for the measurement setting $-\mathbf{a}$ ($-\mathbf{b}$). This property can be realized in experiment simply by rotating the reference frame such that $\mathbf{a}\rightarrow-\mathbf{a}$ ($\mathbf{b}\rightarrow-\mathbf{b}$).

In the literature \cite{Pop}, the Bell-like scenarios are usually described as abstract blackbox models with abstract labels denoting the measurement settings and outcomes. Thus, it may seem that Assumption $1$ cannot be justified as relevant axiom in the blackbox scheme. As pointed out earlier at the definition of a complete physical theory, Assumption $1$ as an experimental evidence, must also have an impact and relevance even in any blackbox model. This experimental evidence is, in fact, one of inherent properties of quantum theory and also implicitly exist in the framework of generalized probabilistic theories \cite{GPT3,GPT4,GPT5}, see appendixes.



\textbf{Assumption $2$}: In any physical theory/model, the flat space isotropy and homogeneity principles satisfy the rotation and translation symmetry group structures, respectively. The $3$-dimensional flat space has the Euclidean group structure $\mathrm{E^{+}(3)}$, which comprises of all rotations and translations of $\mathrm{\mathbb{E}^{3}}$ and arbitrary finite combinations thereof.

The group structure of Euclidean frames requires the following conditions to be met: close under group transformation, identity transformation, inverse transformation and composition law. Hence, there are two points which be noted ($i$) The space symmetries respect Assumption $1$ in that two opposite directions are transformed to two opposite directions under space symmetry group transformations. ($ii$) We have excluded time symmetry group transformations because time does not appear explicitly in the Bell-type experiments.

\textbf{Assumption $3$}: Laws of physics are covariant and observable quantities are invariant under the Euclidean space symmetry group transformations. In particular, the physical laws transform under a certain representation of the group elements between admissible frames. Here, the observable quantities are the probability distributions, the correlation functions, and the upper bound of violation of the CHSH inequality.


The invariance of correlation functions under the isotropy/homogeneity symmetry transformations was implicitly proved in an electronic analogue of the Michelson-Morley experiment \cite{LoMM}. In this experiment, where the isotropy of the electron's dispersion relation was verified at the level of one part in $10^{18}$, the superposition of energy levels can be described as an entangled Bell state wherein the phase of energy shift plays the same role as a relative phase of the Alice and Bob measurement settings.

\begin{figure}
\centering
\includegraphics[scale=.62]{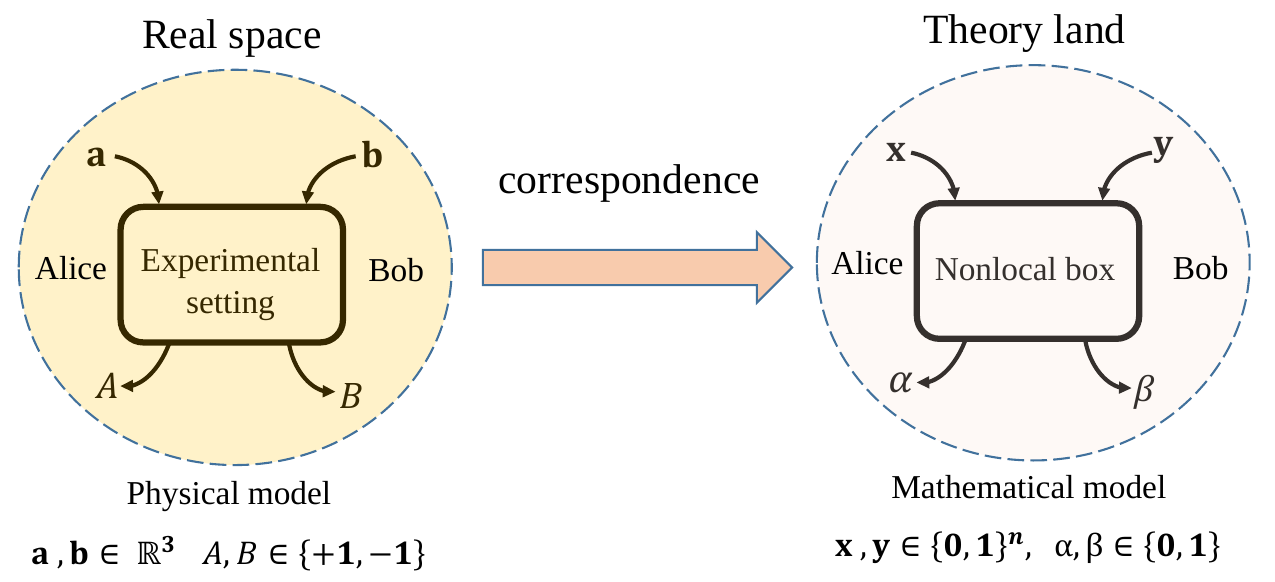}
\caption{Schematic correspondence between a physical experiment (performed in the real space) and an NL-box mathematical model.
In the Bell-like experiments, Alice and Bob hold one part of a shared system and each party chooses one arbitrary measurement direction ($\mathbf{a}$ and $\mathbf{b}$) to apply the action of a physical apparatus (e.g., magnetic fields) on their systems and detect the state of the binary measurement outcomes $A$ and $B$, respectively. The parties repeat the process many times to estimate the probability distribution $P_{\mathrm{exp}}(A, B|\mathbf{a},\mathbf{b})$ by using the measurement inputs and recorded outputs. Any experiment can be viewed an abstract black box, which is more convenient for mathematical analyzes. As pointed out in one of the EPR axioms \cite{EPR} (and its reverse in the no-restriction hypothesis \cite{Alm1}), a mathematical model is complete if every observable quantity in the experiment has a counterpart in the physical model (and viceversa). To make the NL-box model represent a complete physical model, we propose correspondence relations among the measurement inputs/outputs in the real space $(\mathbf{a},\mathbf{b})/(A, B)$ and the inputs/outputs of corresponding the NL-box models $(\mathbf{x},\mathbf{y})/(\alpha, \beta)$, where $A=(-1)^{\alpha}$ and $B=(-1)^{\beta}$. For details, see the main text.}\label{Cor}
\end{figure}


In the following, we describe nonlocality in a generic manner and introduce the abstract NL-box models in the Boolean algebra terminology, and next, revisit Assumptions $1-3$ in the NL-box framework. Similar to the Bell-like scenario, two distant parties, Alice and Bob, share a NL-box machine and choose inputs $\mathbf{x}$ and $\mathbf{y}$, and the box produces outputs $\alpha$ and $\beta$, respectively. The box is characterized by the joint conditional probability as
\begin{equation}\label{nb}
P_{\mathrm{NLB}}(\alpha,\beta|\mathbf{x},\mathbf{y})= \left\{
\begin{array}{ll}
1/2 &: \alpha\oplus\beta=f(\mathbf{x},\mathbf{y}) \hspace{.2 cm}(\mathrm{mod}\hspace{.1 cm}2)  \\
0 &:\text{ otherwise }.
\end{array}        \right.
\end{equation}
In earlier versions of the NL-boxes, the inputs $\mathbf{x}$ and $\mathbf{y}$ are binary numbers $x, y\in\left\{0,1\right\}$ \cite{Bra2,Dam} or integer numbers $x,y\in\{0,1,...,m-1\}$ \cite{BP1,JM}, and $f(\mathbf{x},\mathbf{y})$ is usually the product of inputs $xy$. The outputs are binary numbers $\alpha,\beta\in\left\{0,1\right\}$, and $\alpha\oplus\beta$ is the addition of outputs calculated modulo $2$. The joint probability (\ref{nb}) satisfies the no-signaling condition $\sum_{\beta}P(\alpha,\beta|\mathbf{x},\mathbf{y})=P(\alpha|\mathbf{x})$. A similar relation holds for $P(\beta|\mathbf{y})$.
But these models give maximum possible value to the CHSH parameter
\begin{eqnarray}\label{CHSH}
\mathcal{S}=|E(\mathbf{x}_{1},\mathbf{y}_{1})-E(\mathbf{x}_{1},\mathbf{y}_{2})|+
|E(\mathbf{x}_{2},\mathbf{y}_{1})+E(\mathbf{x}_{2},\mathbf{y}_{2})|,
\end{eqnarray}
where $-1\leqslant E(\mathbf{x}_{i},\mathbf{y}_{j})\leqslant1$ and $E(\mathbf{x}_{i},\mathbf{y}_{j})=\sum_{\alpha,\beta=0,1}(-1)^{\alpha\oplus\beta}P(\alpha,\beta|\mathbf{x}_{i},\mathbf{y}_{j}),\, i,j=1,2$,
are correlation functions between measurement outputs. Different theoretical frameworks give different upper bounds on the CHSH parameter $\mathcal{S}$,
which can be used to quantify the strength of correlations $E$.
The NL-box models reach the maximum possible value $\mathcal{S}_{\mathrm{NLB}}\leqslant 4$, which corresponds to $0\leqslant E_{\mathrm{NLB}}\leqslant 1$. Local realistic theories/models give the upper bound $\mathcal{S}_{\mathrm{L}}\leqslant2$ (the very CHSH inequality \cite{CHSH}), which corresponds to $0\leqslant E_{\mathrm{L}}\leqslant 1/2$. The upper bound of local models is violated once $E>1/2$, and quantum correlations exceed the Tsirelson bound $\mathcal{S}_{\mathrm{Q}}\leqslant2\sqrt{2}$ with $0\leqslant E_{\mathrm{Q}}\leqslant\sqrt{2}/2$, which is attained by performing suitable measurements on pair of qubits in the singlet state.

As indicated in Fig. \ref{Cor}, we apply the correspondence relation between the measurement inputs/outputs in the real space and inputs/outputs in the NL-box models to reproduce Assumptions $1-3$.

\textbf{NL-box Assumption $1$}: The perfect correlation (anti-correlation) conservation law needs to be satisfied by the NL-box models. Without referring to exact form of input spaces, we present reverse (which will be specified later) of Alice's and Bob's inputs $\mathbf{x}$ and $\mathbf{y}$ by $\bar{\mathbf{x}}$ and $\bar{\mathbf{y}}$, respectively. Now, if Alice (Bob) reverses her (his) measurement input, the output changes from $\alpha$ to $\alpha\oplus1$ ($\beta$ to $\beta\oplus1$):
\begin{eqnarray}\label{nl1}
&&\alpha\oplus\beta=\mathbf{x}\cdot\mathbf{y},\hspace{1.7cm}(\alpha\oplus1)\oplus\beta=\bar{\mathbf{x}}\cdot\mathbf{y},\\\nonumber
&&\alpha\oplus(\beta\oplus1)=\mathbf{x}\cdot \bar{\mathbf{y}},\hspace{.7cm}(\alpha\oplus1)\oplus(\beta\oplus1)=\bar{\mathbf{x}}\cdot \bar{\mathbf{y}}. \hspace{.2 cm}(\mathrm{mod}\hspace{.1 cm}2)
\end{eqnarray}
Due to deterministic structure of the NL-box models (\ref{nb}), $\alpha\oplus\beta$ is exactly determined by a given pair of inputs, i.e., $P_{\mathrm{NLB}}(\alpha\oplus\beta|\mathbf{x},\mathbf{y})=0,1, \hspace{.2 cm}\forall \hspace{.1cm}\alpha, \beta, \mathbf{x}, \mathbf{y}$. Therefore, if one of the parties flips her/his input, the $\alpha/\beta$ shall deterministically change to $\alpha\oplus1/\beta\oplus1$. Otherwise, these models are not representing a complete physical model.

One can see that the binary-input--output of the NL-box models \cite{Bra2,Dam} cannot explain the NL-box Assumption $1$. For example, if Alice selects the input $x=0$, the Bob outcome does not change from $\beta$ to $\beta\oplus1$ independently of flipping his input from $y$ to $\bar{y}$. Similarly, in the integer-input--binary output NL-box models \cite{BP1,JM,BP2}, if Alice selects even numbers as input,  $x=2k$, then reversing Bob's input from $y$ to $\bar{y}$ does not affect the Bob's outcome from $\beta$ to $\beta\oplus1$. The correlations (\ref{nl1}) restrict the input space to specific form satisfied by following relations:
\begin{eqnarray}\label{nl12}
&&\mathbf{x}\cdot(\mathbf{y}\oplus \bar{\mathbf{y}})=1,\hspace{.3cm}
(\mathbf{x}\oplus \bar{\mathbf{x}})\cdot \mathbf{y}=1,\hspace{.3cm}\nonumber\\
&&\bar{\mathbf{x}}\cdot\mathbf{y}=\mathbf{x}\cdot \bar{\mathbf{y}}, \hspace{1cm}\mathbf{x}\cdot \mathbf{y}=\bar{\mathbf{x}}\cdot \bar{\mathbf{y}}, \hspace{1 cm}(\mathrm{mod}\hspace{.1 cm}2) \hspace{.5cm}\forall \hspace{.2cm}\mathbf{x}, \mathbf{y}.
\end{eqnarray}

To include the relations (\ref{nl12}) in the NL-box models, we increase the dimension of the inputs spaces to include strings of $n$ binary digits. We refer to  this version of the perfect NL-box model as their extended version. Here, the input spaces are represented by $\mathbf{x}=(x_{1},\cdots,x_{n})$ and $\mathbf{y}=(y_{1},\cdots,y_{n})$ with $x_{i},y_{j}\in\{0,1\}$, in which the outputs are binary numbers, $\alpha,\beta\in\{0,1\}$. In addition, the function $f(\mathbf{x},\mathbf{y})$ in Eq. (\ref{nb}) becomes the inner product function of inputs $f(\mathbf{x},\mathbf{y})=\mathrm{IP}_{\mathrm{n}}(\mathbf{x}, \mathbf{y})=\mathbf{x}\cdot \mathbf{y}=x_{1}y_{1}\oplus\cdots\oplus x_{n}y_{n}$,
where $x_{i}y_{i}$ is the usual product of the input elements and the summation is calculated modulo $2$. Without loss of generality, we assume that the dimension of the input spaces to be even ($n=2k$) and the number of characters $1$ ($0$) to be odd,
e.g., $(1,0,1,1,0,0)$ for $n=6$.
We represent input spaces by $\{\mathcal{X}\}$ and $\{\mathcal{Y}\}$, which are closed under the Boolean summation of any odd number of inputs.
Moreover, we define the reverse of inputs by $\bar{\mathbf{x}}=\mathbf{1}\oplus\mathbf{x}$ and $\bar{\mathbf{y}}$ as $\mathbf{1}\oplus\mathbf{y}$, where $\mathbf{1}$ is the string of $n$ $1$'s, $\mathbf{1}=(1,1,\cdots,1)$. These nonlocal models satisfy the conditions (\ref{nl12}) for all strings of the input spaces with $n\geqslant6$. For example, $(\mathbf{x}\oplus \bar{\mathbf{x}})\cdot\mathbf{y}=\mathbf{1}\cdot\mathbf{y}=\bigoplus_{i=1}^{n}y_{i}\equiv1, \hspace{.1cm}\forall \hspace{.1cm}\mathbf{x}, \mathbf{y}$.


As we will see latter in this article, $n$ dose not paly any role in the derivation of symmetry conditions. This shows that any finite set of measurement inputs in the real space can be represented in the NL-box framework by suitable choice of $n$. This correspondence enables Alice and Bob to freely and independently select the measurement inputs from the input spaces $\{\mathcal{X}\}$ and $\{\mathcal{Y}\}$, respectively.


\textbf{NL-box Assumption $2$}: The set of transformation operators $F\in\left\{\mathcal{F}\right\}$ satisfies the group structure.
The Boolean transformations $F$ transform multivariate polynomials over $n$-tuple of the inputs to another $n$-tuple of the inputs $F: \{0,1\}^{n}\rightarrow \{0,1\}^{n}$. The $n$-variable functions $F$ can be viewed as a vector of $n$ functions defined as the component functions, i.e., $F(\mathbf{x})=\left( f^{1}(\mathbf{x}),\cdots,f^{l}(\mathbf{x}),\cdots,f^{n}(\mathbf{x})\right)$. The component functions $f^{l}: \{0,1\}^{n}\rightarrow \{0,1\}$ can be represented in the standard form which is called the algebraic normal form \cite{ANF,Re} which is comprised of the summation of all distinct $k$-th order products of the variables $ 1\leqslant k \leqslant n$.

\textbf{NL-box Assumption $3$}: The observable quantities are invariant under the Boolean symmetry transformations $F$. Here after we refer to the larger symmetry group which includes both the boolean symmetry transformation as well as the space isotropy/homogenouty transformations as the \textit{fundamental symmetry group}. Similar to Assumption $3$, the observable quantities in the NL-box models are the probability distributions, the correlation functions, and the upper bound of violation of the CHSH inequality.




We now apply the NL-box Assumption $3$ to the NL-box correlation functions (\ref{nb}),
\begin{eqnarray}\label{inv}
\alpha\oplus\beta=\mathbf{x}\cdot \mathbf{y}=F(\mathbf{x})\cdot F(\mathbf{y}).
\hspace{2cm}(\mathrm{mod}\hspace{.1cm} 2)
\end{eqnarray}

After imposing the NL-box Assumptions $2$ and $3$ on the extended version of the NL-box correlation function, the Boolean transformations $F$ reduce to the simple form $F(\mathbf{x})=\mathbf{R}\mathbf{x}\oplus\mathbf{T}$, where $\mathbf{R}$ and $\mathbf{T}$ are $n\times n$ and $n\times 1$ matrices, respectively, with elements $\mathbf{R}_{ij},\mathbf{T}_{k1}\in \{0,1\}$. In order for the correlation functions to remain invariant under symmetries $F$, the following conditions shall be respected by the inputs and the transformation matrices:
\begin{eqnarray}\label{con}
&&F(\mathbf{x})\cdot F(\mathbf{y})
=\mathbf{x}^{t}\mathbf{R}^{t}\mathbf{R}\mathbf{y}\oplus\mathbf{T}\cdot[\mathbf{R}(\mathbf{x}\oplus\mathbf{y})\oplus\mathbf{T}]=\mathbf{x}\cdot \mathbf{y}\nonumber\\
&&\mathbf{R}^{t}\mathbf{R}=\mathbf{I}, \hspace{.5cm}\mathbf{T}\cdot[\mathbf{R}(\mathbf{x}\oplus\mathbf{y})\oplus\mathbf{T}]=0,
\hspace{.3cm}(\mathrm{mod}\hspace{.3cm} 2),\forall\hspace{.1cm}\mathbf{x}, \mathbf{y}, \mathbf{R}, \mathbf{T},
\end{eqnarray}
where the superscript $t$ represents transposition and $\mathbf{I}$ is the identity matrix ($\mathbf{I}_{ij}=\delta_{ij}$). We refer to  Eq. (\ref{con}) as the \textit{symmetry condition}. This condition divides the set of the inputs and the transformation matrices $\{\mathrm{\mathcal{H}}\}$ in to $l$ subsets $\{\mathcal{H}_{m}\}\equiv\left\{\{\mathcal{X}_{m}\},\{\mathcal{Y}_{m}\},\{\mathcal{R}_{m}\},\{\mathcal{T}_{m}\}\right\}, m=1,\cdots,l$ such that every subset respects the symmetry condition (\ref{con}).


In each subset, say the $m$-th, we select a pair of inputs $(\mathbf{x}_{1}, \mathbf{x}_{2})$ in Alice's part and another pair $(\mathbf{y}_{1}, \mathbf{y}_{2})$ in Bob's part, and have $(\mathbf{x}_{1}\oplus\mathbf{x}_{2})\cdot(\mathbf{y}_{1}\oplus\mathbf{y}_{2})=0$, deities of calculations is given in the appendixes. To make a closer connection with the CHSH parameter $\mathcal{S}$ (\ref{CHSH}), we define a parameter $\mathcal{W}$ as
\begin{eqnarray}\label{cond3}
\mathcal{W}=\frac{1}{4}\left[1-(-1)^{\mathbf{x}_{1}\cdot(\mathbf{y}_{1}\oplus\mathbf{y}_{2})}\right]
\left[1+(-1)^{\mathbf{x}_{2}\cdot(\mathbf{y}_{1}\oplus\mathbf{y}_{2})}\right],
\end{eqnarray}
which measures the value of the symmetry of the space or nonlocality in the NL-box model framework. This parameter vanishes ($\mathcal{W}=0$) for every subset $\{\mathcal{H}_{m}\},\hspace{.1cm} \forall\hspace{.1cm} m$. In another perspective, the four pairs of the measurement inputs $(\mathbf{x}_{i}, \mathbf{y}_{j}),\hspace{.1cm}i,j=1,2$, violate the CHSH inequality at the maximum value of $4$, if the inputs satisfy $\mathbf{x}_{1}\cdot\mathbf{y}_{1}=\mathbf{x}_{1}\cdot\mathbf{y}_{2}\oplus1$ and $\mathbf{x}_{2}\cdot\mathbf{y}_{1}=\mathbf{x}_{2}\cdot\mathbf{y}_{2}$. For these inputs we have $\mathcal{W}=1$, which indicates the maximum value for nonlocality. Thus, in the NL-box framework, the representation of two experimental observable phenomena, the maximum violation of the CHSH inequality ($\mathcal{W}=1$) and the symmetry condition of the space ($\mathcal{W}=0$), are incompatible. This is one of the main results of this work.

It is a natural expectation which any \textit{complete} physical model should simultaneously display both nonlocal effects \cite{Han,Sha, Ant,Win,Pan,Sup} and the isotropy/homogeneity of the space \cite{MM,LoM1,LoMM}, as two fundamental physical properties of the nature. However, as we argue there exists an inconsistency between these properties in the NL-box framework. To remedy this issue, we consider an imperfect version of the NL-box models (see Refs. \cite{BP2,JM,BP1,Mas,IC}) and will show that the mentioned fundamental inconsistency is lifted exactly at the Tsirelson bound \cite{Tsi}. This is a threshold which retrieves the internal consistency to the imperfect NL-box models. Moreover, we will go beyond the standard definition of ``imperfection" in the NL-box models and will analytically derive the Tsirelson bound as the natural consequence of the NL-box Assumption $3$ without reference to or need for the probabilistic interpretation of the NL-box models.

As pointed out in the NL-box Assumption $1$, the perfect NL-box models are deterministic with the probability distribution
$P_{\text{NLB}}(\alpha\oplus\beta|\mathbf{x},\mathbf{y})=0,1$. The imperfections in the NL-box models are usually parameterized by adding local- or white-noise randomization procedures to the measurement outputs \cite{BP2,JM,BP1,Mas,IC}.
It is not difficult to see that we can instead use the following probability distribution,
\begin{equation}\label{pnb}
P_{\mathrm{imp-NLB}}(\alpha,\beta|\mathbf{x},\mathbf{y})= \left\{
\begin{array}{ll}
p/2 &: \alpha\oplus\beta=\mathbf{x}\cdot\mathbf{y} \hspace{.2 cm}(\mathrm{mod}\hspace{.1 cm}2)  \\
(1-p)/2 &:\text{ otherwise  \hspace{.2cm} $0\leqslant p\leqslant1$,}
\end{array}        \right.
\end{equation}
rather than the some introduced in Refs. \cite{BP2,JM,BP1,Mas,IC} to represent such imperfect NL-box models, where $p=1$ represents the perfect the NL-box models (\ref{nb}).


In addition, to have a complete description of imperfect NL-box models, the symmetry Assumptions $1-3$ shall be revisited for this new scenario. If we impose the NL-box Assumptions $1$ and $2$ on the imperfect NL-box models, we find that the inputs respect Eqs. (\ref{nl12}). Therefore, the inputs can be represented by the $n$-tuples $\mathbf{x} $ and $\mathbf{y}$ for all values of the correlation $E$, as indicated in the previous discussions. Moreover, as we show later in the appendixes, the symmetry transformations $F$ here also reduce to the standard form $F(\mathbf{x})=\mathbf{R}\mathbf{x}\oplus\mathbf{T}$.

\begin{figure}
\centering
\includegraphics[scale=.5]{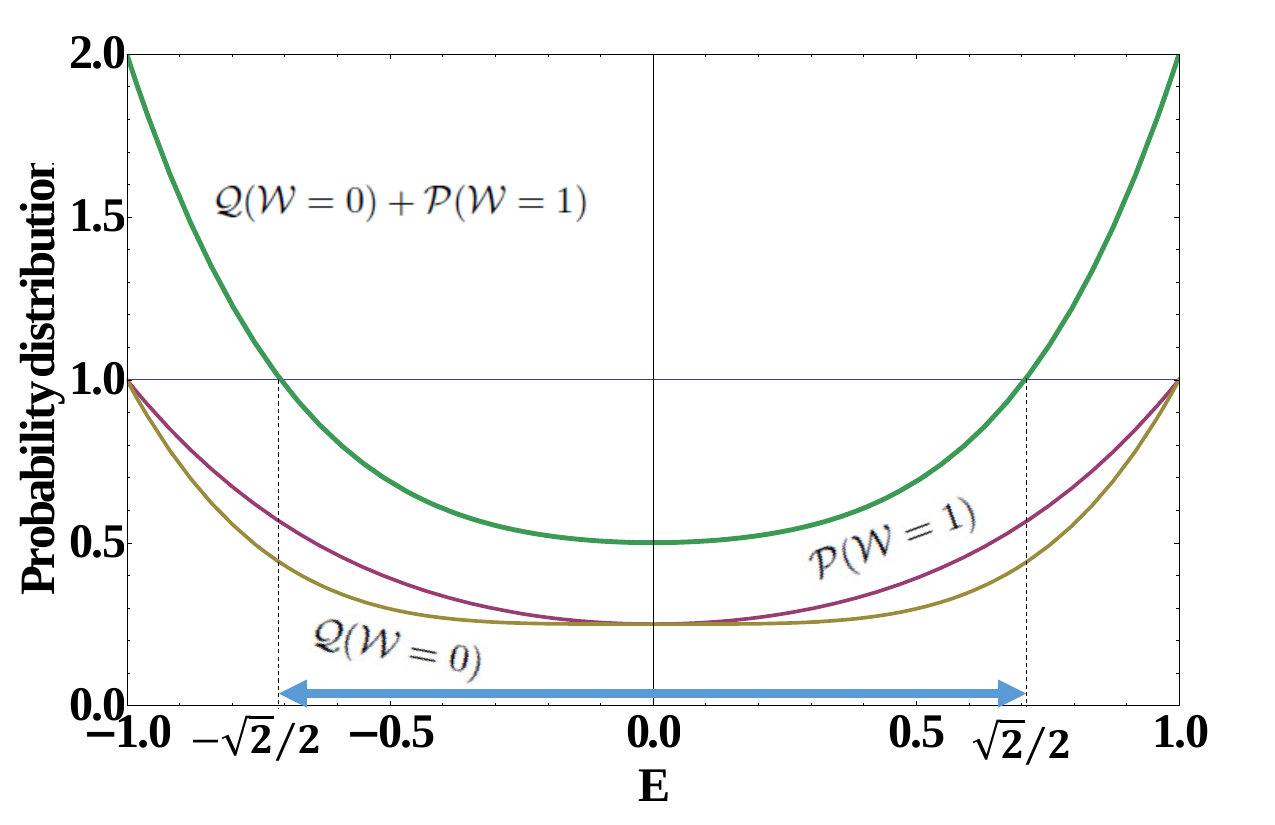}
\caption{
The probability distributions for the imperfect version of the NL-box models.
We show later in appendices that probability for $\mathcal{W}$ to vanish is $\mathcal{Q}(\mathcal{W}=0)=(1/4)[1+3E^{4}]$ and probability for $\mathcal{W}$ to be unity is given by $\mathcal{P}(\mathcal{W}=1)=(1/4)[1+E^{2}]^{2}$. This figure depicts $\mathcal{Q}(\mathcal{W}=0)$ (brown), $\mathcal{P}(\mathcal{W}=1)$ (red) and $\mathcal{Q}(\mathcal{W}=0)+\mathcal{P}(\mathcal{W}=1)$ (green) as function of the correlation $E$, which shows the symmetry condition and nonlocality are mutually consistent in the interval $-\frac{\sqrt{2}}{2}\leqslant E\leqslant\frac{\sqrt{2}}{2}$ (blue interval). It exactly coincides with the Tsirelson bound as the threshold for internal consistency of NL-box models. }\label{P+Q}
\end{figure}

We now redescribe the NL-box Assumption $3$ in the imperfect NL-box framework: the probability of given correct outputs ($p$) must
remain invariant under the symmetry group transformations ($F$). That is, if
$\alpha\oplus\beta=\mathbf{x}\cdot\mathbf{y}=c$, with probability $p$, we request that
\begin{eqnarray*}\label{ucon}
\alpha'\oplus\beta'=\mathbf{x}\cdot\mathbf{y}\oplus[\mathbf{R}(\mathbf{x}\oplus\mathbf{y})]\cdot\mathbf{T}=c,
\end{eqnarray*}
with the same probability $p$, for all $\mathbf{R}, \mathbf{T}$ and $c\in \{0,1\}$ as a correct output.


If two pairs of correlations mutually give the correct outputs ($\alpha\oplus\beta=\alpha'\oplus\beta'=c$) or mutually give the incorrect outputs ($\alpha\oplus\beta=\alpha'\oplus\beta'=c\oplus1$), the symmetry condition (\ref{con}) will be respected with probabilities $p^{2}$ and $(1-p)^{2}$, respectively. Hence the symmetry condition is given with probability $p^{2}+(1-p)^{2}$.
We now apply the same procedure to four pairs of input settings $(\mathbf{x}_{i},\mathbf{y}_{j}),\hspace{.1cm} i,j=1,2$ to determine the symmetry parameter $\mathcal{W}=0$. After some algebra, the total probability becomes $\mathcal{Q}(\mathcal{W}=0)=[p^{2}+(1-p)^{2}]^{3}+[1-p^{2}-(1-p)^{2}]^{3}=(1/4)[1+3E^{4}]$. On the other hand, for the same four pairs of input settings, the CHSH inequality is violated at the maximum value $4$ ($\mathcal{W}=1$) with the total probability $\mathcal{P}(\mathcal{W}=1)=p^{4}+(1-p)^{4}+2p^{2}(1-p)^{2}=(1/4)[1+E^{2}]^{2}$, as it will be shown later in the appendixes. Here, we have used different probability symbols $\mathcal{P}$ and $\mathcal{Q}$ to stress that they are derived from independent physical facts and methods, the former from nonlocal property of the imperfect NL-box correlation function (\ref{pnb}) and the latter from the symmetry condition (\ref{con}). These probability distributions reduce to our earlier outcome for the perfect NL-box models ($E=1$) which indicates that the nonlocality ($\mathcal{P}_{E=1}(\mathcal{W}=1)=1$) and symmetry condition ($\mathcal{Q}_{E=1}(\mathcal{W}=0)=1$) cannot be simultaneously compatible, see Fig. \ref{P+Q}.

In order to have consistent NL-box models, it is logical and sufficient that the two incompatible/exclusive properties $\mathcal{W}=1$ and $\mathcal{W}=0$ cannot be detected simultaneously. The exclusive events were mathematically explained in one of Boole's axioms which state that the sum of the probability of mutually exclusive events can never be greater than $1$ \cite{Bol}. In particular, similar to flipping a coin in which the sum of the probabilities of ``heads" and ``tails" is always equal to or less than $1$, regardless of the way we calculate these probabilities, the sum of these $\mathcal{W}$ probabilities should also satisfy $\mathcal{Q}(\mathcal{W}=0)+\mathcal{P}(\mathcal{W}=1)\leqslant1$. This logical inequality was recalled in probability theory as the inclusion-exclusion principle \cite{IE} and was appeared in various physical phenomena, e.g., the wave-particle duality \cite{Eng}.

As depicted in Fig. \ref{P+Q}, the probability distributions are mutually consistent in the interval $-\frac{\sqrt{2}}{2}\leqslant E\leqslant\frac{\sqrt{2}}{2}$ (the very Tsirelson bound), which exactly coincides with experimental observations \cite{Han,Sha, Ant,Win,Pan,Sup} and predictions of quantum physics \cite{Tsi}. Interestingly, unlike the information-theoretic principles \cite{Bra2,IC,ML,ML1,LO} where the dimension of input/output spaces or the number of NL-boxes play a crucial role in the derivation of the Tsirelson bound, we have derived this bound for arbitrary-input and binary-output NL-box models without reference to the number of NL-boxes. Moreover, in the information-theoretic approaches \cite{IC,ML1,LO}, the correlation function $E$ asymptotically/approximately converges to $\sqrt{2}/2$ as the number of NL-boxes increases; whereas in our study we have determined this value \textit{exactly} (with no approximation).

We remind that in our scenario the nonlocality and symmetry condition are assumed as \textit{independent} features of a physical system, which could yield conflicting results. Nevertheless, nature behaves such that these features indeed coexist and are always consistent. Thus, it is naturally expected that these features should appear inherently and consistently in any complete physical model, e.g., by manifesting as the invariance of observable quantities under the fundamental symmetry group transformations. We also note that the unpredictability of measurement outputs is one of the essential assumptions of the information-theoretic scenarios \cite{Bra2,IC,ML,ML1,LO}, which is manifested in the imperfect NL-box Eq. (\ref{pnb}). It is then of fundamental relevance and importance to investigate whether imperfect effects in the NL-box models are really necessary to derive the Tsirelson bound.






Now, we want to prove that it is possible to put both nonlocality and symmetry condition in a coherent framework and analytically derive the Tsirelson bound as the natural outcome of the NL-box Assumption $3$ without need to invoke imperfection effects within the NL-box models. Since the CHSH parameters are not invariant under the symmetry group elements $F\in\left\{\mathcal{F}\right\}$, $\mathcal{S}(\mathbf{x}_{1},\mathbf{x}_{2},\mathbf{y}_{1},\mathbf{y}_{2})\neq \mathcal{S}(F(\mathbf{x}_{1}),F(\mathbf{x}_{2}),F(\mathbf{y}_{1}),F(\mathbf{y}_{2}))
\equiv\mathcal{S}(\mathbf{R}^{t}\mathbf{T},\mathbf{x}_{1},\mathbf{x}_{2},\mathbf{y}_{1},\mathbf{y}_{2}),\, \forall\, \mathbf{R},\, \mathbf{T}$ (hence they are not observable quantities), rather than deriving an invariant form for them, we determine the upper bound on $\mathcal{S}$, which has been widely verified in experiments \cite{Han,Sha, Ant,Win,Pan,Sup}.





To this end, we employ variance of the CHSH parameter
$(\triangle\mathcal{S})^{2}=\langle\mathcal{S}^{2}\rangle-\langle\mathcal{S}\rangle^{2}\geqslant0$ and the determine maximum $\langle\mathcal{S}^{2}\rangle_{\mathrm{max}}=\sum_{\mathbf{R}^{t}\mathbf{T}}
D(\mathbf{R}^{t}\mathbf{T})\mathcal{S}_{\mathrm{max}}^{2}(\mathbf{R}^{t}\mathbf{T},\mathbf{x}_{1},\mathbf{x}_{2},\mathbf{y}_{1},\mathbf{y}_{2})$, where the expectation value is taken on all symmetry transformations $\mathbf{R}^{t}\mathbf{T}$ with the same probability distribution $D(\mathbf{R}^{t}\mathbf{T})=1/2^{n-1},\,\forall\,\mathbf{R},\,\mathbf{T}$. In fact, this uniformity of the probability distribution $D$ indicates that the CHSH parameters $\mathcal{S}(\mathbf{R}^{t}\mathbf{T},\mathbf{x}_{1},\mathbf{x}_{2},\mathbf{y}_{1},$ $\mathbf{y}_{2})$ are equivalent for all $\mathbf{R}^{t}\mathbf{T}$. After simple algebra, we obtain $\langle\mathcal{S}^{2}\rangle_{\mathrm{max}}=8$, which is the very Tsirelson bound $|\langle\mathcal{S}\rangle|\leqslant 2\sqrt{2}$. This result exactly distinguishes \textit{physical} correlations from \textit{nonphysical} ones in the sense that every NL-box with stronger than quantum correlations violate the symmetry condition (\ref{con}).


The physical/invariant form of the observable variables---here the correlation function---can be generated by employing linear combination of the perfect NL-box correlations with appropriate coefficients (probabilities) which is conditioned to be invariant under the symmetry group transformations $\left\{\mathcal{F}\right\}$. For the NL-box models (\ref{nb}), it is expected that the correlation function be equal to the quantum correlation function for a pair of qubits in the singlet state. We have relegated the details in the appendixes.

The impact of the symmetry group structure on derivation of the upper bound of the CHSH parameter (the Tsirelson bound) has far-reaching implications on fundamental aspects of various physical theories, from quantum physics to nonlocal-realistic models, e.g. quantum theory of motion \cite{QTM}. Due to Ref. \cite{OW}, the degree of nonlocality of any theory can be characterized by the strength of the fine-grained uncertainty relation (a new measure for determining degree of uncertainty/certainty in any theory) and the strength of the steerability of the theory \cite{Ste1,Ste2}. Now, we apply our scenario to calculate the upper bound of the fine-grained uncertainty relation which, helps better understand the implication of the flat-space symmetry group structure on the uncertainty/certainty and probablilistic/deterministic interpretation of the measurement outcomes in quantum/nonlocal-realistic theory.

The fine-grained uncertainty relation, for a fixed set of measurement inputs, consists of a set of inequalities
$\left\{P^{\mathrm{cer}}(\sigma_{\alpha, \mathbf{x}}, \beta)=\sum_{\mathbf{y}}p(\mathbf{y})p(\beta|\mathbf{y})_{\sigma_{\alpha, \mathbf{x}}}\leqslant \zeta_{\sigma_{\alpha, \mathbf{x}}, \beta}\,|\,\forall\, \alpha,\, \beta,\, \mathbf{x}\right\}$,
which involve the probability of detecting a specific output $\beta$ when the system is prepared in the state $\sigma_{\alpha, \mathbf{x}}$ (obtained from a post-selected measurement of value $\alpha$ which is performed with  the measurement input $\mathbf{x}$) and next is measured on the direction $\mathbf{y}\in \{\mathcal{Y}\}$ chosen from the distribution $\{p(\mathbf{y})\}$. The maximum amount of the uncertainty (the ``strength") $\zeta_{\sigma_{\alpha, \mathbf{x}}, \beta}$ is obtained by maximizing $P^{\mathrm{cer}}(\sigma_{\alpha, \mathbf{x}}, \beta)$ over all initial states which are allowed in a given theory. Although the upper bound of the CHSH parameter already characterizes the value of $\zeta_{\sigma_{\alpha, \mathbf{x}}, \beta}$ \cite{OW}, we show that if we alternatively calculate the upper bound $\zeta_{\sigma_{\alpha, \mathbf{x}}, \beta}$ by directly imposing the symmetry-group transformations on $P^{\mathrm{cer}}$, we obtain the same value to this symmetry-only consideration $\zeta_{\sigma_{\alpha, \mathbf{x}}, \beta}=1/2+1/(2\sqrt{2})$. We have relegated the details in the appendixes. This implies that the uncertainty $P^{\mathrm{cer}}(\sigma_{\alpha, \mathbf{x}}, \beta)\leqslant 1/2+1/(2\sqrt{2})$ can be understood as a \textit{universal} relation which is respected by every physical theory in the flat space. Moreover and more interestingly, for this result of the fine-grained uncertainty relation that ``one cannot certainly predict a measurement outcome for all simultaneous measurements whenever the $\zeta_{\sigma_{\alpha, \mathbf{x}}, \beta}<1$" \cite{OW}, the imperfections in the NL-box models, i.e., the probabilistic interpretation of the measurement outcomes, is an emergent consequence of the symmetries of space. Remarkably, the unpredictability of measurement outcomes is indirectly established in generating the physical/invariant form of correlation functions which they are given by probabilistic combination of the perfect NL-box correlations. In this terminology, quantum physics is a probabilistic and optimal nonlocal model which respects the flat-space symmetries.

Unifying general relativity and quantum physics as two cornerstones of modern physics, is still elusive despite long efforts. The former is recognized by, e.g., quantum entanglement, existence of nonlocality in nature, probabilistic nature of observable quantities, and the Heisenberg uncertainty relation, while the latter is based on equivalence of the structure of the spacetime and the mass distribution. Recently, a fundamental connection was proposed between the spacetime and quantum entanglement, in which the spacetime is an emergent property of quantum entanglement \cite{Ra}. On the other hand, in our work in this paper we have demonstrated the crucial impact of the symmetry-group structure of the flat space on determining the maximum value of nonlocality in any theory. Our result characterizes the upper bound of the fine-grained uncertainty relation \cite{OW} (as an extended version of the Heisenberg uncertainty principle) and also implies that the probabilistic interpretation of measurement outcomes is considered as emergent property of the flat space symmetry-group structure. This shows how the structure of spacetime, as a continuous, deterministic, and causal structure, creates probabilistic and uncertain properties for any consistent theory.

\vspace{.1cm}

\textit{Acknowledgements}.---We would like to thank A. T. Rezakhani for interesting discussions and valuable comments.














\newpage


\vspace{.5cm}

\centerline{\large \bf {Appendixes}}


\vspace{.5cm}

\noindent \textbf{\ref{I}.~~\hyperref[I]{The isotropy and the homogeneity principles in the flat space}}\dotfill\textbf{\pageref{I}}

\noindent \textbf{\ref{IA}.~~\hyperref[IA]{Mathematical definition of the isotropy and the homogeneity in classical physics and quantum mechanics}}\dotfill\textbf{\pageref{IA}}

\noindent \textbf{\ref{IB}.~~\hyperref[IB]{Representation of the flat-space isotropy and homogeneity principles in the Bell-type experiments,
a brief review}}\dotfill\textbf{\pageref{IB}}

\noindent \textbf{\ref{II}.~~\hyperref[II]{Postquantum models}}\dotfill\textbf{\pageref{II}}

\noindent \textbf{\ref{IIA}.~~\hyperref[IIA]{A brief review of Postquantum models}}\dotfill\textbf{\pageref{IIA}}

\noindent \textbf{\ref{IIB}.~~\hyperref[IIB]{Every element of the physical reality must have a counterpart in complete physical model}}\dotfill\textbf{\pageref{IIB}}

\noindent \textbf{\ref{IIC}.~~\hyperref[IIC]{Functions and mathematical symmetry group structure in Boolean algebra}}\dotfill\textbf{\pageref{IIC}}

\noindent \textbf{\ref{IID}.~~\hyperref[IID]{The symmetry group assumptions in the perfect NL-box framework}}\dotfill\textbf{\pageref{IID}}

\noindent \textbf{\ref{IIE}.~~\hyperref[IIE]{Derivation of explicit form of symmetry group transform $F$}}\dotfill\textbf{\pageref{IIE}}

\noindent \textbf{\ref{IIF}.~~\hyperref[IIF]{Derivation of symmetry parameter $\mathcal{W}=0$ in the perfect NL-box models}}\dotfill\textbf{\pageref{IIF}}

\noindent \textbf{\ref{III}.~~\hyperref[III]{The probabilistic version of NL-box model}}\dotfill\textbf{\pageref{III}}

\noindent \textbf{\ref{IIIA}.~~\hyperref[IIIA]{The NL-box Assumption $1, 2$ and $3$ in the imperfect NL-box models}}\dotfill\textbf{\pageref{IIIA}}

\noindent \textbf{\ref{IIIB}.~~\hyperref[IIIB]{Derivation of the Tsirelson bound in the imperfect NL-box models}}\dotfill\textbf{\pageref{IIIB}}

\noindent \textbf{\ref{IV}.~~\hyperref[IV]{Probability and uncertainty as emergent outcomes of symmetry group assumptions in the NL-box framework}}\dotfill\textbf{\pageref{IV}}

\noindent \textbf{\ref{IVA}.~~\hyperref[IVA]{The Tsirelson bound as a natural outcome of fundamental symmetry group transformations}}\dotfill\textbf{\pageref{IVA}}

\noindent \textbf{\ref{IVB}.~~\hyperref[IVB]{The symmetry group structure determines the upper bound of the fine-grained uncertainty relation}}\dotfill\textbf{\pageref{IVB}}

\noindent \textbf{\ref{IVC}.~~\hyperref[IVC]{The invariant form of the NL-box correlation function}}\dotfill\textbf{\pageref{IVC}}

\noindent \textbf{\ref{IVD}.~~\hyperref[IVD]{Multipartite and high-dimension post-quantum models}}\dotfill\textbf{\pageref{IVD}}

\noindent \textbf{\ref{V}.~~\hyperref[V]{An explicit representation of measurement input spaces and transformation matrices $\mathcal{R}$ and $\mathcal{T}$ in $n=6$}}\dotfill\textbf{\pageref{V}}

\vspace{.5cm}

In this appendix, we formally state and prove the results mentioned in the main text. It is organized as follows. In Sec. \textbf{\ref{I}}, we briefly review definition of the isotropy and the homogeneity principles in the classical physics and relativity theory. Moreover, we describe the assumptions which are concluded from these axioms and their representation in the quantum mechanics. In Sec. \textbf{\ref{II}}, we review postquantum models and study mathematical group structure in the Boolean algebra. We derive the symmetry condition and symmetry parameter as criteria of symmetry in the perfect NL-box models. In Sec. \textbf{\ref{III}}, we consider the probabilistic version of nonlocal box correlation functions and derive the Tsirelson bound as threshold for internal consistency of nonlocal models. In Sec. \textbf{\ref{IV}}, we determine the upper bounds of the CHSH parameter (the Tsirelson bound) and the fine-grained uncertainty relation as natural outcome of symmetry group structure. Finally, in Sec. \textbf{\ref{V}}, we present a explicit form of inputs,outputs space and symmetry transformation matrices for $n=6$. Besides, there are some points which are not directly related to main text message, meanwhile, they shed light to better understanding our scenario and answer to some questions which may arise.

\section{The isotropy and the homogeneity principles in the flat space}\label{I}

In this section, we briefly review presentation of the isotropy and the homogeneity principles various physical theories, include classical physics, relativity theory, and quantum mechanics frameworks. Afterwards, we illustrate these principles in wide range of the Bell experiments.

\subsection{Mathematical definition of the isotropy and the homogeneity in classical physics and quantum mechanics}\label{IA}

In classical physics, the isotropy and homogeneity principles of the flat space are manifested at the definition of the absolute space. Newton believed to reality of three-dimensional Euclidean space with the geometrical structure of $\mathrm{\mathbb{E}^{3}}$ (isotropic and homogenous) \cite{Mau}. He introduced absolute space and time to define velocity \cite{Bar}. The Euclidean group $\mathrm{E(\emph{n})}$ comprises all translations, rotations, and reflections of $\mathrm{\mathbb{E}^{\emph{n}}}$ and arbitrary finite combinations of them \cite{Ec}. The Euclidean group can be seen as the symmetry group of the space itself, and contains the group of symmetries of any figure (subset) of that space. The Euclidean group has two subgroups, the orthogonal group $\mathrm{O(\emph{n})}$ and the translational group $\mathrm{T(\emph{n})}$. In particular, $\mathrm{E(\emph{n})}$ is the semidirect product of $\mathrm{O(\emph{n})}$ extended by $\mathrm{T(\emph{n})}$, $\mathrm{E(\emph{n})=O(\emph{n})\times T(\emph{n})}$. Any element of $\mathrm{E(\emph{n})}$ is a translation followed by an orthogonal transformation or viceversa. There is a special orthogonal group $\mathrm{SO(\emph{n})}$ which in the matrix representation of group, the determinant of matrices are equal to one. The group $\mathrm{SO(\emph{n})}$ is a subgroup of $\mathrm{O(\emph{n})}$ which shows $\mathrm{E^{+}(\emph{n})=SO(\emph{n})\times T(\emph{n})}$.

In the special relativity and quantum mechanics, the flat spacetime usually is considered as an independent object in which physical events take place. The homogeneity and the isotropy assumptions mean that the results of a measurement of a length or time interval of a specific event shall not depend on location, direction and time of the interval happens to be in our reference frame \cite{SR,SR1}.

In the general relativity, spacetime is a dynamical variable which affected by matter distribution. In modern physical cosmology, the isotropy and homogeneity principles are manifested in the cosmological principle which state that on the largest cosmic scales, the Universe is both homogeneous and isotropic. These characterization of geometrical structure of the three-dimensional space determines the form of the spacetime metric whit constant the scalar curvature. In the spherical coordinate, the spacetime metric takes specific form which called the Friedmann metric:
\begin{eqnarray}\label{}
ds^{2}=-dt^{2}+a^{2}(t)\left[\frac{dr^{2}}{1-kr^{2}}+r^{2}(d\theta^{2}+\sin^{2}\theta d\phi^{2})\right]
\end{eqnarray}
with $k = 0,\pm1$ and the pre-factor $a(t)$ determines the overall scale of the spatial metric. This metric explains a universe that is spatially homogeneous and isotropic at each instant of time. The spatial hyper surfaces of the Friedmann universe have positive, zero and negative curvatures for $k = +1, 0$ and $-1$ respectively. In the other hand, in the flat space a vector will not change if it is parallel transported around a closed curve. In fact, this is a test for the flatness of spacetime and we can expect the change in a vector, on being parallel transported around a closed curve, to provide an insight into the curvature of the spacetime. In a general framework, the change in a vector, when moved around an infinitesimal closed loop, is proportional with the four-indexed object $R^{a}_{bcd}$ which is called the curvature tensor. In the flat spacetime, we can always choose an inertial coordinate system in which the $R^{a}_{bcd}$ vanish at all events. But since this is a tensor equation, it will be valid in any other coordinate system. Thus we find that, in flat spacetime, we must have
$R^{a}_{bcd}=0$ \cite{Pad}.

The isotropy and homogeneity principles is convincingly supported by the all of observable evidences. The well-known experiments of these important cornerstones of physics are the Michelson-Morley type experiments \cite{MM,MM1,MM2} which verify the isotropy of the speed of light, the Hughes-Drever type experiments \cite{LoM1,LoM3,LoM4,LoM5,LoM6,LoM7,LoM8} which test the isotropy of the kinetic energy of particles so that their dispersion relations are independent of the direction of their velocity, and an electronic analogue of a Michelson-Morley experiment \cite{LoMM,LoMM9}.


As pointed out in main text, the isotropy and homogeneity principles can be considered as natural criteria to distinguish quantum physics from NL-box models. We explain in detail some assumptions about the flat space symmetries which every physical theory shall respect. The typical Bell-like experiment setup is defined with two parties (conventionally called Alice and Bob) which share a bipartite system and perform local measurements. Alice measures along $\mathbf{a}$, Bob measures along $\mathbf{b}$, and obtain outcomes $A\in\{+1,-1\}$ and $B\in\{+1,-1\}$, respectively. Here, $\mathbf{a}, \mathbf{b}\in\mathbb{R}^{3}$, $A$ and $B$ indicate whether the spin is pointing along ($+1$) or opposite ($-1$) directions with respect to chooses to measure. We denote $P(A,B|\mathbf{a},\mathbf{b},\sigma)$ the probability of obtaining outcomes $A$ and $B$ when performing measurements $\mathbf{a}$ and $\mathbf{b}$ when the system is prepared in the state $\sigma$. In quantum theory, $\sigma$ is a density matrix (or hidden variable in the realistic interpretation) which is simply an abstract representation of a state for a general physical theory. In what follows, we present the assumptions which are concluded from the isotropy and homogeneity principles, and characterizes the assumptions  in the quantum mechanics scheme:

\vspace{.2cm}

\textbf{Assumption $1$}: The spin perfect correlation (anti-correlation) conservation law. Here, in the Bell-type experiments, the probability to obtain output $A$ ($B$) for the measurement setting $\mathbf{a}$ ($\mathbf{b}$) is equal to the probability to obtaining output $-A$ ($-B$) for the measurement setting $-\mathbf{a}$ ($-\mathbf{b}$), or especially, $P(A, B|\mathbf{a}, \mathbf{b})=P(-A, B|-\mathbf{a}, \mathbf{b})$. Similar relation holds for Bob.

\begin{figure}
\centering
\includegraphics[scale=.7]{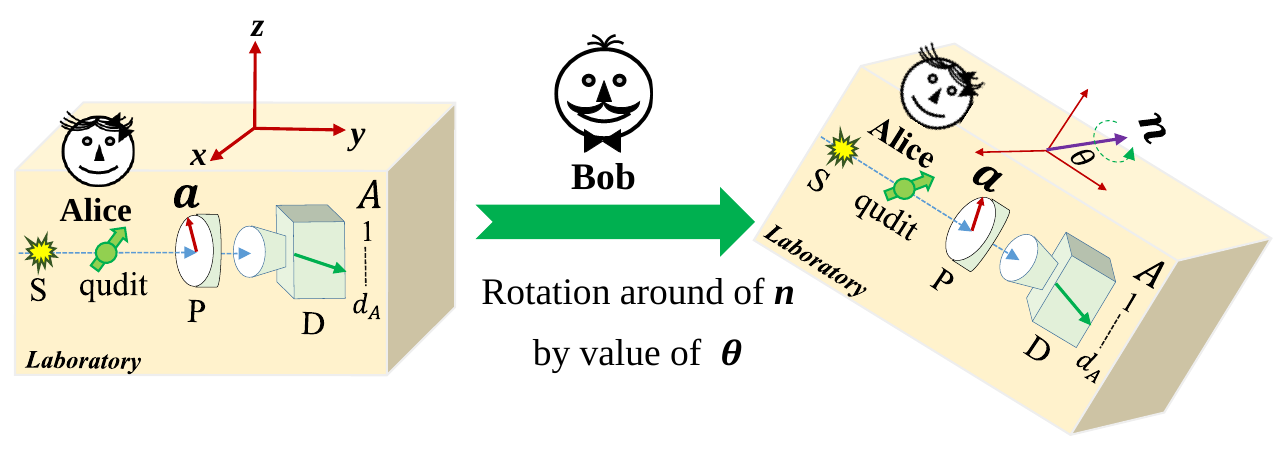}
\caption{Schematic presentation of the rotation symmetry transformations in the real space.
After receiving the qudit to laboratory, internal party (Alice) chooses an arbitrary direction $\mathbf{a}\in\mathcal{R}^{3}$ and obtains one of outcomes $A\in\left\{1,\cdots,d_{A}\right\}$. The external (Bob) can text the isotropy/homogeneity of space by rotation/translation of all elements of each experimental setup, including the source (S), qudit, the polarizer (P), detector (D), and the hole lab by the same amount and asks from Alice about detecting any impact on the value of observable quantities. We represent arbitrary rotation/translation direction of the lab by $\mathbf{n}$ and the value $\theta$.
}\label{RT}
\end{figure}

This assumption can be realized as fundamental logical law which is named \textit{non-contradiction} principle, which state that
``contradictory statements cannot be true in the same sense and at the same time" which is at harmony with the inclusion-exclusion principle \cite{IE}, as indicated in main text. Assumption $1$ can simply be understood in experiment by rotating the Alice reference frame around a direction which is orthogonal to $\mathbf{a}$ by value of $\pi$ in which the Alice new measurement inputs and new outputs will be $-A$ and $-\mathbf{a}$ in the new frame. For an out of lab observer, Bob in Fig. \ref{RT}, Assumption $1$ is a logical law.

In the quantum mechanics framework, the spin perfect correlation (anti-correlation) conservation law is represented by the spin measurement operators in the $\mathbf{a}$ and $-\mathbf{a}$ directions ($\bm{\sigma}\cdot\mathbf{a}$ and $\bm{\sigma}\cdot(-\mathbf{a})$) in Alice site and similar relations for Bob site. There is a one-to-one relation between the eigenvalues of the two operators. Thus, it is redundant to perform two separate measurement because the result of one of them (e.g., $\bm{\sigma}\cdot\mathbf{a}$) will uniquely identify the results of the other one. Explicitly, the experimenter preforms the measurement in the $\mathbf{a}$ direction and concludes that the outcome of the measurement in the $-\mathbf{a}$ direction to be the opposite of $\mathbf{a}$. In the realistic models, Assumption $1$ holds for any value of hidden variables $\lambda\in\{\Lambda\}$, in which $P(A, B|\mathbf{a}, \mathbf{b}, \lambda)=P(-A, B|-\mathbf{a}, \mathbf{b}, \lambda), \, \forall\,\lambda$.

\vspace{.1cm}

\textit{Remark 1}. The perfect correlation (anti-correlation) conservation law is one of fundamental properties of any physical theory. This property can be realized in generalized probabilistic theories \cite{GPT3,GPT4,GPT5,GPT,GPT1,GPT2} framework, a popular approach for studying any physical theory. In this framework a physical procedure is defined by preparing states of system $\omega_{n},\, n=1,\cdots,N$ and measurement operators $e_{m},\, m=1,\cdots,M$ in which the probability a measurement effect $e_{m}$ on the state $\omega_{n}$ is given $p(e_{m}|\omega_{n})=e_{m}(\omega_{n})$. The measurement operators $\{e_{m},\, m=1,\cdots,M\}$ represents a complete set $\sum_{m=1}^{M}e_{m}=u$ which the $u$ is unit effect of measurement operator with $p(u|\omega_{n})=u(\omega_{n})=1,\, \forall\, n$. To realize the perfect correlation (anti-correlation) conservation law in generalized probabilistic theories framework, we define measurement operators in opposite direction with $e_{m}=e_{M+1-m}, \, \forall\,m$, or equivalently $e_{M+1-m}=e_{m}, \, \forall\,m$. Here, two fundamental probability tables of $p(e_{m}|\omega_{n})=e_{m}(\omega_{n})$ and $p(e_{M+1-m}|\omega_{n})=e_{M+1-m}(\omega_{n})$ are equal, $p(e_{m}|\omega_{n})=p(e_{M+1-m}|\omega_{n}),\,\forall\,m,\,n$.



\vspace{.1cm}

In any physical theory/model, the symmetry transformations shall respect axioms of symmetry group structures. In the $3$-dimensional flat space, the isotropy and homogeneity principles have Euclidean symmetry group structure $\mathrm{E^{+}(3)=SO(3)\times}\mathrm{T(3)}$ which comprises all rotations and translations of $\mathrm{\mathbb{E}^{3}}$ and arbitrary finite combinations of them. We characterize this property in Assumption $2$.


\vspace{.1cm}

\textbf{Assumption $2$}: The isotropy and homogeneity symmetry transformations satisfy the Euclidean symmetry group structure $\mathrm{E^{+}(3)=SO(3)\times}\mathrm{T(3)}$. In the three dimension Euclidean space, the rotation and translation operators are represented by rotation matrix $\mathcal{R}(\theta)$ (noncommutative group) and translate vector $\mathcal{T}$ (commutative group), respectively $\mathbf{a}'=R(\theta)\mathbf{a}+\mathcal{T}$. Similar to set of orthogonal rotation matrix $R(\Theta)$, there exists set of rotation operator $U(\Theta)\in\left\{\mathbf{U}\right\}$ which group elements satisfy symmetry group axioms in any physical model. In particular, a non-empty set $\left\{\mathbf{U}\right\}$ together with a binary operation on $\left\{\mathbf{U}\right\}$, which we denote by ``$\star$", respect four group axioms as following:

(\textit{i}) Closed under product of group elements. The combines any two elements $U(\Theta_{i})$ and $U(\Theta_{j})$ of $\left\{\mathbf{U}\right\}$ to form an element of $\left\{\mathbf{U}\right\}$, denoted as $U(\Theta_{i})\star U(\Theta_{j})\in\left\{\mathbf{U}\right\}$,

(\textit{ii}) Existence of an identity transformation. There exists a unique element $\openone\in\left\{\mathbf{U}\right\}$ such that any element $U(\Theta_{i})\in\left\{\mathbf{U}\right\}$ satisfies $U(\Theta_{i}) \star \openone=\openone \star U(\Theta_{i})=U(\Theta_{i})$.



(\textit{iii}) Existence of inverse transformations. For each $U(\Theta_{i})\in\left\{\mathbf{U}\right\}$, there exists an element $\bar{U}(\Theta_{i})\in\left\{\mathbf{U}\right\}$ such that $U(\Theta_{i})\star\bar{U}(\Theta_{i})=\bar{U}(\Theta_{i})\star U(\Theta_{i})=\openone$. For each $U(\Theta_{i})$, the element $\bar{U}(\Theta_{i})$ is unique; it is called the inverse of $U(\Theta_{i})$ and is commonly denoted by $U^{-1}(\Theta_{i})$.

(\textit{iv}) Associativity law. Any three elements of the group elements $U(\Theta_{i}), U(\Theta_{j}), U(\Theta_{k})\in\left\{\mathbf{U}\right\}$ satisfy the following associativity condition: $\left[U(\Theta_{i})\star U(\Theta_{j})\right]\star U(\Theta_{k})
=U(\Theta_{i})\star\left[U(\Theta_{j})\star U(\Theta_{k})\right]$.


In quantum mechanics, the explicit forms of the rotation/translation operators are, respectively, as follows:
\begin{eqnarray}\label{Rot}
&&U_{\mathrm{rotation}}(\theta,\mathbf{k})=\exp\left(\frac{-i\mathbf{J}\cdot\mathbf{k}\theta}{\hbar}\right),\\\nonumber
&&U_{\mathrm{translation}}(\theta,\mathbf{k})=\exp\left(\frac{-i\mathbf{P}\cdot\mathbf{k}\theta}{\hbar}\right),
\end{eqnarray}
where $\mathbf{J}$ and $\mathbf{P}$ are, respectively, the representations of the angular momentum and the liner momentum in three dimensions.

\vspace{.2cm}

\textbf{Assumption $3$}: The laws of physics are covariant and the observable quantities are invariant under symmetry transformations. Assumption $3$ can be realized in experiment by transforming all elements of experiment, systems, sources, channels, or other measurement settings of the lab and detecting any impact on the value of any observable quantities. As depicted in Fig. \ref{RT}, an external observer, for example Bob, can observe transformations, but an internal observer, for example Alice, cannot detect any things.

Assumption $3$ can mathematically be described as: the physical laws transform under a certain representation of the group transformations (where, the rotation and the translation in the flat space) between admissible frames. For example, in the Bell experiment scenario, the probability distributions are invariant under rotation and translation transformations of initial state of system $\sigma$ and measurement inputs $\mathbf{a}_{i}$ and $\mathbf{b}_{j}$
\begin{eqnarray*}\label{TR}
&&P(a,b|\mathbf{a},\mathbf{b},\sigma)=P(a,b|\mathbf{a}',\mathbf{b}',\sigma')\\
&&\sigma\rightarrow\sigma', \hspace{.3cm}
\mathbf{a}_{i}\rightarrow R(\theta)\mathbf{a}_{i}+T,\hspace{.2cm} \mathbf{b}_{j}\rightarrow R(\theta)\mathbf{b}_{i}+T,
\hspace{.1cm}\forall\hspace{.1cm} \mathbf{a}_{i}, \mathbf{b}_{i},\hspace{.1cm} i,j=1,2.
\end{eqnarray*}
where $\sigma'$ is new state of system after transformation operators.

In the quantum mechanics framework, the above transformations can be described by initial state of system by $\rho_{AB}$, measurement operators $M_{a_{i}}$ and $M_{b_{j}}$, the corresponding measurement outputs are represented by $a_{i}$ and $b_{j}$, respectively. Assumption $3$ is explained in quantum physics as:   \begin{eqnarray}\label{inv2}
&&P(a_{i},b_{j}|\mathbf{a},\mathbf{b},\rho_{AB})=Tr\left(M_{a_{i}}M_{b_{j}}\rho_{AB}\right)
=Tr\left(\left[U_{A}(\mathbf{n},\theta)M_{a_{i}}U^{\dagger}_{A}(\mathbf{n},\theta)\right]\right.\\\nonumber
&&\hspace{4.5cm}\left.\left[U_{B}(\mathbf{n},\theta)M_{b_{j}}U^{\dagger}_{B}(\mathbf{n},\theta)\right]
\left[U_{B}(\mathbf{n},\theta)U_{A}(\mathbf{n},\theta)\rho_{AB} U^{\dagger}_{A}(\mathbf{n},\theta)U^{\dagger}_{B}(\mathbf{n},\theta)\right]\right)
\end{eqnarray}
where, $U_{k}(\mathbf{n},\theta),\hspace{.1cm} k=A,B$ was represented in Eq. (\ref{Rot}) for the angular-momentum (spin) or the liner momentum
in three dimensions.

Assumption $3$
was implicitly proved in an electronic analogue of Michelson-Morley experiment \cite{LoMM}. In this experiment, energy dispersion relation of two bounded electrons, inside a pair of calcium ions, are used to detect anisotropy of space. The energy levels of the electrons are prepared at the superposition state which undergo Michelson-Morley type experiment \cite{MM}. The wave packet of electrons split into two parts with different orientations and recombine them after a time evolution. As the Earth rotates,
anisotropy of the absolute spatial orientation of the two parts of the wave packets modify relative phase of the entangled energy levels.
After removing noise effects,
the isotropy of the electron's dispersion relation was verified at the level of one part in $10^{18}$. This result guarantees invariance of the Bell correlation functions by describing the superposition energy levels as entangled Bell state so that the phase of energy shift plays the same role as relative phase of Alice and Bob measurement settings.

\textit{Remark 2}. There is another self-evident phenomenon in real space in which the opposite measurement directions $\mathbf{a}$ and $-\mathbf{a}$ transform to the opposite measurement directions $\mathbf{a}'$ and $-\mathbf{a}'$ under the Euclidean space symmetry group transformations. Moreover, in the Bell type experiments which the source emits two subsystems with total angular-momentum/ spin zero, if Alice and Bob take same/opposite measurement inputs, the parties obtain opposite/equal outputs in the each rounds of experiment. In the hidden variable models, for any one of the hidden variable $\lambda$, we have $P(A=-B|\mathbf{a}=\mathbf{b},\lambda)=1$ and $P(A=B|\mathbf{a}=-\mathbf{b},\lambda)=1, \forall \hspace{.1cm}\lambda$). This properties play crucial roles in characterizing correlation functions and symmetry transformations in abstract mathematical models, e.g., the nonlocal box models, see Sec. \ref{IVC}.






\subsection{Representation of the flat-space isotropy and homogeneity principles in the Bell-type experiments,
a brief review}\label{IB}

For explanation of an experiment with a physical model, we usually make a correspondence relation between physical objects (in experiment) and mathematical symbols (in physical model). In quantum mechanic description of the Bell-type inequalities, physical objects are represented with quantum states (qubits), and operation of external devices by unitary transformations or measurement operators. This correspondence enables us to imply that the isotropy/homogeneity symmetry transformations should have counterpart in experimental setting.

Realization of the Bell-types inequalities in experiment have been addressed in diverse platforms and experimental settings \cite{Sha,Ant,Pan,Han,Win,Sup}. In realization of the CHSH inequality by entangled photons \cite{Sha,Ant,Pan,Ph,Ph1}, the qubits are represented by photon polarization state and the external devices are optical devices such as quarter-wave/half-wave plates which control photon polarization direction. In another type of realization of the CHSH inequality in experiment, the qubits are represented by two-level trapped ions in the ground and excited states \cite{Han,Win,NV,NV1,NV2,NV3,NV4}, for example electronic spin in the center of a nitrogen vacancy \cite{Han} or a single rubidium-$87$ atom in an optical dipole trap \cite{Win,NV}. In these experiments, the external fields, such as microwave pulses \cite{Han} or the polarization of lasers \cite{Win,NV}, control and rotate the state of the atomic systems. Similar to schemes \cite{Han,Win}, in presentation of qubit with superconducting circuit, the qubit is rotated and controlled by external microwave pulses \cite{Sup1,Sup2}.

In all of experimental test of the Bell-type inequalities \cite{Sha,Ant,Pan,Han,Win,Sup}, there is one-to-one correspondence between system symbols in physical model (for example, state vectors in quantum mechanics formalism) and experimental objects (polarization of photon, atomic state, state of superconducting circuit). Similar one-to-one correspondence holds between action of mathematical operations (unitary transformations and measurement operators in quantum mechanics formalism) on the system symbols and controlling/experimenting devices (such as optical devices, lasers, microwave pulses, single-photon detector, spin-dependent fluorescence, gated microwave tone). Then, it is logical that the isotropy/homogeneity principle should have a counterpart in any physical theory, e.g., Eq. (\ref{inv2}) represent the isotropy/homogeneity principle in quantum mechanics framework. These transformations are realized by rotating/translating physical systems (qubits) by external devices (optical devices, lasers, microwave pulses) and rotating/translating experimental apparatus (sources, channels, or other measurement settings of the lab) by experimenters.




\section{postquantum models}\label{II}

Quantum nonlocality is at the heart of wide range phenomena, from fundamental quantum paradoxes \cite{EPR,Cat1,Cat2,WF} to device-independent quantum information processing \cite{Ran1,Ran2,Di1,Di2,Di3,Di5,Di6,Di7} and the game theory \cite{Gam}. Although quantum nonlocality ($\mathcal{Q}$) is stronger than their local/classical ($\mathcal{L}$) counterpart, it cannot be arbitrarily strong just by imposing the no-signaling principle ($\mathcal{NS}$) \cite{PR1} or even in conjunction with other principles \cite{Dam,Bra2,ML,ML1,IC,MBL,LO,Alm1}, see Sec. \ref{IVD} for more information. In particular, there is a well-known strict inclusions on the degree of nonlinearity in various nonlocal scenarios $\mathcal{L}\subseteqq \mathcal{Q}\subseteqq \mathcal{NS}$. Intensive efforts have been devoted to search for principles which precisely demarcate the boundary of the quantum set $\mathcal{Q}$ of correlations and close this discrepancy \cite{BCP,Pop}. As pointed out in man text, we have proposed a natural principle to explain why quantum correlations are strictly weaker than no-signaling correlations and predict the quantum set $\mathcal{Q}$.

\subsection{A brief review of postquantum models}\label{IIA}

The postquantum models were originally proposed in continuous variables \cite{PR1,Sv,Non1,Non5,Non3}. Afterwards, the generalized probabilistic theory \cite{GPT3,GPT4,GPT5,GPT,GPT1,GPT2} has introduced wide range of models, from classical and quantum models to postquantum correlations \cite{Non2,Non7}. In these models, the strength of nonlocal correlations crucially depend on geometric property of the local state spaces. Moreover, there is another type of continuous-variable postquantum models which correlation functions are invariant under spatial rotation transformations \cite{Non4,Non6}. In particular, they considered general form of correlation function and tried to simulate quantum correlation by imposing space symmetry transformations on them.

The Boolean nonlocal box models \cite{Bra2,Dam,BP1,JM} are proposed to describe nonlocality in an abstract and generic manner which have practical application in context of quantum information theory as a resource for device-independent quantum information processing. In these scenarios, the violation of the Bell-type inequalities, e.g., the CHSH inequality \cite{CHSH}, can be certified the intrinsic randomness \cite{Ran1,Ran2} or security of distributed messages \cite{Di1,Di2} independent of internal operations and experimental apparatus \cite{Di3,Di5,Di6,Di7}. Due to abstract properties of the nonlocal box models, they can be applied in narrow range of physical models. Meanwhile, the NL-box models are more appropriate to discover fundamental relations between concepts of physics.

In addition to the abstract NL-box models \cite{PR1,BCP,JM,BP1,BP2,Mas}, there is also another type of postquantum models which is refereed to the almost quantum correlations model \cite{Alm}. This model satisfies all kinematic postulates of standard quantum correlations except one: the postulate of commutation of two measurement operators for two different parties is replaced by the equality of arbitrary permutations of the measurement operators on a specific normalized vector state. Interestingly, the set of the almost quantum correlations is strictly larger than the standard quantum set and violates the bound maximum quantum correlations.

As pointed out in main text, all of proposed principles \cite{Dam,Bra2,ML,ML1,IC,MBL,LO,Alm1} have provided only partial/approximate answers to single out quantum mechanics. Hence, these principles \cite{Dam,Bra2,ML,ML1,IC,MBL,LO,Alm1} are unsuccessful to distinguish the multipartite or high-dimension (input/output) postquantum models from the multipartite or high-dimension (input/output) quantum correlations \cite{NIP,NIP1}. Therefore, the extension of postquantum models to multipartite or high-dimension (input/output) systems shed light to search for principle(s) which characterize quantum correlations. In Sec. \ref{IVD}, we will show ability of the flat space symmetry-group structure to exactly distinguish physical correlations from nonphysical in one of multipartite systems.

\subsection{Every element of the physical reality must have a counterpart in complete physical model}\label{IIB}

Symmetry transformations are specific operations on the configuration space of a given theory. These properties can be used to classify symmetries. One general  classification of symmetry transformations is based on the internal or external symmetries. In the internal symmetries do not act on the spacetime variables and only act on the internal degree of freedom physical systems $\phi(x)\rightarrow \tilde{\phi}(x)$. The external symmetries usually refer to transformation which involve change in external spacetime structure $\phi(x)\rightarrow \tilde{\phi}(\tilde{x}),\hspace{.2cm} x\rightarrow \tilde{x}$.

As pointed out in Sec. \ref{IB}, in every  physical theory/model we always propose correspondence relation between experiment systems/apparatus/operations and mathematical symbols in physical model. In quantum mechanic, we represent the systems, apparatus, and operations with the density matrices, measurement operators, and unitary transformations, respectively. This correspondence enables us predicts predicts measurement outcomes for testing mathematical models. For example, in realization of the CHSH experiment by entangled photons \cite{Sha,Ant,Pan,Ph,Ph1}, the photon polarizations are rotated by applying optical devices such as quarter-wave/half-wave plates. These rotations are realized by unitary transformations in quantum mechanics framework.

Similar to quantum mechanics scenario, we have tried to present experiment transformations in the abstract Boolean nonlocal models. A question may arises: why can Boolean symmetry group transformations be described as the isotropy and the homogeneity symmetries in the flat space? As pointed out in main text, to give a complete explanation of every physical theories ``every element of the physical reality must have a counterpart in the physical theory" \cite{EPR}. To do, we imply a correspondence relation between the measurement inputs/outputs, transformations, symmetries and invariant variables in the experiment, and the NL-box inputs/outputs, Boolean transformations, mathematical symmetry group structure and invariant variables in the NL-box models. We firstly identified the inputs/outputs of real experiment with the inputs/outputs of NL-box models. This is first step to describe the experiment in any mathematical model. Secondly, any transformations on the external variables, for example the measurement directions $\mathbf{a}$ and $\mathbf{b}$, are corresponded to mathematical transformations $F$ which control and change the inputs $\mathbf{x}$ and $\mathbf{y}$ in the NL-box models. Afterwards, we imply a correspondence between the mathematical structure of flat space symmetry transformations and the mathematical structure of the Boolean symmetry group transformations. Finally, similar to real space which the correlation functions or the probability distributions shall be remain invariant under space symmetry transformations, we require that the correlation functions or the probability distributions remain invariant under Boolean symmetry transformations.


Due to Sec. \ref{IB}, it shall note that the internal degree of freedoms do not paly any role in the Bell type experiments. Besides, in other problems, we may apply the Boolean algebra for internal degree of freedoms. In these cases, the symmetry transformations may be explained in other manners. Here we emphasis that symmetry transformations in postquantum models do not represent their counterparts in real space. For example, in the almost quantum correlations \cite{Alm} framework, we used mathematical tools (kets, projectors, unitary transformation) as quantum mechanics notations, although difference is evident.


\subsection{Functions and mathematical symmetry group structure in Boolean algebra}\label{IIC}

A Boolean function $F$ is a element of Boolean transformation space $F\in\{\mathcal{F}\}$ so that $F: \{0,1\}^{n}\rightarrow \{0,1\}^{n}$ or equivalently $F: V_{2}^{n}\rightarrow V_{2}^{n}$, where $V_{2}^{n}=\mathbf{x}$ are $n$-tuple of inputs ($\mathbf{x}=(x_{1},...,x_{n}),  x_{i}\in\{0,1\}$). The $n$-variable Boolean function $F(\mathbf{x})$ can be considered to be a multivariate polynomial over $V_{2}^{n}$:
\begin{eqnarray}\label{F1p}
F(\mathbf{x})=\left( f^{1}(\mathbf{x}),...,f^{l}(\mathbf{x}),...,f^{n}(\mathbf{x})\right)
\end{eqnarray}

The component functions $f^{l}(\mathbf{x})$ can be expressed in standard form, which is called the algebraic normal form \cite{ANF,Re}, the summation of all distinct $k$-th order products ($ 1\leqslant k \leqslant n$) of the variables:
\begin{eqnarray}\label{F2p}
f^{l}(x_{1},...,x_{n})=&&C_{f}^{l,0}\oplus\bigoplus_{i}C_{f,i}^{l,1}x_{i}\oplus\bigoplus_{i<j}C_{f,ij}^{l,2}x_{i}x_{j}\oplus\bigoplus_{i<j<k}C_{f,ijk}^{l,3}x_{i}x_{j}x_{k}
\\\nonumber&&\oplus...\oplus C_{f,1,...,n}^{l,n}x_{1}x_{2}...x_{n} \hspace{.3cm}(\text{mod}\hspace{.1cm} 2), \hspace{.2cm} \forall \hspace{.1cm} l.
\end{eqnarray}
where, the symbols $\bigoplus$ and $\oplus$ represent summation of the Boolean variables which are calculate in module $2$, and the coefficients $C_{f}^{l,0}, C_{f,i}^{l,1}, C_{f,ij}^{l,2}, C_{f,ijk}^{l,3},...,C_{f,1,...,n}^{l,n}\in\{0,1\}$ are vectors of the polynomial representing the Boolean function $f^{l}$, all of summations are calculated in module $2$. The number of variables in the highest order product term with nonzero coefficient is called the algebraic degree, or simply the degree of $f^{l}$ and denoted by $deg(f^{l})$.

\vspace{.1cm}

Similar to symmetry group structure in real space, the Boolean transformations $F\in\left\{\mathcal{F}\right\}$ shall respect group $G$ ($F,V_{2}^{n},\ast$) which is defined by a nonempty set $V_{2}^{n}$ together with transformations $F$ on $V_{2}^{n}$ such that every inputs $\mathbf{x}$ transform under $F$ as $F(\mathbf{x})=(f^{1}(\mathbf{x}),...,f^{n}(\mathbf{x}))$ and an operator $\ast$ between two transformations $F$. The Boolean transformations $F(\mathbf{x})=(f^{1}(\mathbf{x}),...,f^{n}(\mathbf{x}))$ shall respect The Boolean group axioms as:


(\textit{i}) Closed under product of group elements. The product of any two elements $F_{1}, F_{2}\in\{\mathcal{F}\}$ is another element of the symmetry transformations, $F_{1}\ast F_{2}, F_{2}\ast F_{1} \in\{\mathcal{F}\}$.

(\textit{ii}) Existence of an identity transformation. There exists a unique element $I\in\{\mathcal{F}\}$ such that any element $F\in\{\mathcal{F}\}$ satisfies $F\ast I=I\ast F=F$. The identity transformation $I$ satisfies $I(\mathbf{x})=(I^{1}(\mathbf{x}),...,I^{n}(\mathbf{x}))=(\mathbf{x}), \hspace{.2cm}\forall \mathbf{x} \in V_{2}^{n}$.

This condition can be satisfied by choice of $C_{I,l}^{l,1}=1,\hspace{.2cm}l=1,...,n$ and other coefficients are equal to zero.

(\textit{iii}) Existence of inverse transformations. For any element $F\in\{\mathcal{F}\}$, there exists a unique element $F^{-1}\in\{\mathcal{F}\}$ such that $F\ast\left(F^{-1}(\mathbf{x})\right)= F^{-1}\ast\left(F (\mathbf{x})\right)=I(\mathbf{x}),\hspace{.2cm}\forall \mathbf{x} \in V_{2}^{n}$.

An $n$-variable Boolean function is reversible if all its output patterns map uniquely to an input pattern and vice-versa. It can be expressed as an $n$-input, $n$-output bijection \cite{Re}.

(\textit{iv}) Associativity law. Any three group elements $F_{1}, F_{2}, F_{3}\in\{\mathcal{F}\}$ satisfy the following associativity condition: $\left[F_{1}\ast F_{2}\right]\ast F_{3}(\mathbf{x})= F_{1}\ast\left[F_{2}\ast F_{3}\right](\mathbf{x}),\hspace{.2cm}\forall \mathbf{x} \in V_{2}^{n}$.




\subsection{The symmetry group assumptions in the perfect NL-box framework}\label{IID}

In the extended version of perfect NL-box model, Alice and Bob choice $n$-tuple of inputs $\mathbf{x} $ and $\mathbf{y}$ from input spaces $\in V_{2}^{n}$ so that $\mathbf{x}=(x_{1},...,x_{n}), \mathbf{y}=(y_{1},...,y_{n}), x_{i},y_{j}\in\{0,1\}, \hspace{.1cm} \forall \hspace{.1cm}i$, and the parties outputs are represented by ($\alpha,\beta$), where $\alpha,\beta\in\{0,1\}$. In this scenario, the correlation of measurement outcomes is given by the inner product function of inputs $\mathrm{IP}_{\mathrm{n}}(\mathbf{x}, \mathbf{y})$
\begin{eqnarray}\label{nlbp}
\alpha\oplus\beta=\mathrm{IP}_{\mathrm{n}}(\mathbf{x}, \mathbf{y})=\bigoplus_{i=1}^{n} x_{i}y_{i}=\mathbf{x}\cdot \mathbf{y}=x_{1}y_{1}\oplus...\oplus x_{n}y_{n}
\hspace{.3cm}(\text{mod}\hspace{.1cm} 2),
\end{eqnarray}
where $x_{i}y_{i}$ is usual product of inputs and summations are calculated in modulo $2$. Similar to the Bell-like experiment setup, which Alice and Bob measurement inputs are arbitrary unit vector $\mathbf{a}$ and $\mathbf{b}$ form three dimensional space $\mathbb{R}^{3}$ with binary outputs $A, B\in\{-1,1\}$, in the NL-box model (\ref{nlbp}) the parties inputs are arbitrary $n$-tuple of inputs $\mathbf{x}$ and $\mathbf{y}$ from input space $\in V_{2}^{n}$ with binary outputs $\alpha, \beta\in\{0,1\}$ which makes the NL-box model (\ref{nlbp}) more appropriate to simulate the Bell-type experiments.

\vspace{.1cm}

\textit{Remark 3}. As pointed out in Eq. (\ref{nb}) of main text, right hand side of the general NL-box models $\alpha\oplus\beta=f(\mathbf{x},\mathbf{y})$ can be arbitrary function of inputs. Thus, it may seem that the present outcomes just hold for specific form of correlation functions (\ref{nlbp}). In another work in progress, we add another NL-box Assumption to NL-box models to answer this point. This new assumption which is compatible with symmetries in real space indicates that observable quantities are invariant under re-indexing of space directions $(\mathbf{x},\mathbf{y},\mathbf{z})$. We realize the new assumption in the Boolean algebra as ``arbitrary permutation (re-indexing) of inputs does not affect correlation function" \cite{Cau}. Besides, we show that the inverse operator $F_{\mathrm{inv}}\in\{\mathcal{F}\}$, which transforms any input to its inverse $F_{\mathrm{inv}}(\mathbf{x})=\bar{\mathbf{x}},\,\forall \hspace{.1cm} \mathbf{x}$, in any type of NL-box models $f(\mathbf{x},\mathbf{y})$ is $\mathbf{x}\oplus \mathbf{1}=\bar{\mathbf{x}},\,\forall \hspace{.1cm} \mathbf{x}$, as pointed out in the NL-box Assumption $1$ of main text. We prove that the new assumption and inverse operator $F_{\mathrm{inv}}$ are sufficient to reduce $f(\mathbf{x},\mathbf{y})$ to standard form Eq. (\ref{nb}).




\vspace{.2cm}

The extended version of perfect NL-box models violate strong form of the CHSH inequality which is based on the hidden variable assumptions.
\begin{eqnarray}\label{CHSH1}
\mathcal{S}_{\mathrm{LHV}}=|E(\mathbf{x}_{1},\mathbf{y}_{1})-E(\mathbf{x}_{1},\mathbf{y}_{2})|+
|E(\mathbf{x}_{2},\mathbf{y}_{1})+E(\mathbf{x}_{2},\mathbf{y}_{2})|\leqslant2
\end{eqnarray}
at the maximum value of $4$. To generate homogeneous and isotropic version of the NL-box models, we redefine conditions the Assumptions $1-3$ (in the main text) in the Boolean algebra framework and impose these conditions on the NL-box models.
\begin{figure}
\centering
\includegraphics[scale=.5]{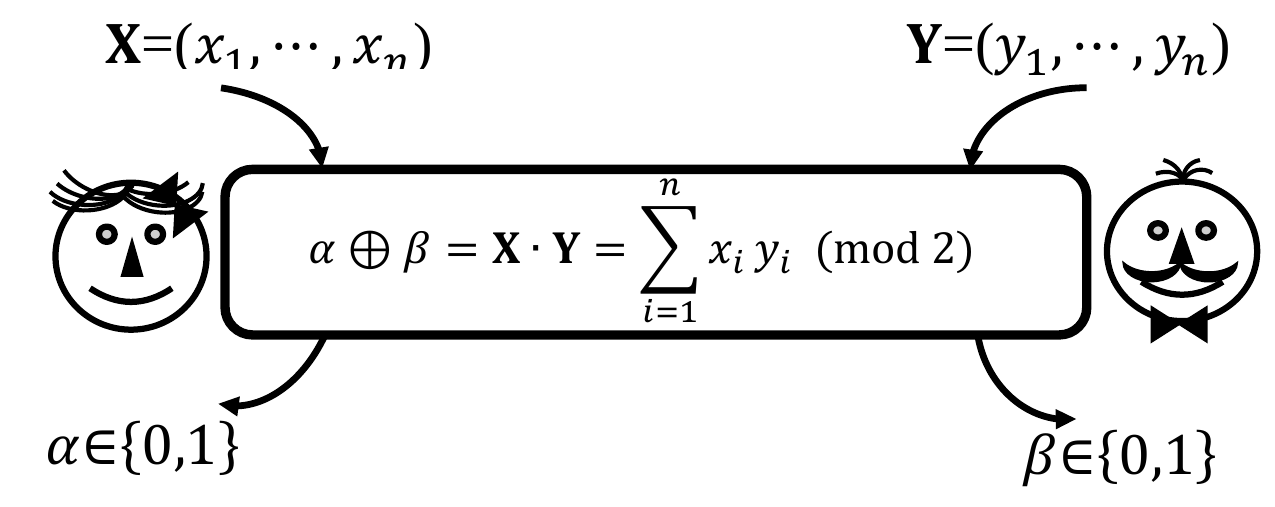}
\caption{Alice and Bob select $n$-tuple of inputs $\mathbf{x} $ and $\mathbf{y}$ from input spaces  $\{\mathcal{X}\}$ and $\{\mathcal{Y}\}$, respectively, where $\mathbf{x}=(x_{1},...,x_{n}), \mathbf{y}=(y_{1},...,y_{n}), x_{i},y_{j}\in\{0,1\}$. The parties outputs are represented by ($\alpha,\beta$), where $\alpha,\beta\in\{0,1\}$. These boxes are characterized by input-output correlation $\alpha\oplus\beta=\mathbf{x}\cdot \mathbf{y}=x_{1}y_{1}\oplus...\oplus x_{n}y_{n},\hspace{.3cm}(\text{mod}\hspace{.1cm} 2)$. The NL-box models are reference frame independent which means that it is sufficient to apply transformation matrices $F$ on the parties measurement inputs without refer to interior structure of NL-boxes.}
\label{unfp}
\end{figure}


\textbf{NL-box Assumption $1$}: The perfect correlation (anti-correlation) conservation law in the NL-box models (Assumption $1$). In the NL-box terminology, we always make correspondence relationes between parties measurement settings ($\mathbf{a}$ and $\mathbf{b}$) in real experiment and inputs ($\mathbf{x}$ and $\mathbf{y}$) in the NL-box model. We denote the opposite measurement settings in the Alice's and Bob's sites by $\bar{\mathbf{x}}$ and $\bar{\mathbf{y}}$ (correspond to $-\mathbf{a}$ and $-\mathbf{b}$), respectively. Therefore, if Alice (Bob) reverses her (his) measurement setting to opposite direction, her (his) outputs flip from $\alpha$ to $\alpha\oplus1$ ($\beta$ to $\beta\oplus1$):
\begin{eqnarray}\label{nl1p}
&&\alpha\oplus\beta=\mathbf{x}\cdot\mathbf{y},\hspace{1.5cm}(\alpha\oplus1)\oplus\beta=\bar{\mathbf{x}}\cdot\mathbf{y},\\\nonumber
&&\alpha\oplus(\beta\oplus1)=\mathbf{x}\cdot \bar{\mathbf{y}},\hspace{.5cm}
(\alpha\oplus1)\oplus(\beta\oplus1)=\bar{\mathbf{x}}\cdot \bar{\mathbf{y}}, \hspace{.2 cm}(\text{mod}\hspace{.1cm}2)
\end{eqnarray}
The above equations restrict the parties' $n$-tuple of inputs to some specific form which shall be satisfied by following relations:
\begin{eqnarray}\label{nl2p}
&&\mathbf{x}\cdot(\mathbf{y}\oplus \bar{\mathbf{y}})=1,\hspace{.6cm}
(\mathbf{x}\oplus \bar{\mathbf{x}})\cdot \mathbf{y}=1,\hspace{.3cm}\nonumber\\
&&\bar{\mathbf{x}}\cdot\mathbf{y}=\mathbf{x}\cdot \bar{\mathbf{y}}, \hspace{1.3cm}\mathbf{x}\cdot \mathbf{y}=\bar{\mathbf{x}}\cdot \bar{\mathbf{y}}, \hspace{.2 cm}(\text{mod}\hspace{.1 cm}2) \hspace{.5cm}\forall \hspace{.2cm}\mathbf{x}, \mathbf{y}.
\end{eqnarray}




\vspace{.1cm}

As indicated in the main text, the original version of NL-box models are not admissible. However, it can be shown that the extended version of
perfect NL-box model (\ref{nlbp}) respects conditions (\ref{nl2p}) by considering subset of input space so that the number of inputs characters ($n$)
is even, $n=2k$; and number of $1$ ($0$) in the inputs are odd $\bigoplus_{i=1}^{n}x_{i}=\bigoplus_{i=1}^{n}y_{i}=1$.
\begin{equation}\label{nup}
\mathbf{x}=(\underbrace{x_{1},x_{2},\hdots,x_{n}}_{\textrm{number of 1 (0) is odd}}), \hspace{.5cm} \mathbf{y}=(\underbrace{y_{1},y_{2},\hdots,y_{n}}_{\textrm{number of 1 (0) is odd}}).
\end{equation}
Hence, as pointed out in Remark $3$, the correlation function (\ref{nlbp}) are invariant under permutation of input indexes $\alpha\oplus\beta=\bigoplus_{i=1}^{n} x_{\pi(i)}y_{\pi(i)}\hspace{.2cm}(\text{mod}\hspace{.1cm} 2)$, where operator $\pi$ represents arbitrary permutation (re-indexing) of inputs \cite{Cau}.

In the extended version of perfect NL-box model, the Alice and Bob input spaces are subset of input space $V_{2}^{n}$ and we call the input subspaces as $\mathbf{x}\in\{\mathcal{X}\}$ and $\mathbf{y}\in\{\mathcal{Y}\}$, respectively. The subspace $\{\mathcal{X}\}$ and $\{\mathcal{Y}\}$ are closed under summation of every odd number of inputs.
\begin{eqnarray}\label{nl14p}
&&\mathbf{x}_{1}\oplus\mathbf{x}_{2}\oplus\mathbf{x}_{3}\in \{\mathcal{X}\},\hspace{.4cm} \forall \hspace{.2cm}\mathbf{x}_{1}, \mathbf{x}_{2}, \mathbf{x}_{3}\in\{\mathcal{X}\},\\\nonumber
&&\mathbf{y}_{1}\oplus\mathbf{y}_{2}\oplus\mathbf{y}_{3}\in \{\mathcal{Y}\},\hspace{.4cm} \forall \hspace{.2cm}\mathbf{y}_{1}, \mathbf{y}_{2}, \mathbf{y}_{3}\in\{\mathcal{Y}\}, \hspace{.7cm}(\text{mod}\hspace{.1cm} 2)
\end{eqnarray}
Moreover, we present $\bar{\mathbf{x}}$ ($\bar{\mathbf{y}}$) as $\mathbf{1}\oplus\mathbf{x}$ ($\mathbf{1}\oplus\mathbf{y}$), where $\mathbf{1}$ is $n$-tuple which all of entities are equal to $1$, $\mathbf{1}=(1,1...,1)$. These set of input space satisfy to the relations (\ref{nl2p}). For example, for every input $\mathbf{x}$, we have $(\mathbf{x}\oplus \bar{\mathbf{x}})\cdot\mathbf{y}=\mathbf{1}\cdot\mathbf{y}=\bigoplus_{i=1}^{n}y_{i}=1$, the final equality is obtained by $(\mathbf{1}\oplus\mathbf{x})\cdot(\mathbf{1}\oplus\mathbf{y})=0\oplus\mathbf{1}\oplus\mathbf{1}\oplus\mathbf{x}\cdot\mathbf{y}$.

\vspace{.1cm}

\textit{Remark 4}. It may seem that the definition of the measurement settings for $\mathbf{x}$ and $\bar{\mathbf{x}}$ (similarly for $\mathbf{y}$ and $\bar{\mathbf{y}}$) are matter of indexing and one can do it in whatever one wants. Every indexing of $\mathbf{x}$ and $\bar{\mathbf{x}}$ shall respect the all of relations in Eq. (\ref{nl2p}) for all of $\mathbf{y}$. This shows that the summation $\mathbf{x}\oplus \bar{\mathbf{x}}$ is constant and independent from $\mathbf{x}$, in otherwise, there are inputs $\mathbf{y}$ which violate the one of relations in Eq. (\ref{nl2p}). For example, if the summation of $\mathbf{x}\oplus \bar{\mathbf{x}}$ be equal to $[1,1,1,1,0,0]$ (for $n=6$), there are many $\mathbf{y}$ which are not consistent with one of relations in Eq. (\ref{nl2p}) such as $\mathbf{y}=[0,0,0,0,0,1]$. Moreover, the reversing transformation matrix $\mathbf{F}_{\mathrm{re}}$, as an elements of symmetry group transformation space $\mathbf{F}_{\mathrm{re}}\in\{\mathcal{F}\}$ (will indicate in the NL-box Assumption $2$), does not revers all of input space elements, $\mathbf{F}_{\mathrm{re}}:\mathbf{x}\rightarrow\bar{\mathbf{x}}, \forall\hspace{.1cm} \mathbf{x}$ as an unique transformation. In other words, the reversing transformation function depends on the input variables $F_{\mathrm{re}}=F(\mathbf{x},\bar{\mathbf{x}})$ which is in conflict with definition of $\mathbf{F}_{\mathrm{re}}$ as one of elements of transformation space $\mathbf{F}_{\mathrm{re}}\in\{\mathcal{F}\}$.

\vspace{.2cm}

\textit{Remark 5}. It may seems that the NL-box correlation (\ref{nlbp}) can be simulated by wiring of the earlier versions of the NL-boxes models \cite{LO,Wir} in which the inputs $\mathbf{x}$ and $\mathbf{y}$ are binary numbers $x, y\in\left\{0,1\right\}$ \cite{Bra2,Dam} or integer numbers $x,y\in\{0,1,...,m-1\}$ \cite{BP1,JM}, and $f(\mathbf{x},\mathbf{y})$ is usually the product of inputs $xy$. But, these models, as building blocks to generate correlation (\ref{nlbp}), do not respect the conditions (\ref{nl2p}), then they have are not acceptable to simulate physical correlations.

\vspace{.2cm}

Interestingly, in the extended version of perfect NL-box model with input settings as (\ref{nup}), there are measurement settings which strongly violate the CHSH inequality (\ref{CHSH1}) at the maximum value $4$ yet. For example, if we choose the measurement inputs $\mathbf{x}_{1}=[1,0,0,0,0,0], \hspace{.1cm}\mathbf{x}_{2}=[0,1,0,0,0,0], \hspace{.1cm}\mathbf{y}_{1}=[0,1,1,1,0,0],$ and $\mathbf{y}_{2}=[0,0,0,1,0,0]$ in $n=6$, the correlation functions will be $\mathbf{x}_{1}\cdot\mathbf{y}_{1}=0,\hspace{.1cm}\mathbf{x}_{1}\cdot\mathbf{y}_{2}=1,\hspace{.1cm}\mathbf{x}_{2}\cdot\mathbf{y}_{1}=0,$ and $\mathbf{x}_{2}\cdot\mathbf{y}_{2}=0,\hspace{.2 cm}(\text{mod}\hspace{.1cm}2)$, which gives $\mathcal{S}_{NLB}=4$. Therefore, there need other symmetry group assumptions which correctly simulate quantum predictions. As pointed out in Sec. \ref{IA}, we redescribe Assumptions $2$ and $3$ in the NL-box scenario.


\vspace{.3cm}

\textbf{NL-box Assumption $2$}: The set of transformation operators $F\in\left\{\mathcal{F}\right\}$ in the NL-box models framework have group structure. As pointed out in Sec. \ref{IIC}, the Boolean transformation $F$ which are actually a permutation and some bit flipping on $n$-tuple of input space, transforms $n$-tuple of inputs to another $n$-tuple of inputs $F: \{0,1\}^{n}\rightarrow \{0,1\}^{n}$. The $n$-variable Boolean function $F(x_{1},...,x_{n})$ can be considered to be a multivariate polynomial over $\{0,1\}^{n}$. This polynomial can be expressed as a vector of $n$ functions called the component functions
\begin{eqnarray}\label{corr}
F(\mathbf{x})=\left( f^{1}(\mathbf{x}),...,f^{l}(\mathbf{x}),...,f^{n}(\mathbf{x})\right)
\end{eqnarray}
More precisely, the component functions $f^{l}: \{0,1\}^{n}\rightarrow \{0,1\}$ can be represented as the algebraic normal form (ANF) \cite{ANF,Re}, sum of products representation of all distinct $k$-th order products ($ 1\leqslant k \leqslant n$) of the variables. Besides, the Boolean transformation $F$ needs to be respected group structure conditions ($i,ii,iii,iv$). Hence, the Boolean transformations $F$ maintain the NL-box Assumption $1$, in particular, two opposite inputs $\mathbf{x}$ and $\bar{\mathbf{x}}$ transform to two opposite directions $F(\mathbf{x})$ and $F(\bar{\mathbf{x}})$ under operation of $F$.

In the next subsection we will show all of group transform function $F(\mathbf{x})=\left( f^{1}(\mathbf{x}),...,f^{n}(\mathbf{x})\right)$ reduce to standard form
\begin{eqnarray}\label{com-f}
f^{l}(x_{1},...,x_{n})=C_{f}^{l,0}\oplus\bigoplus_{i}C_{f,i}^{l,1}x_{i},
\end{eqnarray}
after imposing the NL-box Assumptions $1, 2$ and $3$ on them. Here, the coefficients $C_{f}^{l,0}, C_{f,i}^{l,1}\in\{0,1\}$ are vector of the polynomial representing the Boolean. We write transformation functions $F(\mathbf{x})$ as $F(\mathbf{x})=\mathbf{R}\mathbf{x}\oplus\mathbf{T}$, where $\mathbf{R}$ and $\mathbf{T}$ are $n\times n$ and $1\times n$ matrices, respectively, with elements of $0$ and $1$, $\mathbf{R}_{ij},\mathbf{T}_{1j}\in \{0,1\}, \bigoplus_{i=1}^{n}\mathbf{R}_{ij}=\bigoplus_{j=1}^{n}\mathbf{R}_{ij}=1$.

Although the Boolean matrices $\mathbf{R}$ and $\mathbf{T}$ have similar properties as real space rotation and translation matrices, they shall not be mistaken with them. In the real space, the rotation and the translation operators are independent group structures, however, in the NL-box terminology, we impose group structure on the abstract nonlocal model and obtain the matrices $\mathbf{R}$ and $\mathbf{T}$:

$\vspace{.1cm}$

(\textit{i}) Closed under product of group elements. Two sequence of transformations $(\mathbf{R}_{1}, \mathbf{T}_{1})$ and $(\mathbf{R}_{2}, \mathbf{T}_{2})$ are given by $(\mathbf{R}, \mathbf{T})$, where $\mathbf{R}=\mathbf{R}_{2}\mathbf{R}_{1}$ and $\mathbf{T}=\mathbf{R}_{2}\mathbf{T}_{1}\oplus\mathbf{T}_{2}$.

(\textit{ii}) Existence of an identity transformation. There exists a unique element $I\in\{\mathcal{F}\}$ such that any element $F\in\{\mathcal{F}\}$ satisfies $F I=I F=F$. The identity transformation $I$ satisfies $I(\mathbf{x})=(I^{1}(\mathbf{x}),...,I^{n}(\mathbf{x}))=(\mathbf{x}), \hspace{.2cm}\forall \mathbf{x} \in V_{2}^{n}$ which is given by $\mathbf{R}=\mathbf{I}$ and $\mathbf{T}=0$, where the $\mathbf{I}$ is identity matrix.

(\textit{iii}) Existence of inverse transformations. The inverse transformation $F(\mathbf{x})=\mathbf{R}\mathbf{x}\oplus\mathbf{T}$ are obtain by $F^{-1}(\mathbf{x})=\mathbf{R}'\mathbf{x}\oplus\mathbf{T}'$, where $\mathbf{R}'=\mathbf{R}^{t}$ and $\mathbf{T}'=\mathbf{R}^{t}\mathbf{T}$.

(\textit{iv}) Associativity law. Any three group elements $F_{1}=:(\mathbf{R}_{1}, \mathbf{T}_{1})$, $F_{2}=:(\mathbf{R}_{2}, \mathbf{T}_{2})$ and $F_{3}=:(\mathbf{R}_{3}, \mathbf{T}_{3})$ satisfy the following associativity condition: $\left[F_{1}F_{2}\right]F_{3}(\mathbf{x})= F_{1}\left[F_{2}F_{3}\right](\mathbf{x}),\hspace{.2cm}\forall \mathbf{x} \in V_{2}^{n}$.

\vspace{.1cm}

This show that all rows (columns) of the Boolean rotation operators $\mathbf{R}\in\{\mathcal{R}\}$ are given by $n$-tuples of the inputs space $\{\mathcal{X}\}$:
\begin{eqnarray}\label{Rp}
&&\mathbf{R}_{i,j=1,...,n}\equiv\mathbf{R}_{i,(n)}=(\underbrace{R_{i,1},R_{i,2},\hdots,R_{i,n}}_{\textrm{number of 1 (0) is odd}})\in\{\mathcal{X}\}, \hspace{.1cm}\text{the $i$-th row of matrix} \mathbf{R}\\\nonumber
&&\mathbf{R}_{i=1,...,n,j}\equiv\mathbf{R}_{(n),j}=(\underbrace{R_{1,j},R_{2,j},\hdots,R_{n,j}}_{\textrm{number of 1 (0) is odd}})\in\{\mathcal{X}\}, \hspace{.1cm}\text{the $j$-th column of matrix} \mathbf{R}.
\end{eqnarray}

The Boolean translation operators $\mathbf{T}\in\{\mathcal{T}\}$ are given by sequence of even number of $1$ and $0$,
\begin{equation}\label{Tp}
\mathbf{T}=(\underbrace{T_{1,1},T_{1,2},\hdots,T_{1,n}}_{\textrm{number of 1 (0) is even}}),
\end{equation}
a explicit example of the matrixes $\mathbf{R}$ and $\mathbf{T}$ are represented in Sec. \ref{V} for $n=6$.


$\vspace{.2cm}$

As a hint supporting NL-box Assumption $3$, we have considered the almost-quantum correlations model \cite{Alm} and have defined local symmetry transformations where the state vector and the measurement operator of each of subsystem undergo an arbitrary local rotation/translation transformation \cite{Almsy}. In the NL-box models, we apply the same symmetry transformations $F$ for all subsystems but since the correlation functions are reference frame independent, the interior structure (e.g., shared state) does not appear in the NL-box models.

\textbf{NL-box Assumption $3$}: The observable quantities, include correlation functions, probability distributions and the upper bound of the CHSH inequality,
are invariant under the Boolean transformations $F$ (Assumption $3$).
\begin{eqnarray}\label{invp}
\alpha\oplus\beta=\mathbf{x}\cdot \mathbf{y}=F(\mathbf{x})\cdot F(\mathbf{y}) \hspace{.2cm}(\text{mod}\hspace{.1cm} 2), \hspace{.3cm} \forall \hspace{.2cm}\mathbf{x}, \mathbf{y}, F,
\end{eqnarray}
the symmetry condition in the NL-box models will be obtained:
\begin{eqnarray}\label{symp}
&&\alpha\oplus\beta=\mathbf{x}\cdot \mathbf{y}=F(\mathbf{x})\cdot F(\mathbf{y})=(\mathbf{R}\mathbf{x}\oplus\mathbf{T})\cdot(\mathbf{R}\mathbf{y}\oplus\mathbf{T})\\\nonumber
&&\hspace{5.2cm}=(\mathbf{R}\mathbf{x}\oplus\mathbf{T})^{t}(\mathbf{R}\mathbf{y}\oplus\mathbf{T})\\\nonumber
&&\hspace{5.2cm}=\mathbf{x}^{t}\mathbf{R}^{t}\mathbf{R}\mathbf{y}\oplus\mathbf{x}^{t}\mathbf{R}^{t}\mathbf{T}\oplus\mathbf{T}^{t}\mathbf{R}\mathbf{y}\oplus\mathbf{T}^{t}\mathbf{T}\\\nonumber
&&\hspace{5.2cm}=\mathbf{x}^{t}\mathbf{R}^{t}\mathbf{R}\mathbf{y}\oplus\mathbf{T}\cdot(\mathbf{R}\mathbf{x}\oplus\mathbf{R}\mathbf{y}\oplus\mathbf{T})
\hspace{.2cm}(\text{mod}\hspace{.1cm} 2)
\end{eqnarray}
the up-index $t$ in the second line represents transpose of matrices. In the last line, we used $\mathbf{x}^{t}\mathbf{R}^{t}\mathbf{T}=\mathbf{T}^{t}\mathbf{R}\mathbf{x}=\mathbf{T}\cdot\mathbf{R}\mathbf{x}$, $\hspace{.1cm}\mathbf{T}^{t}\mathbf{R}\mathbf{y}=\mathbf{T}\cdot\mathbf{R}\mathbf{y}$, and $\hspace{.1cm}\mathbf{T}^{t}\mathbf{T}=\mathbf{T}\cdot\mathbf{T}\hspace{.2cm}(\text{mod}\hspace{.1cm} 2)$.
In order that the NL-box be invariant under Boolean transformations, the following relations need to be satisfied by transformation matrices:
\begin{eqnarray}\label{conp}
\mathbf{R}^{t}\mathbf{R}=\mathbf{I}, \hspace{.2cm}\mathbf{T}\cdot[\mathbf{R}(\mathbf{x}\oplus\mathbf{y})\oplus\mathbf{T}]=0\hspace{.2cm}(\text{mod}\hspace{.1cm} 2), \hspace{.1cm}\forall\hspace{.1cm}\mathbf{x}, \mathbf{y}, \mathbf{R}, \mathbf{T}
\end{eqnarray}
the $\mathbf{I}$ is identity matrix.

Therefore, the matrices $\mathbf{R}$ satisfy to $\mathbf{R}_{i,(n)}\cdot\mathbf{R}_{l,(n)}=\bigoplus_{e=1}^{n} R_{i,e}R_{l,e}=\delta_{il}$ and $\mathbf{R}_{(n),j}\cdot\mathbf{R}_{(n),k}=\bigoplus_{e=1}^{n} R_{e,j}R_{e,k}=\delta_{jk}$, where all of summations are determined at module 2. We refer to Eqs. (\ref{conp}) as the ``symmetry conditions" which guarantee isotropy and homogeneity of the NL-box models.

$\vspace{.1cm}$



\subsection{Derivation of explicit form of symmetry group transform $F$}\label{IIE}

In this subsection, we use the NL-box correlation functions (\ref{invp}) to derive exact form of the transform function $F$.

\textit{Remark 6}. The exact form of the transform function $F$ can also be derived from a Boolean description of Remark 2. In particular, the transformation function $F$ should have this property that opposite directions should be transformations opposite directions too. That is, $\mathbf{x}$ and $\bar{\mathbf{x}}$ should be transformed to the opposite directions $F(\mathbf{x})$ and $F(\bar{\mathbf{x}})$:
$F(\mathbf{x}\oplus \mathbf{1})=F(\bar{\mathbf{x}})=\overline{F(\mathbf{x})}=F(\mathbf{x})\oplus \mathbf{1},\hspace{.1cm}\forall\hspace{.1cm}\mathbf{x}, F$.
Otherwise, there would be an observable quantity which does not remain invariant under the symmetry transformations.




In the correlation function framework, the Boolean transformation $F$ are automorphism on the input spaces $\{\mathcal{X}\}$ and $\mathcal{Y}$. We have assumed that the inner product of the inputs $\mathbf{x}\cdot \mathbf{y}$, are mapped to the inner product of the outputs $F(\mathbf{x})\cdot F(\mathbf{y})$. In addition, for every odd number of inputs, e.g., three inputs $\mathbf{x_{1}},\mathbf{x_{2}},\mathbf{x_{3}}$, which belong to the input space $\{\mathcal{X}\}$, the sum of these inputs also belongs to same space, $\mathbf{x}=\mathbf{x_{1}}\oplus\mathbf{x_{2}}\oplus\mathbf{x_{3}}\in\mathcal{X}$. Therefore, we have the following relations:
\begin{eqnarray*}\label{}
&&F(\mathbf{x})\cdot F(\mathbf{y})=\mathbf{x}\cdot \mathbf{y}\\
&&\hspace{2.2cm}=\left[\mathbf{x}_{1}\oplus\mathbf{x}_{2}\oplus\mathbf{x}_{3}\right]\cdot \mathbf{y}\\
&&\hspace{2.2cm}=\left[(\mathbf{x}_{1}\cdot \mathbf{y})\oplus(\mathbf{x}_{2}\cdot \mathbf{y})\oplus(\mathbf{x}_{3}\cdot \mathbf{y})\right]\\
&&\hspace{2.2cm}=\left[F(\mathbf{x}_{1})\cdot F(\mathbf{y})\right]\oplus \left[F(\mathbf{x}_{2})\cdot F(\mathbf{y})\right] \oplus \left[F(\mathbf{x}_{3})\cdot  F(\mathbf{y})\right]\\
&&\hspace{2.2cm}=\left[F(\mathbf{x}_{1})\oplus F(\mathbf{x}_{2})\oplus F(\mathbf{x}_{3})\right]\cdot F(\mathbf{y}).\hspace{.4cm} \forall \hspace{.2cm}\mathbf{x}, \mathbf{y}, F, \hspace{1.7cm}(\text{mod}\hspace{.1cm} 2)
\end{eqnarray*}
The last line shows that the $F$ is a liner function on the input spaces up to freedom $K(\mathbf{x})$ such that
\begin{eqnarray}\label{sum1}
F(\mathbf{x}_{1}\oplus\mathbf{x}_{2}\oplus\mathbf{x}_{3})=F(\mathbf{x}_{1})\oplus F(\mathbf{x}_{2})\oplus F(\mathbf{x}_{3})\oplus K(\mathbf{x}),\hspace{.4cm} \forall \hspace{.2cm}\mathbf{x}_{1}, \mathbf{x}_{2}, \mathbf{x}_{3}, F, \hspace{1.7cm}\\\nonumber
K(\mathbf{x})\cdot F(\mathbf{y})=0,\hspace{.4cm} \forall \hspace{.2cm}\mathbf{x}, \mathbf{y}, F, \hspace{.7cm}(\text{mod}\hspace{.1cm} 2)
\end{eqnarray}
where the last line implies $K(\mathbf{x})\equiv \mathbf{0}$.

By using (\ref{F2p}) and (\ref{sum1}), we have
\begin{eqnarray*}\label{}
&&C_{f}^{l,0}\oplus\bigoplus_{i}C_{f,i}^{l,1}(x_{i,1}\oplus x_{i,2}\oplus x_{i,3})\oplus\bigoplus_{i<j}C_{f,ij}^{l,2}(x_{i,1}\oplus x_{i,2}\oplus x_{i,3})(x_{j,1}\oplus x_{j,2}\oplus x_{j,3})
\\
&&\hspace{3cm}\oplus\bigoplus_{i<j<k}C_{f,ijk}^{l,3}(x_{i,1}\oplus x_{i,2}\oplus x_{i,3})(x_{j,1}\oplus x_{j,2}\oplus x_{j,3})
(x_{k,1}\oplus x_{k,2}\oplus x_{k,3})
\\
&&\hspace{3cm}\oplus...\oplus C_{f,1,...,n}^{l,n}(x_{1,1}\oplus x_{1,2}\oplus x_{1,3})(x_{2,1}\oplus x_{2,2}\oplus x_{2,3})
...(x_{n,1}\oplus x_{n,2}\oplus x_{n,3})
\\
&&=C_{f}^{l,0}\oplus\bigoplus_{i}C_{f,i}^{l,1}x_{i,1}\oplus\bigoplus_{i<j}C_{f,ij}^{l,2}x_{i,1}x_{j,1}\oplus\bigoplus_{i<j<k}C_{f,ijk}^{l,3}x_{i,1}x_{j,1}x_{k,1}
\oplus...\oplus C_{f,1,...,n}^{l,n}x_{1,1}x_{2,1}...x_{n,1}
\\
&&\oplus C_{f}^{l,0}\oplus\bigoplus_{i}C_{f,i}^{l,1}x_{i,2}\oplus\bigoplus_{i<j}C_{f,ij}^{l,2}x_{i,2}x_{j,2}\oplus\bigoplus_{i<j<k}C_{f,ijk}^{l,3}x_{i,2}x_{j,2}x_{k,2}
\oplus...\oplus C_{f,1,...,n}^{l,n}x_{1,2}x_{2,2}...x_{n,2}
\\
&&\oplus C_{f}^{l,0}\oplus\bigoplus_{i}C_{f,i}^{l,1}x_{i,2}\oplus\bigoplus_{i<j}C_{f,ij}^{l,2}x_{i,3}x_{j,3}\oplus\bigoplus_{i<j<k}C_{f,ijk}^{l,3}x_{i,3}x_{j,3}x_{k,3}
\oplus...\oplus C_{f,1,...,n}^{l,n}x_{1,3}x_{2,3}...x_{n,3}\\
&&\hspace{13cm}(\text{mod}\hspace{.1cm} 2), \hspace{.2cm} \forall \hspace{.1cm} l.
\end{eqnarray*}
which means that
\begin{eqnarray}\label{}
C_{f,ij}^{l,2}=C_{f,ijk}^{l,3}=...= C_{f,1,...,n}=0, \hspace{.2cm} \forall \hspace{.1cm} l.
\end{eqnarray}

Therefore, every element of the transform function can be written as
\begin{eqnarray}\label{}
f^{l}(x_{1},...,x_{n})=C_{f}^{l,0}\oplus\bigoplus_{i}C_{f,i}^{l,1}x_{i}, \, \forall\, l
\end{eqnarray}
where the coefficients $C_{f}^{l,0}, C_{f,i}^{l,1}\in\{0,1\}$ are vectors and matrices of the polynomial representing the Boolean variables. Now, due to Eq. (\ref{F1p}) the symmetry transformation is given by $F(\mathbf{x})=\mathbf{R}\mathbf{x}\oplus\mathbf{T}$.


\subsection{Derivation of symmetry parameter $\mathcal{W}=0$ in the perfect NL-box models}\label{IIF}

As mentioned in the main text, the Boolean transformation matrices $\mathbf{R}$ and $\mathbf{T}$ are $n\times n$ and $n\times 1$ matrices with elements $\mathbf{R}_{ij},\mathbf{T}_{k1}\in \{0,1\}$. Moreover, we have collected the input spaces $\{\mathcal{X}\}$ and $\{\mathcal{Y}\}$, and the the Boolean transformation matrix spaces $\{\mathcal{R}\}$ and $\{\mathcal{T}\}$ in the total space $\{\mathcal{H}\}\equiv\left\{\{\mathcal{X}\},\{\mathcal{Y}\},\{\mathcal{R}\},\{\mathcal{T}\}\right\}$. The matrices of space $\{\mathcal{H}\}$ shall be consistence with the symmetry condition
\begin{eqnarray*}\label{}
\mathbf{T}\cdot[\mathbf{R}(\mathbf{x}\oplus\mathbf{y})\oplus\mathbf{T}]&=&\mathbf{T}^{t}\mathbf{R}(\mathbf{x}\oplus\mathbf{y})\oplus\mathbf{T}^{t}\mathbf{T}\\
&=&\mathbf{T}^{t}\mathbf{R}(\mathbf{x}\oplus\mathbf{y})\\
&=&[\mathbf{R}^{t}\mathbf{T}]\cdot(\mathbf{x}\oplus\mathbf{y})\\
&=&\mathbf{T}'\cdot(\mathbf{x}\oplus\mathbf{y})=0,
\hspace{.1cm}(\mathrm{mod}\hspace{.1cm} 2),\forall\hspace{.1cm}\mathbf{x}, \mathbf{y}, \mathbf{R}, \mathbf{T},
\end{eqnarray*}
in the last line, we defined $\mathbf{R}^{t}\mathbf{T}=\mathbf{T}'\in \{\mathcal{T}\}$, and used Eq. (\ref{Tp}) to determine $\mathbf{T}^{t}\mathbf{T}=0\hspace{.1cm}(\mathrm{mod}\hspace{.1cm} 2)$ in the second line. Besides, we know that the addition of any pair of input space belong to $\{\mathcal{T}\}$ space, $\mathbf{x}\oplus\mathbf{y}\in \{\mathcal{T}\}$.


Therefore, the symmetry condition for any pair of input spaces $\mathbf{x}$ and $\mathbf{y}$, and any element of  $\mathbf{T}\in \{\mathcal{T}\}$ space takes simple form $\mathbf{T}\cdot(\mathbf{x}\oplus\mathbf{y})=0,\hspace{.1cm}(\mathrm{mod}\hspace{.1cm} 2)$. Now, we divide set of translation matrices $\{\mathcal{T}\}$ to subsets $\{\mathcal{T}_{m}\}, \, m=1,\cdots,l$ in which for any pairs of $\mathbf{T}_{i}, \mathbf{T}_{j}\in \{\mathcal{T}_{m}\}$, $\mathbf{T}_{i}\cdot\mathbf{T}_{j}=0$ (mod $2$). Besides, the space $\{\mathcal{R}\}$ is divided to subsets $\{\mathcal{R}_{m}\}, \, m=1,\cdots,l$ in which if $\mathbf{T}_{i}\in \{\mathcal{T}_{m}\}$ and $\mathbf{R}_{j}\in \{\mathcal{R}_{m}\}$, then $\mathbf{R}_{j}\mathbf{T}_{i}\in \{\mathcal{T}_{m}\}$. In fact, the Boolean rotation and translation sets are divided to subsets $\mathcal{R}_{m}$ and $\mathcal{T}_{n}$ so that $\bigcup_{m=1}^{l}\mathcal{R}_{m}\equiv\{\mathcal{R}\}$, $\bigcup_{m=1}^{l}\mathcal{T}_{m}\equiv\{\mathcal{T}\}$ and $\bigcap_{m=1}^{l}\mathcal{T}_{m}\equiv\{\textbf{0},\textbf{1}\}$. In addition, we divide spaces $\{\mathcal{X}\}$ and $\{\mathcal{Y}\}$ to subsets $\{\mathcal{X}_{m}\}$ and $\{\mathcal{Y}_{m}\}$ in which for all of $\mathbf{x}\in\{\mathcal{X}_{m}\}$ and $\mathbf{y}\in\{\mathcal{Y}_{m}\}$, we have $\mathbf{x}\oplus\mathbf{y}\in\{\mathcal{T}_{m}\}$.

By summarizing the above statements, we find that the symmetry condition divides the total space $\{\mathcal{H}\}$ to $l$ subsets of the inputs and the transformation matrices $\{\mathcal{H}_{m}\}\equiv\left\{\{\mathcal{X}_{m}\},\{\mathcal{Y}_{m}\},\{\mathcal{R}_{m}\},\{\mathcal{T}_{m}\}\right\}, m=1,\cdots,l$ such that every subset respects the symmetry condition. An explicit representation of matrices $\{\mathcal{H}_{m}\}, m=1,\cdots,l$ are given Sec. \ref{V} for $n=6$.



In one of subsets, for example $m$th subset, we consider symmetry condition for four pairs of inputs $(\mathbf{x}_{i}, \mathbf{y}_{j}), i,j=1,2$ and define
\begin{eqnarray}\label{con3p}
&&\mathbf{R}(\mathbf{x}_{1}\oplus\mathbf{y}_{1})=\mathbf{T}_{x_{1}y_{1}},\hspace{.3cm} (\textit{i})\hspace{2.2cm}
\mathbf{R}(\mathbf{x}_{1}\oplus\mathbf{y}_{2})=\mathbf{T}_{x_{1}y_{2}},\hspace{.3cm} (\textit{ii})
\nonumber\\
&&\mathbf{R}(\mathbf{x}_{2}\oplus\mathbf{y}_{1})=\mathbf{T}_{x_{2}y_{1}},\hspace{.3cm} (\textit{iii})\hspace{2cm}
\mathbf{R}(\mathbf{x}_{2}\oplus\mathbf{y}_{2})=\mathbf{T}_{x_{2}y_{2}},\hspace{.3cm} (\textit{iv})
\end{eqnarray}
where, $\mathbf{T}_{x_{i}y_{j}}\in\{\mathcal{T}_{m}\}, \, i,j=1,2$, and any summation of $\mathbf{T}_{x_{i}y_{j}}, \, i,j=1,2$ belong to subspace $\mathbf{T}_{x_{i}y_{j}}\oplus\mathbf{T}_{x_{k}y_{l}}\in\{\mathcal{T}_{m}\}, \, i,j,k,l=1,2$. Hence, we know that for any $\mathbf{T}\in\{\mathcal{T}_{m}\}$, we have $\mathbf{T}\cdot\mathbf{R}(\mathbf{x}_{i}\oplus\mathbf{y}_{j})=0, \forall \, i,j,k,l=1,2$.

By summing Eqs. (\textit{i}) and (\textit{iii}), we get the first line of below relation, and by summing Eqs. (\textit{i}) and Eq. (\textit{ii}), we get the second line as:
\begin{eqnarray}\label{con3i}
&&i\oplus iii=\mathbf{R}(\mathbf{x}_{1}\oplus\mathbf{x}_{2})=\mathbf{T}_{x_{1}y_{1}}\oplus\mathbf{T}_{x_{2}y_{1}}\equiv\mathbf{T}_{x} ,\nonumber\\
&&i\oplus ii=\mathbf{R}(\mathbf{y}_{1}\oplus\mathbf{y}_{2})=\mathbf{T}_{x_{1}y_{1}}\oplus\mathbf{T}_{x_{1}y_{2}}\equiv\mathbf{T}_{y},
\end{eqnarray}
where, we again have $\mathbf{T}_{x_{1}y_{1}}\oplus\mathbf{T}_{x_{2}y_{1}},\hspace{.1cm} \mathbf{T}_{x_{1}y_{1}}\oplus\mathbf{T}_{x_{1}y_{2}}\in \{\mathcal{T}_{m}\}$. We now product the l.h.s. and the r.h.s. of Eqs. (\ref{con3i}) to each other, and get:
\begin{eqnarray*}\label{condp}
\left[\mathbf{R}(\mathbf{x}_{1}\oplus\mathbf{x}_{2})\right]\cdot\left[\mathbf{R}(\mathbf{y}_{1}\oplus\mathbf{y}_{2})\right]=\mathbf{T}_{x}\cdot\mathbf{T}_{y}=0 \hspace{.2cm}(\text{mod}\hspace{.1cm} 2),
\end{eqnarray*}
or equivalently:
\begin{eqnarray}\label{cond1p}
(\mathbf{x}_{1}\oplus\mathbf{x}_{2})\cdot(\mathbf{y}_{1}\oplus\mathbf{y}_{2})=0 \hspace{.2cm}(\text{mod}\hspace{.1cm} 2), \hspace{.1cm} \forall \hspace{.1cm}\mathbf{x}_{1}, \mathbf{x}_{2}\in \{\mathcal{X}_{m}\},\hspace{.5cm} \forall \hspace{.1cm}\mathbf{y}_{1}, \mathbf{y}_{2}\in \{\mathcal{Y}_{m}\},
\end{eqnarray}
where, we used the orthogonal relation of the rotation matrices (\ref{conp}). Therefore, as pointed out in main text, the symmetry parameter
\begin{eqnarray*}\label{}
\mathcal{W}=\frac{1}{4}\left[1-(-1)^{\mathbf{x}_{1}\cdot(\mathbf{y}_{1}\oplus\mathbf{y}_{2})}\right]
\left[1+(-1)^{\mathbf{x}_{2}\cdot(\mathbf{y}_{1}\oplus\mathbf{y}_{2})}\right],
\end{eqnarray*}
is equal to zero $\mathcal{W}=0$ as measure of symmetry of space in the NL-box framework.

\section{The probabilistic version of NL-box model}\label{III}

A general form of probability distribution for the imperfect NL-box models are usually parameterized as
\begin{equation}\label{nlx}
P(\alpha,\beta\mid\mathbf{x},\mathbf{y})=\xi P_{\mathrm{NLB}}(\alpha,\beta\mid\mathbf{x},\mathbf{y})+ \gamma P_{\mathrm{L}}(\alpha,\beta\mid\mathbf{x},\mathbf{y})+(1-\xi-\gamma)P_{\mathbf{I}},
\end{equation}
where $P_{\mathrm{L}}(\alpha,\beta\mid\mathbf{x},\mathbf{y})=\delta_{0\alpha}\delta_{0\beta}$ is local noise and $P_{\mathbf{I}}=\frac{1}{d_{\alpha}}\frac{1}{d_{\beta}}\frac{1}{d_{x}}\frac{1}{d_{y}}$ represents classical random noise such that box outputs completely random bits, regardless of the inputs \cite{ML1,LO}, where $d_{\alpha},\, d_{\beta},\, d_{x}$ and $d_{y}$ refer to dimensions of the parties outputs and inputs spaces, respectively. The noise parameters $\xi$ and $\gamma$ are constant and inherent noises which we do not specify what has caused them. Additionally, there are other types of imperfect NL-box models \cite{ML1,LO,MBL,QV,QV1,QV2,QV3,QC1,QC2,QC3,QC4,QC5,QC6,QC7,QC8} which have stronger correlation than quantum mechanics.

The general imperfect correlation (\ref{nlx}) has been applied to determine upper bound of various bipartite quantum correlations, from maximally quantum states to nonmaximally/mixed quantum states \cite{ML1,LO,MBL,QV,QV1,QV2,QV3,QC1,QC2,QC3,QC4,QC5,QC6,QC7,QC8}. In these models, the parameter $\gamma$ characterizes value of local marginal correlations $\langle A\rangle=\langle(-1)^{\alpha}\rangle=\sum_{\alpha}(-1)^{\alpha}P(\alpha\mid\mathbf{x})$ which is nonzero ($\langle A\rangle\neq0$) for any values of $\gamma\neq0$, similar relation holds for Bob $\langle B\rangle=\langle(-1)^{\beta}\rangle=\sum_{\beta}(-1)^{\beta}P(\beta\mid\mathbf{y})\neq0$. Here, $\sum_{\beta} P(\alpha,\beta\mid\mathbf{x},\mathbf{y})=P(\alpha\mid\mathbf{x})$ and similar relation holds for Bob. The upper bound of these type of postquantum correlations ($\gamma\neq0$) lie below of the Tsirelson bound which usually is not determinable by the information-theoretic principles \cite{LO,QV1,QV3}. Determining the upper bound of the nonmaximally/mixed quantum correlations is out of scope of present article, meanwhile, we expect that our approach sheds light on the simulating nonmaximally/mixed quantum correlations, see Sec. \ref{IVD} for determining upper bound of some types of the multipartite postquantum correlations.


As we will show in Sec. \ref{IVA}, we will analytically derive the Tsirelson bound as the natural consequence of the NL-box Assumption $3$ without reference to or need for the probabilistic interpretation of the NL-box models. However, we employ the imperfect NL-box models which are suitable to determine the Tsirelson bound and respect $\langle A\rangle=\langle B\rangle=0$.

Any bipartite probabilistic NL-box models, with local marginal probability of zero, can be simulated by the joint conditional probability distributions as:
\begin{equation}\label{nlp}
P_{\mathrm{imp-NLB}}(\alpha,\beta|\mathbf{x},\mathbf{y})= \left\{
\begin{array}{ll}
p/2 &: \alpha\oplus\beta=f(\mathbf{x},\mathbf{y}) \hspace{.2 cm}(\mathrm{mod}\hspace{.1 cm}2)  \\
(1-p)/2 &:\alpha\oplus\beta\neq f(\mathbf{x},\mathbf{y}) \hspace{.2 cm}(\mathrm{mod}\hspace{.1 cm}2).
\end{array}        \right.
\end{equation}
Here, the the white noises are \textit{absorbed} in the definition of probability $p$. Similar to perfect NL-box models, in the earlier versions of the imperfect NL-box models, the inputs $\mathbf{x}$ and $\mathbf{y}$ are binary numbers $x, y\in\left\{0,1\right\}$ or integer numbers $x,y\in\{0,1,...,m-1\}$ \cite{IC,ML1,LO}, and $f(\mathbf{x},\mathbf{y})$ is usually the product of inputs $xy$. The outputs are binary numbers $\alpha,\beta\in\left\{0,1\right\}$, and $\alpha\oplus\beta$ is the addition of outputs calculated modulo $2$. The joint probability (\ref{nlp}) satisfies the no-signaling condition $\sum_{\beta}P(\alpha,\beta|\mathbf{x},\mathbf{y})=P(\alpha|\mathbf{x})$. A similar relation holds for $P(\beta|\mathbf{y})$.

In this scenario the correlation functions is given by:
\begin{eqnarray}\label{Ep}
E(\mathbf{x},\mathbf{y})=P(\alpha=\beta| \mathbf{x},\mathbf{y})-P(\alpha\neq\beta| \mathbf{x},\mathbf{y})=2P(\alpha=\beta| \mathbf{x},\mathbf{y})-1,
\hspace {.0cm}\text{with}\hspace {.0cm} -1\leqslant E \leqslant1.
\end{eqnarray}
Different theoretical frameworks give different upper bounds on the CHSH parameter $\mathcal{S}$,
which can be used to quantify the strength of correlations $E$.
The NL-box models reach the maximum possible value $\mathcal{S}_{\mathrm{NLB}}\leqslant 4$, which corresponds to $0\leqslant E_{\mathrm{NLB}}\leqslant 1$. Local realistic theories/models give the upper bound $\mathcal{S}_{\mathrm{L}}\leqslant2$ (the very CHSH inequality \cite{CHSH}), which corresponds to $0\leqslant E_{\mathrm{L}}\leqslant 1/2$. The upper bound of local models is violated once $E>1/2$, and quantum correlations exceed the Tsirelson bound $\mathcal{S}_{\mathrm{Q}}\leqslant2\sqrt{2}$ with $0\leqslant E_{\mathrm{Q}}\leqslant\sqrt{2}/2$, which is attained by performing suitable measurements on pair of qubits in the singlet state.

Similar to the perfect NL-box models and we will show in the next subsection, the earlier versions of the imperfect NL-box models are not admissible. Therefore, to remedy this issue, we increase the dimension of the inputs spaces to include strings of $n$ binary digits, $\mathbf{x}=(x_{1},\cdots,x_{n})$ and $\mathbf{y}=(y_{1},\cdots,y_{n})$ with $x_{i},y_{j}\in\{0,1\}$. In addition, the function $f(\mathbf{x},\mathbf{y})$ in Eq. (\ref{nlp}) becomes the inner product function of the inputs,
$f(\mathbf{x},\mathbf{y})=\mathrm{IP}_{\mathrm{n}}(\mathbf{x}, \mathbf{y})=\bigoplus_{i=1}^{n} x_{i}y_{i}=\mathbf{x}\cdot \mathbf{y}=x_{1}y_{1}\oplus...\oplus x_{n}y_{n} \hspace{.1cm}(\text{mod}\hspace{.1cm} 2)$, where $x_{i}y_{i}$ is the usual product of the input elements and the summation is calculated modulo $2$.

We clarify all rounds of the imperfect NL-box machine (\ref{nlp}) as mixture of ``correct" and ``incorrect" operations in which $n_{p}$ rounds give correct outputs, where $p=\frac{n_{p}}{n_{p}+n_{1-p}}$, and $n_{1-p}$ rounds give incorrect outputs, where $1-p=\frac{n_{1-p}}{n_{p}+n_{1-p}}$. As represented in Eq. (\ref{nlp}), the probability $p$ (or equivalent noises $\xi$ and $\gamma$ in Eq. (\ref{nlx})) is constant parameter which will be determined by imposing a principle. In the next section, we impose the NL-box Assumption $1, 2$ and $3$ on the $P_{\mathrm{imp-NLB}}(\alpha,\beta|\mathbf{x},\mathbf{y})$ and require the probability distributions remain invariant. The distribution $P_{\mathrm{imp-NLB}}$.




\subsection{The NL-box Assumption $1, 2$ and $3$ in the imperfect NL-box model}\label{IIIA}

Similar to perfect NL-box scenario, the symmetry Assumptions $1-3$ shall be revisited in the imperfect NL-box models scenario.


As pointed out in Assumption $1$, the probability distribution of opposite directions is defined as $P(A=B|\mathbf{a},\mathbf{b})=P(-A= B|-\mathbf{a},\mathbf{b})=1-P(-A=B|\mathbf{a},\mathbf{b})=1-P(A=B|-\mathbf{a},\mathbf{b})$, where $A, B\in\{-1, 1\}$ and $\mathbf{a},\mathbf{b}\in\mathrm{\mathbb{R}^{3}}$.

\textbf{Imperfect NL-box Assumption $1$}: The perfect correlation (anti-correlation) conservation law shall be held in the imperfect NL-box models as:
\begin{eqnarray*}\label{}
P(\alpha=\beta|\alpha\oplus\beta=\mathbf{x}\cdot\mathbf{y}, \mathbf{x},\mathbf{y})&=& 1-P(\alpha\oplus1=\beta|\alpha\oplus\beta=\mathbf{x}\cdot\mathbf{y}, \mathbf{x},\mathbf{y})\\
&=&1-P(\alpha=\beta|(\alpha\oplus1)\oplus\beta=\mathbf{x}\cdot\mathbf{y}, \mathbf{x},\mathbf{y})\\
&=&1-P(\alpha=\beta|\alpha\oplus\beta=1\oplus\mathbf{x}\cdot\mathbf{y}, \mathbf{x},\mathbf{y}),\\
P(\alpha=\beta|\alpha\oplus\beta=\mathbf{x}\cdot\mathbf{y}, \mathbf{x},\mathbf{y})&=&1-P(\alpha=\beta|\alpha\oplus\beta=\bar{\mathbf{x}}\cdot\mathbf{y}, \bar{\mathbf{x}},\mathbf{y}),
\end{eqnarray*}
in the second line, we renamed $\alpha\oplus1$ as $\alpha$. The probability distributions on the third and the fourth lines are equal
\begin{eqnarray*}\label{Pro1p}
P(\alpha=\beta|\alpha\oplus\beta=1\oplus\mathbf{x}\cdot\mathbf{y}, \mathbf{x},\mathbf{y})=
P(\alpha=\beta|\alpha\oplus\beta=\bar{\mathbf{x}}\cdot\mathbf{y}, \bar{\mathbf{x}},\mathbf{y}), \hspace{.2cm}(\text{mod}\hspace{.1cm} 2), \hspace{.1cm} \forall \hspace{.1cm}\mathbf{x}, \mathbf{y}.
\end{eqnarray*}
It shows that $1\oplus\mathbf{x}\cdot\mathbf{y}=\bar{\mathbf{x}}\cdot\mathbf{y}, \hspace{.1cm}(\text{mod}\hspace{.1cm} 2), \hspace{.1cm} \forall \hspace{.1cm}\mathbf{x}, \mathbf{y}$ which is the same as Eq. (\ref{nl2p}) in the perfect NL-box framework. Similarly, other relations in Eq. (\ref{nl2p}) can be derived. A similar relations hold for Bob. This indicates that the parties input spaces are subset of input spaces $\{0,1\}^{n}$ which we represent the input subspaces by $\mathbf{x}\in\{\mathcal{X}\}$ and $\mathbf{y}\in\{\mathcal{Y}\}$. Therefore, Eqs. (\ref{nup}) and (\ref{nl14p}) hold for the measurement inputs in the imperfect NL-box models.

As indicated in the main text, the original version of the perfect NL-box models are not admissible. Now, according to Eq. (\ref{nl2p}) which holds in the imperfect NL-box models, the original version of \textit{imperfect NL-box} models \cite{IC,ML1,LO} are not admissible. As an interesting point, for bipartite systems with zero total angular momentum (singlet states in quantum mechanics framework), the probability of obtain same outputs for opposite measurement inputs is equal to one $P(\alpha,\beta|\alpha\oplus\beta=\bar{\mathbf{x}}\cdot\mathbf{x}, \bar{\mathbf{x}},\mathbf{x})=1$, where $\bar{\mathbf{x}}\cdot\mathbf{x}=0, \hspace{.1cm} (\text{mod}\hspace{.1cm} 2), \hspace{.1cm} \forall \hspace{.1cm}\mathbf{x}$. This property indicates that  models admissible to describe zero total angular momentum systems.

\textbf{Imperfect NL-box Assumption $2$}: The imperfect NL-boxes shall respect homogeneous and isotropic symmetry group structures. In particular, the set of transformation operators $F\in\left\{\mathcal{F}\right\}$ in the imperfect NL-box models framework shall be respect group axioms. The same as Eq. (\ref{corr}),  the Boolean symmetry transformations $F\in\left\{\mathcal{F}\right\}$ are multivariate polynomial over $\{0,1\}^{n}$ with the component functions $f^{l}: \{0,1\}^{n}\rightarrow \{0,1\}$. Similar to the perfect NL-box Assumption $2$, the imperfect NL-box models shall respect symmetry group axioms ($i,ii,iii,iv$). Hence, the Boolean transformations $F$ maintain the imperfect NL-box Assumption $1$, in which two opposite inputs $\mathbf{x}$ and $\bar{\mathbf{x}}$ transform to two opposite directions $F(\mathbf{x})$ and $F(\bar{\mathbf{x}})$ under operation of $F$.

\textbf{Imperfect NL-box Assumption $3$}: The probability distributions remain invariant under symmetry transformation $F$. One of properties of NL-box Assumption $3$ is that the Boolean transformations $F$ maintain the NL-box Assumption $1$, in which two opposite inputs $\mathbf{x}$ and $\bar{\mathbf{x}}$ transform to two opposite directions $F(\mathbf{x})$ and $F(\bar{\mathbf{x}})$ under operation of $F$. We prove this property in the imperfect NL-box models by considering following relations:
\begin{eqnarray*}\label{}
P(\alpha,\beta|\alpha\oplus\beta=\bar{\mathbf{x}}\cdot\mathbf{y})&=&P(\alpha,\beta|\alpha\oplus\beta=F(\bar{\mathbf{x}})\cdot F(\mathbf{y})),\\
P(\alpha,\beta|\alpha\oplus\beta=(1\oplus\mathbf{x})\cdot\mathbf{y})&=&P(\alpha,\beta|\alpha\oplus\beta=F(1\oplus\mathbf{x})\cdot F(\mathbf{y})),\\
P(\alpha,\beta|(1\oplus\alpha)\oplus\beta=\mathbf{x}\cdot\mathbf{y})&=&P(\alpha,\beta|\alpha\oplus\beta=F(1\oplus\mathbf{x})\cdot F(\mathbf{y})),\\
P(\alpha,\beta|(1\oplus\alpha)\oplus\beta=F(\mathbf{x})\cdot F(\mathbf{y}))&=&P(\alpha,\beta|\alpha\oplus\beta=F(1\oplus\mathbf{x})\cdot F(\mathbf{y})),\\
P(\alpha,\beta|\alpha\oplus\beta=1\oplus F(\mathbf{x})\cdot F(\mathbf{y}))&=&P(\alpha,\beta|\alpha\oplus\beta=F(1\oplus\mathbf{x})\cdot F(\mathbf{y})),\\
P(\alpha,\beta|\alpha\oplus\beta=(1\oplus F(\mathbf{x}))\cdot F(\mathbf{y}))&=&P(\alpha,\beta|\alpha\oplus\beta=F(1\oplus\mathbf{x})\cdot F(\mathbf{y})),
\end{eqnarray*}
where, we have omitted measurement inputs for reduce of redundancy and on the fourth line, we have used of the NL-box Assumption $3$ and invariance of the probabilities under symmetry transformations. The last line means that $F(\bar{\mathbf{x}}) =F(1\oplus\mathbf{x})=1\oplus F(\mathbf{x}),\hspace{.1cm} \forall \hspace{.1cm}\mathbf{x}, F, \hspace{.1cm}(\text{mod}\hspace{.1cm} 2)$. This property is sufficient to derive explicit form of the transform function as $F(\mathbf{x})=\mathbf{R}\mathbf{x}\oplus\mathbf{T}$. However, we derive the symmetry transformations from another way.

The symmetry transform function $F(\mathbf{x})=\mathbf{R}\mathbf{x}\oplus\mathbf{T}$ can be derived in the same way which is used in the perfect NL-box scheme. In particular, we assumed that inner-product of inputs $\mathbf{x}\cdot \mathbf{y}$ is saved and mapped to transformed inputs $F(\mathbf{x})\cdot F(\mathbf{y})$. Besides, input space $\{\mathcal{X}\}$ is close under summation of every odd number of inputs. For example, three inputs $\mathbf{x}_{1}, \mathbf{x}_{2}$, and $\mathbf{x}_{3}$ which belong to input space $\mathbf{x}_{1}, \mathbf{x}_{2}, \mathbf{x}_{3}\in\{\mathcal{X}\}$, their summation belongs to input space again, $\mathbf{x}=\mathbf{x_{1}}\oplus\mathbf{x_{2}}\oplus\mathbf{x_{3}}$. This indicate that:
\begin{eqnarray*}\label{}
P(\alpha,\beta|\alpha\oplus\beta=\mathbf{x}\cdot\mathbf{y})&=&P(\alpha,\beta|\alpha\oplus\beta=F(\mathbf{x})\cdot F(\mathbf{y}))\\
P(\alpha,\beta|\alpha\oplus\beta=\left[\mathbf{x}_{1}\oplus\mathbf{x}_{2}\oplus\mathbf{x}_{3}\right]\cdot\mathbf{y})&=&P(\alpha,\beta|\alpha\oplus\beta
=F(\mathbf{x})\cdot F(\mathbf{y}))\\
P(\alpha,\beta|\alpha\oplus\beta=\mathbf{x}_{1}\cdot \mathbf{y}\oplus\mathbf{x}_{2}\cdot \mathbf{y}\oplus\mathbf{x}_{3}\cdot \mathbf{y})&=&P(\alpha,\beta|\alpha\oplus\beta=F(\mathbf{x})\cdot F(\mathbf{y}))\\
P(\alpha,\beta|\alpha\oplus\beta=F(\mathbf{x}_{1})\cdot F(\mathbf{y})\oplus F(\mathbf{x}_{2})\cdot F(\mathbf{y})\oplus F(\mathbf{x}_{3})\cdot F(\mathbf{y}))&=&P(\alpha,\beta|\alpha\oplus\beta=F(\mathbf{x})\cdot F(\mathbf{y})),\\
P(\alpha,\beta|\alpha\oplus\beta=\left[F(\mathbf{x}_{1})\oplus F(\mathbf{x}_{2})\oplus F(\mathbf{x}_{3})\right]\cdot F(\mathbf{y}))&=&P(\alpha,\beta|\alpha\oplus\beta=F(\mathbf{x})\cdot F(\mathbf{y})),\\
&&\hspace{1.5cm} \forall \hspace{.1cm}\mathbf{x}, \mathbf{y}, F, \hspace{.2cm}(\text{mod}\hspace{.1cm} 2),
\end{eqnarray*}
in the fourth line we have used symmetry relations $\mathbf{x}_{i}\cdot\mathbf{y}=F(\mathbf{x}_{i})\cdot F(\mathbf{y}),\, i=1,2,3$. The last line shows that the $F$ is a liner function of inputs up to a function $K(\mathbf{x})$
\begin{eqnarray}\label{sum}
F(\mathbf{x}_{1}\oplus\mathbf{x}_{2}\oplus\mathbf{x}_{3})=F(\mathbf{x}_{1})\oplus F(\mathbf{x}_{2})\oplus F(\mathbf{x}_{3})\oplus K(\mathbf{x}),\hspace{.4cm} \forall \hspace{.2cm}\mathbf{x}_{1}, \mathbf{x}_{2}, \mathbf{x}_{3}, F, \hspace{1.7cm}\\\nonumber
K(\mathbf{x})\cdot F(\mathbf{y})=0,\hspace{.4cm} \forall \hspace{.1cm}\mathbf{x}, \mathbf{y}, F, \hspace{.2cm}(\text{mod}\hspace{.1cm} 2),
\end{eqnarray}
the last line means $K(\mathbf{x})\equiv \mathbf{0}$.

Now, we can repeat Sec. (\ref{IIE}) and Sec. (\ref{IIF}), and re-derive the symmetry transformations $F(\mathbf{x})=\mathbf{R}\mathbf{x}\oplus\mathbf{T}$ for all of $F\in\{\mathcal{F}\}$, in the same way which was performed in the perfect NL-box scheme.

\subsection{Derivation of the Tsirelson bound in the imperfect NL-box models}\label{IIIB}

As pointed out at the main text, the symmetry condition (\ref{conp}) is not consistence with violation of the CHSH inequality at the maximum value of $4$. However, the nonlocality and the isotropy/homogeneity of space are two physical facts of nature which these features indeed coexist and are always consistent. Therefore, a complete NL-box model shall exhibit both features at the same time. To remedy inconsistency, we take imperfect version of the NL-box models (\ref{nlp}) which simultaneously relaxes the symmetry conditions (\ref{conp}) and perfect nonlocal correlations so that the symmetry conditions become compatible with degree of nonlocality as has predicted in experiments.

In what follows, we will determine probability of obtaining the symmetry condition (\ref{conp}) and the probability of obtaining maximum value of nonlocality $4$, afterwards, we will show that inconsistency is lifted exactly at the Tsirelson bound. To determine probability of obtaining the symmetry condition, we take four pairs of the CHSH experiment settings $(\mathbf{x}_{i}, \mathbf{y}_{j})\, i,j=1,2$ in which the correct measurement outputs of any pair is given by probability $p$,
\begin{eqnarray}\label{Be12p}
&&\alpha_{11}\oplus\beta_{11}=\mathbf{x}_{1}\cdot\mathbf{y}_{1}=0, \hbox{is given correct outputs with probability} \hspace{.1cm} p,\\\nonumber
&&\alpha_{12}\oplus\beta_{12}=\mathbf{x}_{1}\cdot\mathbf{y}_{2}=1, \hbox{is given correct outputs with probability} \hspace{.1cm} p,\\\nonumber
&&\alpha_{21}\oplus\beta_{21}=\mathbf{x}_{2}\cdot\mathbf{y}_{1}=0, \hbox{is given correct outputs with probability} \hspace{.1cm} p,\\\nonumber
&&\alpha_{22}\oplus\beta_{22}=\mathbf{x}_{2}\cdot\mathbf{y}_{2}=0, \hbox{is given correct outputs with probability} \hspace{.1cm} p,\hspace{.3cm}(\mathrm{mod}\hspace{.1cm} 2).
\end{eqnarray}


\begin{table}
\begin{tabular}{|c||c|c|}
  \hline
   The symmetry conditions $\rightarrow$ & $[\mathbf{R}(\mathbf{x}\oplus\mathbf{y})]\cdot\mathbf{T}=0$ &  $[\mathbf{R}(\mathbf{x}\oplus\mathbf{y})]\cdot\mathbf{T}=1$ \\
   The probability of symmetry conditions $\downarrow$  &   &      \\
    \hline\hline
$P([\mathbf{R}(\mathbf{x}_{1}\oplus\mathbf{y}_{1})]\cdot\mathbf{T})$ & $p^{2}+(1-p)^{2}$ & $1-p^{2}-(1-p)^{2}$  \\
    \hline
$P([\mathbf{R}(\mathbf{x}_{1}\oplus\mathbf{y}_{2})]\cdot\mathbf{T})$ & $p^{2}+(1-p)^{2}$ & $1-p^{2}-(1-p)^{2}$ \\
      \hline
$P([\mathbf{R}(\mathbf{x}_{2}\oplus\mathbf{y}_{1})]\cdot\mathbf{T})$ & $p^{2}+(1-p)^{2}$ & $1-p^{2}-(1-p)^{2}$ \\
      \hline
$P([\mathbf{R}(\mathbf{x}_{2}\oplus\mathbf{y}_{2})]\cdot\mathbf{T}\,|\, I,II,III)$&       $1$          &         $$               \\
\hline
$P([\mathbf{R}(\mathbf{x}_{2}\oplus\mathbf{y}_{2})]\cdot\mathbf{T}\,|\, I',II',III')$&       $$          &         $1$               \\
\hline
$\mathcal{Q}(\mathcal{W}_{k}=0)=\prod_{i,j=1,2}P([\mathbf{R}(\mathbf{x}_{i}\oplus\mathbf{y}_{j})]\cdot\mathbf{T})$                                                  & $[p^{2}+(1-p)^{2}]^{3}$ & $[1-p^{2}-(1-p)^{2}]^{3}$ \\
  \hline
\end{tabular}
\begin{tabular}{|c|}
   $\hspace{1.04cm}\mathcal{Q}(\mathcal{W}=0)=\sum_{k=1,2}P(\mathcal{W}_{k}=0)= [p^{2}+(1-p)^{2}]^{3}+[1-p^{2}-(1-p)^{2}]^{3}\hspace{.42cm}$\\
    \hline
  \end{tabular}
\caption{Derivation of $\mathcal{Q}(\mathcal{W}=0)$. In the imperfect NL-box models, the symmetry condition (\ref{unlp}) will be a probabilistic quantity which may respect or disrespect with probabilities $P(\mathbf{T}\cdot[\mathbf{R}(\mathbf{x}_{i}\oplus\mathbf{y}_{j})]=0)$ and $P(\mathbf{T}\cdot[\mathbf{R}(\mathbf{x}_{i}\oplus\mathbf{y}_{j})]=1)$, respectively.
The first row presents the respected and disrespected symmetry conditions and the first column represents probability distributions for measurement inputs $(\mathbf{x}_{i}\cdot\mathbf{y}_{j})\, i,j=1,2$. In the last row, we represent the total probability $\mathcal{Q}(\mathcal{W}=0)$ which is equal to summation of the subsets $\mathcal{Q}(\mathcal{W}_{k}=0), \, k=1,2$.}\label{table1}
\end{table}


Now, for any pairs of measurement inputs $(\mathbf{x}_{i}, \mathbf{y}_{j})\, i,j=1,2$, the correlation functions (\ref{Be12p}) are transformed under symmetry transformation $F\in\left\{\mathcal{F}\right\}$ to:
\begin{eqnarray}\label{unlp}
&&\alpha'_{11}\oplus\beta'_{11}=\mathbf{x}_{1}\cdot\mathbf{y}_{1}\oplus\mathbf{T}\cdot[\mathbf{R}(\mathbf{x}_{1}\oplus\mathbf{y}_{1})]=0, \hbox{is given correct outputs with probability} \hspace{.1cm} p,\nonumber\\\nonumber
&&\alpha'_{12}\oplus\beta'_{12}=\mathbf{x}_{1}\cdot\mathbf{y}_{2}\oplus\mathbf{T}\cdot[\mathbf{R}(\mathbf{x}_{1}\oplus\mathbf{y}_{2})]=1, \hbox{is given correct outputs with probability} \hspace{.1cm} p,\\\nonumber
&&\alpha'_{21}\oplus\beta'_{21}=\mathbf{x}_{2}\cdot\mathbf{y}_{1}\oplus\mathbf{T}\cdot[\mathbf{R}(\mathbf{x}_{2}\oplus\mathbf{y}_{1})]=0, \hbox{is given correct outputs with probability} \hspace{.1cm} p,\\\nonumber
&&\alpha'_{22}\oplus\beta'_{22}=\mathbf{x}_{2}\cdot\mathbf{y}_{2}\oplus\mathbf{T}\cdot[\mathbf{R}(\mathbf{x}_{2}\oplus\mathbf{y}_{2})]=0, \hbox{is given correct outputs with probability} \hspace{.1cm} p,\nonumber\\
&&\hspace{10cm}(\mathrm{mod}\hspace{.1cm} 2), \hspace{.1cm} \forall\hspace{.1cm} \mathbf{R}, \mathbf{T}.
\end{eqnarray}

\textit{Remark 7}. In Eqs. (\ref{unlp}), we have used $\mathbf{x}\cdot\mathbf{y}$ instead of $\mathbf{x}\mathbf{R}^{t}\cdot\mathbf{R}\mathbf{y}$. As we know, the probabilities of obtaining correct outcomes are equal for these two sets of measurement inputs. In particular, we could consider four pairs of measurement settings $(\mathbf{R}\mathbf{x}_{i}, \mathbf{R}\mathbf{y}_{j})\, i,j=1,2$ instead measurement settings $(\mathbf{x}_{i}, \mathbf{y}_{j})\, i,j=1,2$ in the statements (\ref{Be12p}). Thus, the rest of calculations are not affected.

In relations (\ref{unlp}), we have employed the NL-box Assumption $3$ in which the probability $p$ has remained invariant under the symmetry group transformations $F\in\left\{\mathcal{F}\right\}$. We now use the statements (\ref{Be12p}) and (\ref{unlp}) to determine probability of obtaining the symmetry conditions (\ref{conp}). Specifically, for pair of the measurement setting $(\mathbf{x}_{i}, \mathbf{y}_{j})$, if the correlation functions Eq. (\ref{nlp}) simultaneity be correct
\begin{eqnarray*}
\left\{
  \begin{array}{ll}
    \mathbf{x}_{i}\cdot\mathbf{y}_{j}=c, & \hbox{is given correct outputs $c$ with probability $p$;}\\
    \mathbf{x}_{i}\cdot\mathbf{y}_{j}\oplus\mathbf{T}\cdot[\mathbf{R}(\mathbf{x}_{i}\oplus\mathbf{y}_{j})]=c, & \hbox{is given correct outputs $c$ with probability $p$; \hspace{.3cm}$\forall\hspace{.1cm} \mathbf{R}, \mathbf{T}$,}
  \end{array}
\right.\label{}
\end{eqnarray*}
the symmetry condition ($\mathbf{T}\cdot[\mathbf{R}(\mathbf{x}_{i}\oplus\mathbf{y}_{j})]=0$) is respected with probability $p^{2}$. Moreover, if the correlation functions Eq. (\ref{nlp}) simultaneity be incorrect
\begin{eqnarray*}
\left\{
  \begin{array}{ll}
    \mathbf{x}_{i}\cdot\mathbf{y}_{j}=c\oplus1, & \hbox{is given correct outputs $c\oplus1$ with probability $p$;}\\
    \mathbf{x}_{i}\cdot\mathbf{y}_{j}\oplus\mathbf{T}\cdot[\mathbf{R}(\mathbf{x}_{i}\oplus\mathbf{y}_{j})]=c\oplus1, & \hbox{is given correct outputs $c\oplus1$ with probability $p$; \hspace{.3cm}$\forall\hspace{.1cm} \mathbf{R}, \mathbf{T}$,}
  \end{array}
\right.\label{}
\end{eqnarray*}
we get \textit{the symmetry condition} (\ref{conp}) again, with probability $(1-p)^{2}$. Here, $i,j=1,2$ and the $c\in\{0,1\}$ refers to correct outputs. Therefore, the symmetry condition is respected with probability $p^{2}+(1-p)^{2}$ for a specific pair of inputs.

Similarly, we can determine the symmetry conditions for two couple pairs of measurement inputs $(\mathbf{x}_{i}, \mathbf{y}_{j})\, i,j=1,2$ which give the following relations:
\begin{eqnarray}
\left\{
  \begin{array}{ll}
    \mathbf{T}\cdot[\mathbf{R}(\mathbf{x}_{1}\oplus\mathbf{y}_{1})]=0, & \hbox{is given with probability $p^{2}+(1-p)^{2}$;} \hspace{1.5cm}(I)\\
    \mathbf{T}\cdot[\mathbf{R}(\mathbf{x}_{1}\oplus\mathbf{y}_{2})]=0, & \hbox{is given with probability $p^{2}+(1-p)^{2}$;} \hspace{1.5cm}(II)\\
    \mathbf{T}\cdot[\mathbf{R}(\mathbf{x}_{2}\oplus\mathbf{y}_{1})]=0, & \hbox{is given with probability $p^{2}+(1-p)^{2}$;} \hspace{1.5cm}(III)\\
    \mathbf{T}\cdot[\mathbf{R}(\mathbf{x}_{2}\oplus\mathbf{y}_{2})]=0, & \hbox{is given with probability \hspace{1cm} $1$,}\\
                                                                       & \hspace{.5cm}\hbox{if I,II,and III simultaneously be correct;} \hspace{1.cm}(IV).
  \end{array}
\right.\label{pro22}
\end{eqnarray}

We shall note that the relations (\ref{pro22}) are not independent because the summation of the symmetry conditions is equal to zero $\sum_{i,j=1,2}\mathbf{T}\cdot[\mathbf{R}(\mathbf{x}_{i}\oplus\mathbf{y}_{j})]=0,\hspace{.3cm}(\text{mod}\hspace{.1cm} 2)$. In particular, if the relations I, II, and III simultaneously are respected, then, the relation IV is respected with probability $1$. Therefore, the relations (\ref{pro22}) simultaneously are respected with probability of $[p^{2}+(1-p)^{2}]^{3}$. The relations (\ref{pro22}) can be used to generate similar Eqs. (\ref{con3p}) and (\ref{con3i}) in the imperfect NL-box scenario which gives Eq. (\ref{cond1p}).


\begin{table}
\begin{tabular}{|c||c|c|c|c|}
  \hline

  Correlation functions with $\mathcal{W}_{k}=1\rightarrow$ & $c_{11}\neq c_{12}=1$ & $c_{11}\neq c_{12}=0$ & $c_{11}\neq c_{12}=1$ & $c_{11}\neq c_{12}=0$\\

     Probability distributions $\downarrow$  & $c_{21}=c_{22}=0$ & $c_{21}=c_{22}=1$ & $c_{21}=c_{22}=1$ & $c_{21}=c_{22}=0$\\
    \hline\hline
 $P(\alpha_{11}\oplus\beta_{11}=c_{11}|\mathbf{x}_{1},\mathbf{y}_{1})$& $p$ & $1-p$ & $p$ & $1-p$\\
    \hline
$P(\alpha_{12}\oplus\beta_{12}=c_{12}|\mathbf{x}_{1},\mathbf{y}_{2})$& $p$ & $1-p$ & $p$ & $1-p$  \\
      \hline
$P(\alpha_{21}\oplus\beta_{21}=c_{21}|\mathbf{x}_{2},\mathbf{y}_{1})$& $p$ & $1-p$ & $1-p$ & $p$ \\
      \hline
$P(\alpha_{22}\oplus\beta_{22}=c_{22}|\mathbf{x}_{2},\mathbf{y}_{2})$& $p$ & $1-p$ & $1-p$ & $p$  \\
    \hline
$P(\mathcal{W}_{k}=1)=$ & & & & \\
  $\prod_{i,j=1,2}P(\alpha_{ij}\oplus\beta_{ij}=c_{ij}|\mathbf{x}_{i},\mathbf{y}_{j})$ & $p^{4}$ & $(1-p)^{4}$ & $p^{2}(1-p)^{2}$ & $p^{2}(1-p)^{2}$ \\
  \hline
\end{tabular}
\begin{tabular}{|c|}
   $\hspace{2.35cm}\mathcal{P}(\mathcal{W}=1)=\sum_{k=1}^{4}P(\mathcal{W}_{k}=1)= p^{4}+(1-p)^{4}+2p^{2}(1-p)^{2}\hspace{2.45cm}$\\
    \hline
  \end{tabular}
\caption{Derivation of $\mathcal{P}(\mathcal{W}=1)$. In the first row, we present the correlation functions for all possible value od $c_{ij}\, i,j=1,2$ which give maximum value $\mathcal{S}=4$ for the CHSH parameter, where $\alpha_{ij}\oplus\beta_{ij}=\mathbf{x}_{i}\cdot\mathbf{y}_{j}=c_{ij}$. The first column represents probability of imperfect correlation function (\ref{nlp}) for various possible value of $c_{ij}$. In the last row, we represent the total probability $\mathcal{P}(\mathcal{W}=1)$ which is equal to summation of the subsets  $\mathcal{P}(\mathcal{W}_{k}=1), \, k=1,2,3,4$.}\label{table2}
\end{table}


\begin{figure}
\centering
\includegraphics[scale=.7]{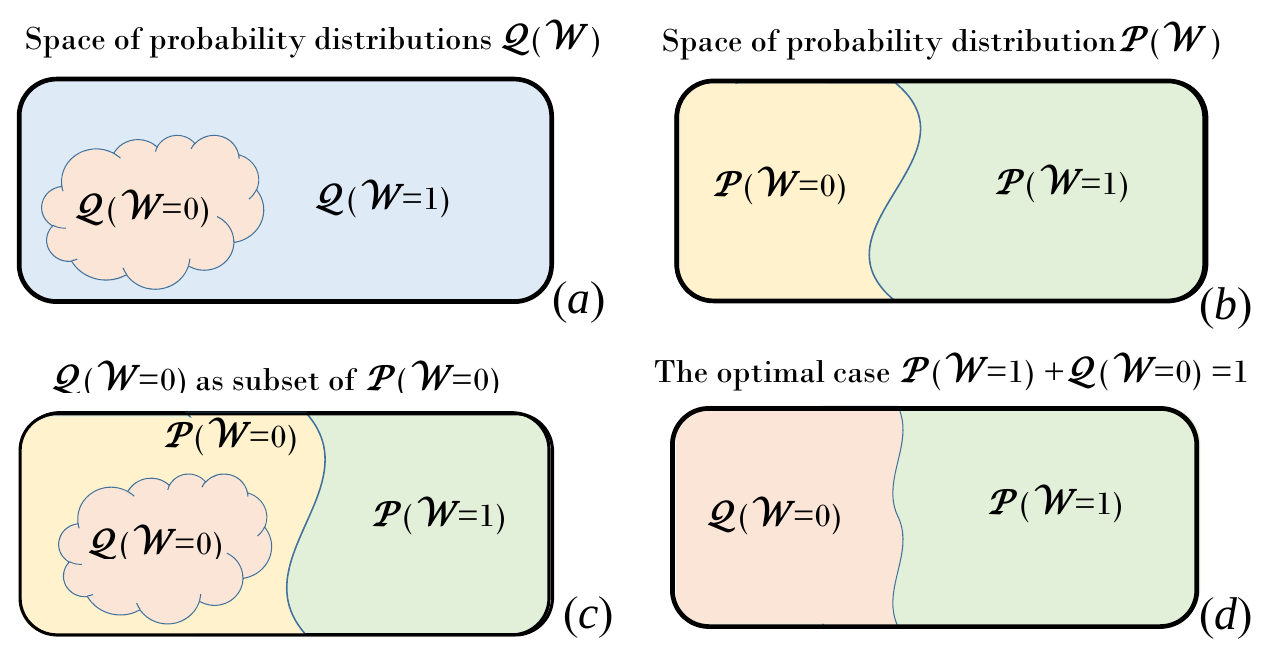}
\caption{($a$) A general representation of the probability distribution $\mathcal{Q}(\mathcal{W})$ for correlation function in interval $-1\leqslant E \leqslant1$. ($b$) Similar representation of the probability distributions $\mathcal{P}(\mathcal{W})$ for an arbitrary $1\leqslant E \leqslant1$. Here, we have used different probability symbols $\mathcal{P}$ and $\mathcal{Q}$ to stress that they are derived from independent methods and have different distributions, the former from the symmetry condition and the latter the imperfect NL-box correlation function. ($c$) In order to have consistent NL-box models, it is sufficient that $\mathcal{Q}(\mathcal{W}=0)$ be a subset of $\mathcal{P}(\mathcal{W}=0)$ in which
the two incompatible properties $\mathcal{W}=0$ and $\mathcal{W}=1$ cannot be detected simultaneously. In other words, $\mathcal{Q}(\mathcal{W}=0)+\mathcal{P}(\mathcal{W}=1)\leqslant 1$ which is equivalent to $-\sqrt{2}/2\leqslant E\leqslant\sqrt{2}/2$. ($d$) The Tsirelson bound $E=\sqrt{2}/2$ is given as $\mathcal{Q}(\mathcal{W}=0)+\mathcal{P}(\mathcal{W}=1)=1$.
}\label{PQ}
\end{figure}


In other hand, we use equations (\ref{Be12p}) and (\ref{unlp}) to determine probability of obtaining \textit{asymmetry conditions} $\mathbf{T}\cdot[\mathbf{R}(\mathbf{x}_{i}\oplus\mathbf{y}_{j})]=1$. We get asymmetry condition, if one of correlation function, for example, (\ref{Be12p}) be correct and (\ref{unlp}) be incorrect or vice-versa
\begin{eqnarray*}\label{}
&&\left\{
  \begin{array}{ll}
    \mathbf{x}_{i}\cdot\mathbf{y}_{j}=c, & \hbox{is given correct outputs with probability} \hspace{.1cm} p, \\
    \mathbf{x}_{i}\cdot\mathbf{y}_{j}\oplus\mathbf{T}\cdot[\mathbf{R}(\mathbf{x}_{i}\oplus\mathbf{y}_{j})]=c\oplus1, & \hbox{is given incorrect outputs with probability} \hspace{.1cm} 1-p,\hspace{.1cm}\forall\hspace{.1cm} \mathbf{R}, \mathbf{T},
  \end{array}
\right.\\
&&\left\{
  \begin{array}{ll}
    \mathbf{x}_{i}\cdot\mathbf{y}_{j}=c\oplus1, & \hbox{is given incorrect outputs with probability} \hspace{.1cm} 1-p, \\
    \mathbf{x}_{i}\cdot\mathbf{y}_{j}\oplus\mathbf{T}\cdot[\mathbf{R}(\mathbf{x}_{i}\oplus\mathbf{y}_{j})]=c, & \hbox{is given correct outputs with probability} \hspace{.1cm} p,\hspace{.3cm}\forall\hspace{.1cm} \mathbf{R}, \mathbf{T},
  \end{array}
\right.
\end{eqnarray*}
where, $i,j=1,2$ and the $c\in\{0,1\}$ refers to correct outputs again. It is equivalent the asymmetry version of symmetry conditions in Eq. (\ref{pro22}) gives the probability distribution equal to $1-p^{2}-(1-p)^{2}$.


Similarly, we can determine the probability of obtaining asymmetry conditions for other pairs of measurement inputs $(\mathbf{x}_{i}, \mathbf{y}_{j})\, i,j=1,2$ and give the probability distributions of asymmetry conditions:
\begin{eqnarray}\label{pro23}
\left\{
  \begin{array}{ll}
   \mathbf{T}\cdot[\mathbf{R}(\mathbf{x}_{1}\oplus\mathbf{y}_{1})]=1, & \hbox{is given with probability $1-[p^{2}+(1-p)^{2}]$;} \hspace{1.5cm}(I')\\
   \mathbf{T}\cdot[\mathbf{R}(\mathbf{x}_{1}\oplus\mathbf{y}_{2})]=1, & \hbox{is given with probability $1-[p^{2}+(1-p)^{2}]$;} \hspace{1.5cm}(II')\\
   \mathbf{T}\cdot[\mathbf{R}(\mathbf{x}_{2}\oplus\mathbf{y}_{1})]=1, & \hbox{is given with probability $1-[p^{2}+(1-p)^{2}]$;} \hspace{1.5cm}(III')\\
   \mathbf{T}\cdot[\mathbf{R}(\mathbf{x}_{2}\oplus\mathbf{y}_{2})]=1, & \hbox{is given with probability \hspace{1cm} $1$,}\\
                                                                       & \hspace{.5cm}\hbox{if I',II',and III' simultaneously be correct;} \hspace{1.3cm}(IV').
  \end{array}
\right.
\end{eqnarray}

Similar to previous step, the above relations (\ref{pro23}) are not independent because the summation of the asymmetry conditions is equal to zero $\sum_{i,j=1,2}\mathbf{T}\cdot[\mathbf{R}(\mathbf{x}_{i}\oplus\mathbf{y}_{j})]=0,\hspace{.3cm}(\text{mod}\hspace{.1cm} 2)$. In particular, if the relations I', II', and III' are simultaneously respected, then, the relation IV' is correct with probability of $1$. Therefore, the relations (\ref{pro23}) simultaneously are correct with probability of $[1-p^{2}-(1-p)^{2}]^{3}$. Similar to previous step, the relations (\ref{pro23}) can be used to generate similar Eqs. (\ref{con3p}) and (\ref{con3i}) in the imperfect NL-box scenario and gives Eq. (\ref{cond1p}).

The relations (\ref{pro23}) are a set of asymmetry conditions, however, they give Eq. (\ref{cond1p}) and hence, the CHSH parameter $\mathcal{S}$ remains invariant under these  conditions. There are other sets of symmetry conditions which the CHSH parameter remains invariant, for example,
\begin{eqnarray*}\label{}
\left\{
  \begin{array}{ll}
   \mathbf{T}\cdot[\mathbf{R}(\mathbf{x}_{1}\oplus\mathbf{y}_{1})]=0, & \hbox{is given with probability $p^{2}+(1-p)^{2}$;}\\
   \mathbf{T}\cdot[\mathbf{R}(\mathbf{x}_{1}\oplus\mathbf{y}_{2})]=0, & \hbox{is given with probability $p^{2}+(1-p)^{2}$;}\\
   \mathbf{T}\cdot[\mathbf{R}(\mathbf{x}_{2}\oplus\mathbf{y}_{1})]=1, & \hbox{is given with probability $1-[p^{2}+(1-p)^{2}]$;}\\
   \mathbf{T}\cdot[\mathbf{R}(\mathbf{x}_{2}\oplus\mathbf{y}_{2})]=1, & \hbox{is given with probability \hspace{1cm} $1$,}
  \end{array}
\right.
\end{eqnarray*}
however, they do not give Eq. (\ref{cond1p}) as measure of inconsistency with nonlocal condition $(\mathbf{x}_{1}\oplus\mathbf{x}_{2})\cdot(\mathbf{y}_{1}\oplus\mathbf{y}_{2})=1$.

Now, due to Eqs. (\ref{pro22}) and (\ref{pro23}), we get the symmetry parameter $\mathcal{W}=0$ with probability of $\mathcal{Q}(\mathcal{W}=0)=[p^{2}+(1-p)^{2}]^{3}+[1-p^{2}-(1-p)^{2}]^{3}=(1/4)[1+3E^{4}]$, see Table \ref{table1}.



\begin{figure}
\centering
\includegraphics[scale=.5]{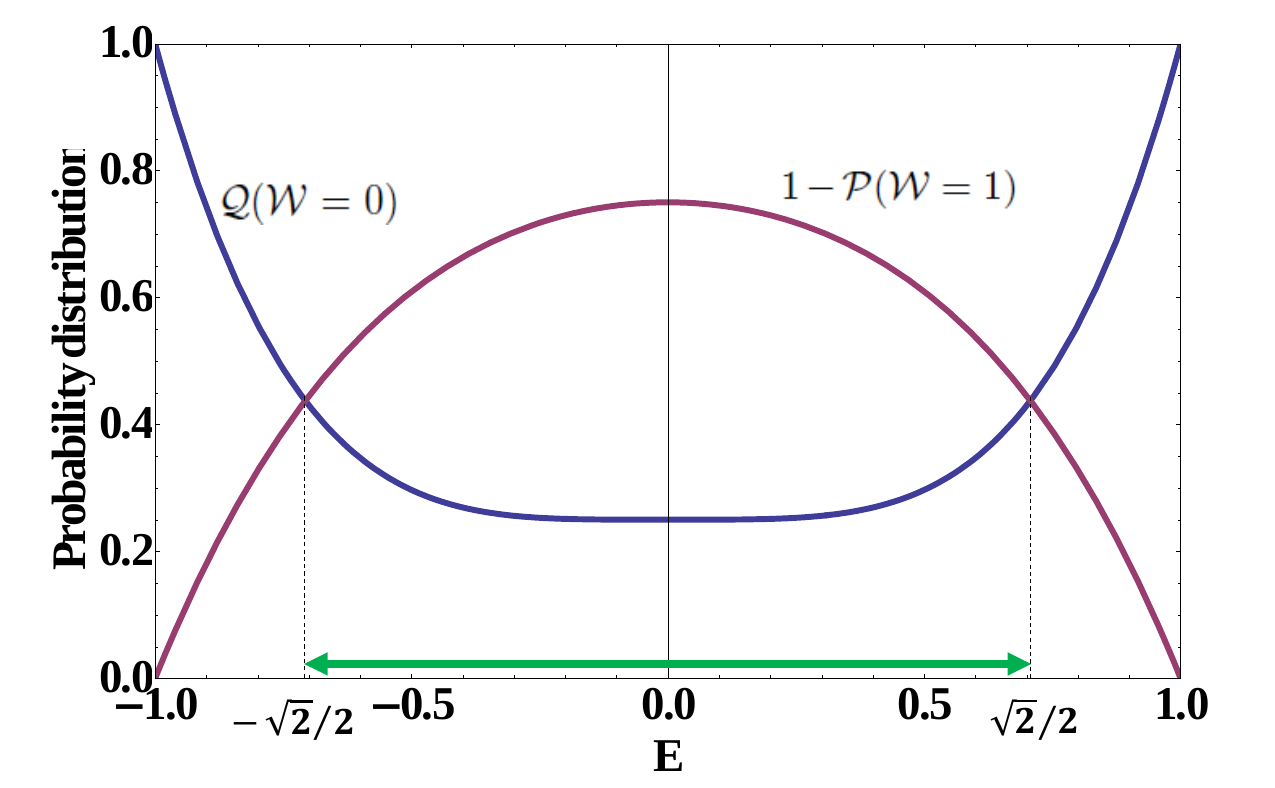}
\caption{
The consistency threshold of the probability distributions for the imperfect version of NL-box models. The probability for $\mathcal{W}$ to vanish is $\mathcal{Q}(\mathcal{W}=0)=(1/4)[1+3E^{4}]$ and probability for $\mathcal{W}$ to be unity is given by $\mathcal{P}(\mathcal{W}=1)=(1/4)[1+E^{2}]^{2}$.
This figure depicts $\mathcal{Q}(\mathcal{W}=0)$ and $1-\mathcal{P}(\mathcal{W}=1)$ as functions of the correlation $E$, which shows the probability distributions are mutually consistent in the interval $-\frac{\sqrt{2}}{2}\leqslant E\leqslant\frac{\sqrt{2}}{2}$ (green interval), which exactly coincides with the Tsirelson bound as the threshold for internal consistency of NL-box models.}
\label{QB}
\end{figure}


In what follows, we use the imperfect NL-box model (\ref{nlp}) to determine the probability of obtaining the upper bound $\mathcal{S}=4$ for the CHSH parameter. Here, due to the NL-box Assumption $3$, the probability $p$ remains under the symmetry group transformations $F\in\left\{\mathcal{F}\right\}$ again. There are four sets of correlation functions which give the maximum value of $4$:
\begin{eqnarray*}
\left\{
  \begin{array}{ll}
    \alpha_{11}\oplus\beta_{11}=\mathbf{x}_{1}\cdot\mathbf{y}_{1}=0, & \hbox{with pro. $p$;} \\
    \alpha_{12}\oplus\beta_{12}=\mathbf{x}_{1}\cdot\mathbf{y}_{2}=1, & \hbox{with pro. $p$;} \\
    \alpha_{21}\oplus\beta_{21}=\mathbf{x}_{2}\cdot\mathbf{y}_{1}=0, & \hbox{with pro. $p$;} \\
    \alpha_{22}\oplus\beta_{22}=\mathbf{x}_{2}\cdot\mathbf{y}_{2}=0, & \hbox{with pro. $p$.}
  \end{array}
\right.\hspace{1cm}
\left\{
  \begin{array}{ll}
    \alpha_{11}\oplus\beta_{11}=\mathbf{x}_{1}\cdot\mathbf{y}_{1}=1, & \hbox{with pro. $1-p$;} \\
    \alpha_{12}\oplus\beta_{12}=\mathbf{x}_{1}\cdot\mathbf{y}_{2}=0, & \hbox{with pro. $1-p$;} \\
    \alpha_{21}\oplus\beta_{21}=\mathbf{x}_{2}\cdot\mathbf{y}_{1}=1, & \hbox{with pro. $1-p$;} \\
    \alpha_{22}\oplus\beta_{22}=\mathbf{x}_{2}\cdot\mathbf{y}_{2}=1, & \hbox{with pro. $1-p$.}
  \end{array}
\right.
\end{eqnarray*}
\begin{eqnarray}
\left\{
  \begin{array}{ll}
    \alpha_{11}\oplus\beta_{11}=\mathbf{x}_{1}\cdot\mathbf{y}_{1}=1, & \hbox{with pro. $1-p$;} \nonumber\\\nonumber
    \alpha_{12}\oplus\beta_{12}=\mathbf{x}_{1}\cdot\mathbf{y}_{2}=0, & \hbox{with pro. $1-p$;} \\\nonumber
    \alpha_{21}\oplus\beta_{21}=\mathbf{x}_{2}\cdot\mathbf{y}_{1}=0, & \hbox{with pro. $p$;} \\\nonumber
    \alpha_{22}\oplus\beta_{22}=\mathbf{x}_{2}\cdot\mathbf{y}_{2}=0, & \hbox{with pro. $p$.}
  \end{array}
\right.\hspace{1cm}
\left\{
  \begin{array}{ll}
    \alpha_{11}\oplus\beta_{11}=\mathbf{x}_{1}\cdot\mathbf{y}_{1}=0, & \hbox{with pro. $p$;} \nonumber\\\nonumber
    \alpha_{12}\oplus\beta_{12}=\mathbf{x}_{1}\cdot\mathbf{y}_{2}=1, & \hbox{with pro. $p$;} \\\nonumber
    \alpha_{21}\oplus\beta_{21}=\mathbf{x}_{2}\cdot\mathbf{y}_{1}=1, & \hbox{with pro. $1-p$;} \\
    \alpha_{22}\oplus\beta_{22}=\mathbf{x}_{2}\cdot\mathbf{y}_{2}=1, & \hbox{with pro. $1-p$.}
  \end{array}
\right.\label{CH4}
\end{eqnarray}
For example, in the first set, the CHSH parameter $\mathcal{S}$ takes maximum value $4$ with probability $P^{4}$. Similar probabilities hold for other set of correlation functions with maximum value $\mathcal{S}=4$. As presented in Table \ref{table2}, the total probability of obtaining maximum value $\mathcal{S}=4$, or equivalently $\mathcal{W}=1$, is equal to $\mathcal{P}(\mathcal{W}=1)=p^{4}+(1-p)^{4}+2p^{2}(1-p)^{2}=(1/4)[1+E^{2}]^{2}$.

As depicted in Figs. \ref{PQ} and \ref{QB} and regardless to way we calculate these probabilities, the NL-box models become consistent if the sum of these two incompatible probabilities properties satisfy $\mathcal{Q}(\mathcal{W}=0)+\mathcal{P}(\mathcal{W}=1)\leqslant 1$. This logical inequality or ``inclusion-exclusion principle" was firstly introduced in the Boole algebra \cite{Bol}. Again, this principle was recalled in other scenarios \cite{Sp,Wri,Cab} and has been defined various forms,e.g., ``if you have several questions and you can answer any two of them, then you can also answer all of them" \cite{Sp}.

The inclusion-exclusion principle has various formulations which including those for counting in combinatorics. This principle mathematically is defined for two events $W_{0}$ and $W_{1}$ in probability space ${\displaystyle (\Omega, {\mathcal{F}}, \mathbb{P})}$ as $P(W_{0}\bigcup W_{1})=P(W_{0})+P(W_{1})-P(W_{0}\bigcap W_{1})$, where, ${\displaystyle (\Omega, {\mathcal{F}}, \mathbb{P})}$ respectively represent the sample space (the set of all possible outcomes of a random process), the event space (set of events which are subset of outcomes in the sample space) and the probability function (assigns to each event in the event space a number (probability) between $0$ and $1$). This logical principle can be redescribed in our scenario as: for two exclusive events $W_{0}$ and $W_{1}$ in probability space ${\displaystyle (\Omega, {\mathcal{F}}, \mathbb{P})}$ in which  $P(W_{0}\bigcap W_{1})=0$, the inclusion-exclusion principle becomes $P(W_{0}\bigcup W_{1})=P(W_{0})+P(W_{1})\leqslant 1$.

The inclusion-exclusion principle was applied in wide range of phenomena, from classical physics to quantum physics. For example, in the wave-particle duality relation, a microscopic physical system travels through a two-path interferometer (double-slit or Mach-Zehnder interferometer), with the paths being equiprobable. The visibility/probability of wave interference/property $\mathcal{V}(\text{wave})$ and the distinguishability/probability of particle trajectory/property $\mathcal{D}(\text{particle})$ are observed in different experimental settings and are determined from independent methods. There is a tradeoff relation between visibility and distinguishability as $\mathcal{V}^{2}(\text{wave})+\mathcal{D}^{2}(\text{particle})\leqslant1$ to present properties of a single system \cite{Eng}.

As another example, the inclusion-exclusion principle holds in generation of a high-resolution quantum image from a partially transmitting object with undetected photons \cite{ima}. A laser light is split at a beam splitter and the created beams illuminate two separate down-conversion nonlinear crystals $\mathrm{NL1}$ and $\mathrm{NL2}$. Every one of the down-conversion nonlinear crystals creates idler and signal photons. The idler photon from $\mathrm{NL1}$ passes through the object with a transmittance coefficient $T$ and phase shift $\delta$, afterwards is over lapped with the idler photon from $\mathrm{NL2}$. The inclusion principle indicates that the source of the idler photons being undefined. The signal photons are combined at beam splitter to create interfere pattern at the detectors. The signal photons and the idler photon from $\mathrm{NL2}$ do not interact with the object, and the idler photon from $\mathrm{NL1}$ are not detected, while images exclusively are visible in detecting the signal photons which never interact with the object. Now, due to the transmittance coefficient $T$ and the phase shift $\delta$ in the different parts of object, the image can be created at the detectors. There is a tradeoff between density of photons in the detectors which is in harmony with the exclusive principle \cite{Rez}.

As pointed out in main text and showed at here, the dimension of input/output spaces or the number of NL-boxes do not play any role in the derivation of the Tsirelson bound, we have exactly derived this bound for arbitrary-input and binary-output NL-box models without reference to the number of NL-boxes.
This type of NL-box correlation is equivalent to the CHSH input/output experiment setup which the parties can choice the arbitrary number of inputs and binary outputs. It would be an important outcome to compare with the information-theoretic principles \cite{Bra2,IC,ML,ML1,LO} which the derivation of the Tsirelson bound completely depend to the dimension of input/output spaces or the number of NL-boxes. For example, in the information causality scenario, the Alice inputs are random and independent bits $\{a_{0}, a_{1}, \cdots, a_{N-1}\}$, with dimension of $2^{N}$, and the Bob inputs $b$ are random variables $\{0, 1, \cdots, N-1\}$. The parties have shared no-signalling black boxes and Alice sends the $m$ classical bits to Bob. Due to the $m$ classical bits and his input $b$, Bob shall guess the value of the Alice input $a_{b}$ \cite{IC}. They have shown, in enough large value $n$ and $E^{2}>\frac{1}{2}=E^{2}_{\mathrm{QM}}$, the NL-box models violate the upper bound of the Shannon mutual information between parties or the information causality condition $I\leqslant m$. In other words, the correlation function $E$ asymptotically converges to $\sqrt{2}/2$ as dimension of the input space $n$ or equivalently the number of NL-boxes increases \cite{ML1}.

\section{Probability and uncertainty as emergent outcomes of symmetry group assumptions in the NL-box framework}\label{IV}

As pointed out in Sec. \ref{III}, the imperfect NL-box models are usually characterized by adding the local noise and the classical random noise such that box outputs completely random bits, regardless of the inputs \cite{ML1,LO}. These noises are parameterized by $\xi$ and $\gamma$, constant and inherent noises whiteout specifying what has caused them. Additionally, we typically expect that noises depend on measurement inputs/outputs and also its source is known. Moreover, there are similar types of imperfect NL-box models \cite{MBL,QV,QV1,QV2} which are generated by adding other types of noises. These models have stronger correlation than quantum mechanics, however, the principles mentioned in Refs. \cite{IC,ML1,LO} cannot rule out such models.

In all of aforementioned models \cite{ML1,LO,MBL,QV,QV1,QV2,QV3,QC1,QC2,QC3,QC4,QC5,QC6,QC7,QC8}, we usually add \textit{free parameters}, e.g., $\xi$ and $\gamma$, in perfect NL-box correlations to derive the Tsirelson bound. It seems derivation of the Tsirelson bound without needing to free parameters would be a fundamental improvement to understanding the relation of space symmetry group and maximum degree of nonlocality. Here, we derive the upper bound of the CHSH parameter $\mathcal{S}$ as natural result of space symmetry. Besides, we derive the nonlocal box correlation functions which are invariant under Boolean symmetry transformations $F$. In this section, we will show that the imperfect effects are not necessary to derive the Tsirelson bound in the NL-box models or the upper bound of the upper bound of the fine-grained uncertainty relation \cite{OW}. However, as pointed out in Sec. \ref{III}, the general imperfect correlation (\ref{nlx}) can be applied for simulating the nonmaximally/mixed quantum states \cite{ML1,LO,MBL,QV,QV1,QV2,QV3,QC1,QC2,QC3,QC4,QC5,QC6,QC7,QC8}.

\subsection{The Tsirelson bound as a natural outcome of fundamental symmetry group transformations}\label{IVA}

As pointed out in the main text, to derive the Tsirelson bound as a natural result of the isotropy and the homogeneity of space in the perfect NL-box models, we apply the NL-box Assumption $3$ to determine upper bound of the CHSH parameter $\mathcal{S}$ as a observable quantity. Here, we consider two couples of inputs $(\mathbf{x}_{1}, \mathbf{x}_{2})$ and $(\mathbf{y}_{1}, \mathbf{y}_{2})$. In the CHSH inequality terminology, the symmetry operations transform measurement inputs by $\mathbf{R}$ and $\mathbf{T}$. Therefore, the value of violation of the CHSH inequality is as function of measurement inputs and symmetry operators  $\mathcal{S}(\mathbf{R}^{t}\mathbf{T},\mathbf{x}_{1},\mathbf{x}_{2},\mathbf{y}_{1},\mathbf{y}_{2})$. It means that the observable value of violation of the CHSH inequality in experiment is given by average on the $\mathcal{S}(\mathbf{R}^{t}\mathbf{T},\mathbf{x}_{1},\mathbf{x}_{2},\mathbf{y}_{1},\mathbf{y}_{2})$ with different $\mathbf{R}$ and $\mathbf{T}$.

\textit{Remark 8}. As indicated in Ref. \cite{Rev}, the Boolean function $F: \{0,1\}^{n}\rightarrow \{0,1\}^{n}$ is reversible if and only if $F$ be
bijective function. Therefore, the column matrices $\mathbf{R}^{t}\mathbf{T},\, \forall\, \mathbf{R},\, \mathbf{T}$ cover all elements of the set  $\{{\mathcal{T}}\}$ and every element $\mathbf{T}\in \{{\mathcal{T}}\}$ has one-to-one relation with another element $\mathbf{T}'\in \{{\mathcal{T}}\}$ for a specific transformation $\mathbf{R}$. This property assures that different $\mathbf{R}^{t}\mathbf{T}$ appears with same probability in the CHSH parameters $\mathcal{S}(\mathbf{R}^{t}\mathbf{T},\mathbf{x}_{1},\mathbf{x}_{2},\mathbf{y}_{1},\mathbf{y}_{2})$ which is consistence with the isotropy and the homogeneity of flat space. We present the probability distribution of different $\mathbf{R}^{t}\mathbf{T}$ with $P(\mathbf{R}^{t}\mathbf{T})$ which are equal to $P(\mathbf{R}^{t}\mathbf{T})=1/2^{n-1},\, \forall\,\mathbf{R}^{t}\mathbf{T}$.

The correlation functions for measurement settings under symmetry transformations are given by
\begin{eqnarray}\label{}
&&\alpha_{11}\oplus\beta_{11}=\mathbf{x}_{1}\cdot\mathbf{y}_{1}\oplus[\mathbf{x}_{1}\oplus\mathbf{y}_{1}]\cdot\mathbf{R}^{t}\mathbf{T},\\\nonumber
&&\alpha_{12}\oplus\beta_{12}=\mathbf{x}_{1}\cdot\mathbf{y}_{2}\oplus[\mathbf{x}_{1}\oplus\mathbf{y}_{2}]\cdot\mathbf{R}^{t}\mathbf{T},\\\nonumber
&&\alpha_{21}\oplus\beta_{21}=\mathbf{x}_{2}\cdot\mathbf{y}_{1}\oplus[\mathbf{x}_{2}\oplus\mathbf{y}_{1}]\cdot\mathbf{R}^{t}\mathbf{T},\\\nonumber
&&\alpha_{22}\oplus\beta_{22}=\mathbf{x}_{2}\cdot\mathbf{y}_{2}\oplus[\mathbf{x}_{2}\oplus\mathbf{y}_{2}]\cdot\mathbf{R}^{t}\mathbf{T}, \hspace{.3cm}(\text{mod}\hspace{.1cm} 2).
\end{eqnarray}
The CHSH parameter (\ref{CHSH1}) for these set of the correlation functions is
\begin{eqnarray}\label{SRT}
\mathcal{S}(\mathbf{R}^{t}\mathbf{T},\mathbf{x}_{1},\mathbf{x}_{2},\mathbf{y}_{1},\mathbf{y}_{2})&=&
\mid(-1)^{\mathbf{x}_{1}\cdot\mathbf{y}_{1}\oplus[\mathbf{x}_{1}\oplus\mathbf{y}_{1}]\cdot\mathbf{R}^{t}\mathbf{T}}
-(-1)^{\mathbf{x}_{1}\cdot\mathbf{y}_{2}\oplus[\mathbf{x}_{1}\oplus\mathbf{y}_{2}]\cdot\mathbf{R}^{t}\mathbf{T}}\mid\nonumber\\
&+&\mid(-1)^{\mathbf{x}_{2}\cdot\mathbf{y}_{1}\oplus[\mathbf{x}_{2}\oplus\mathbf{y}_{1}]\cdot\mathbf{R}^{t}\mathbf{T}}
+(-1)^{\mathbf{x}_{2}\cdot\mathbf{y}_{2}\oplus[\mathbf{x}_{2}\oplus\mathbf{y}_{2}]\cdot\mathbf{R}^{t}\mathbf{T}}\mid.
\end{eqnarray}
To obtain maximum value of $\langle\mathcal{S}\rangle_{\mathrm{max}}$, similar to one of approaches to obtain the Tsirelson bound in the quantum mechanics, we use variance of the CHSH parameter $(\triangle\mathcal{S})^{2}=\langle\mathcal{S}^{2}\rangle-\langle\mathcal{S}\rangle^{2}\geqslant0$. It means that the maximum of the CHSH parameter is given by $\sqrt{\langle\mathcal{S}^{2}\rangle}_{\mathrm{max}}\geqslant\langle\mathcal{S}\rangle$. Therefore, we need to calculate $\sqrt{\langle\mathcal{S}^{2}\rangle}_{\mathrm{max}}$
\begin{eqnarray*}\label{}
\mathcal{S}^{2}(\mathbf{R}^{t}\mathbf{T},\mathbf{x}_{1},\mathbf{x}_{2},\mathbf{y}_{1},\mathbf{y}_{2})
&=&4+2(-1)^{(\mathbf{y}_{1}\oplus\mathbf{y}_{2})\cdot\mathbf{R}^{t}\mathbf{T}}
\left[-(-1)^{\mathbf{x}_{1}\cdot(\mathbf{y}_{1}\oplus\mathbf{y}_{2})}+(-1)^{\mathbf{x}_{2}\cdot(\mathbf{y}_{1}\oplus\mathbf{y}_{2})}\right]\\
&+&2\left|(-1)^{(\mathbf{x}_{1}\oplus\mathbf{x}_{2})\cdot\mathbf{R}^{t}\mathbf{T}}
\left[(-1)^{(\mathbf{x}_{1}\oplus\mathbf{x}_{2})\cdot\mathbf{y}_{1}}
-(-1)^{(\mathbf{x}_{1}\oplus\mathbf{x}_{2})\cdot\mathbf{y}_{2}}\right]\right.\\
&+&\left.(-1)^{(\mathbf{x}_{1}\oplus\mathbf{x}_{2}\oplus\mathbf{y}_{1}\oplus\mathbf{y}_{2})\cdot\mathbf{R}^{t}\mathbf{T}}
\left[(-1)^{\mathbf{x}_{1}\cdot\mathbf{y}_{1}\oplus\mathbf{x}_{2}\cdot\mathbf{y}_{2}}
-(-1)^{\mathbf{x}_{1}\cdot\mathbf{y}_{2}\oplus\mathbf{x}_{2}\cdot\mathbf{y}_{1}}\right]\right|.
\end{eqnarray*}
According to relation among the brackets in the above relations
\begin{eqnarray*}\label{}
u&=&\left[-(-1)^{\mathbf{x}_{1}\cdot(\mathbf{y}_{1}\oplus\mathbf{y}_{2})}+(-1)^{\mathbf{x}_{2}\cdot(\mathbf{y}_{1}\oplus\mathbf{y}_{2})}\right],\\
v&=&\left[(-1)^{(\mathbf{x}_{1}\oplus\mathbf{x}_{2})\cdot\mathbf{y}_{1}}-(-1)^{(\mathbf{x}_{1}\oplus\mathbf{x}_{2})\cdot\mathbf{y}_{2}}\right],\\
w&=&\left[(-1)^{\mathbf{x}_{1}\cdot\mathbf{y}_{1}\oplus\mathbf{x}_{2}\cdot\mathbf{y}_{2}}
-(-1)^{\mathbf{x}_{1}\cdot\mathbf{y}_{2}\oplus\mathbf{x}_{2}\cdot\mathbf{y}_{1}}\right],\hspace{1cm}uv=2w,
\end{eqnarray*}
the maximum value of the bracket sentences are given by $u=v=w=2$. If we select these values, the $\mathcal{S}^{2}$ reduce to the following relation:
\begin{eqnarray*}
\mathcal{S}^{2}(\mathbf{R}^{t}\mathbf{T},\mathbf{x}_{1},\mathbf{x}_{2},\mathbf{y}_{1},\mathbf{y}_{2})
=8\left[1+(-1)^{(\mathbf{y}_{1}\oplus\mathbf{y}_{2})\cdot\mathbf{R}^{t}\mathbf{T}}\right].
\end{eqnarray*}
In the above relation, we used $\left|(-1)^{(\mathbf{x}_{1}\oplus\mathbf{x}_{2})\cdot\mathbf{R}^{t}\mathbf{T}}\right|=1$. The right hand side of above relation is vanish for one half of $\mathbf{R}$ and $\mathbf{T}$ and is $16$ for another one half. Consequently, $\langle\mathcal{S}^{2}\rangle_{\mathrm{max}}=8$ and the maximum value of violation of the CHSH inequality is the Tsirelson bound $2\sqrt{2}\geqslant\langle\mathcal{S}\rangle_{\mathrm{max}}$. It is clear that any combination of $\mathcal{S}(\mathbf{R}^{t}\mathbf{T},\mathbf{x}_{1},\mathbf{x}_{2},\mathbf{y}_{1},\mathbf{y})$ with different $\mathbf{R}$ and $\mathbf{T}$ reduces maximum value of $\langle\mathcal{S}^{2}\rangle$.


Similar to conventional description of the expectation values, which average is taken on the measurement outcomes, we take average on the different values of $|\mathcal{S}^{2}$ which are given by different symmetry transformations $\mathbf{R}, \mathbf{T}\in\{\mathcal{F}\}$.

In follows, we show that any combination of CHSH parameters, as measurement outcomes, reduce the upper bound $\langle\mathcal{S}^{2}\rangle$. As pointed out in main text, $\mathcal{S}^{2}(\mathbf{R}^{t}\mathbf{T},\mathbf{x}_{1},\mathbf{x}_{2},\mathbf{y}_{1},\mathbf{y}_{2})$ are not observable variables, the physical variables shall be invariant under symmetry transformations. Thus, a suggestion may be appropriate combination of $\mathcal{S}(\mathbf{R}^{t}\mathbf{T},\mathbf{x}_{1},\mathbf{x}_{2},\mathbf{y}_{1},\mathbf{y}_{2})$, for example:
\begin{eqnarray*}
\mathcal{\tilde{S}}_{\mathrm{physical}}(\{\mathcal{F}\}_{\mathrm{sub set}},\mathbf{x}_{2},\mathbf{y}_{1},\mathbf{y}_{2})
=\sum_{\mathbf{R}, \mathbf{T}\in\{\mathcal{F}\}_{\mathrm{sub set}}}\mathcal{C}(\mathbf{R}^{t}\mathbf{T},\mathbf{x}_{1},\mathbf{x}_{2},\mathbf{y}_{1},\mathbf{y}_{2})
\mathcal{S}(\mathbf{R}^{t}\mathbf{T},\mathbf{x}_{1},\mathbf{x}_{2},\mathbf{y}_{1},\mathbf{y}_{2})
\end{eqnarray*}
where, the summation is taken on arbitrary subset of symmetry transformations $\{\mathcal{F}\}_{\mathrm{sub set}}\in\{\mathcal{F}\}$ and $\mathcal{C}$ are expansion coefficients in which $\sum_{\mathbf{R}, \mathbf{T}\in\{\mathcal{F}\}_{\mathrm{sub set}}}\mathcal{C}(\mathbf{R}^{t}\mathbf{T},\mathbf{x}_{1},\mathbf{x}_{2},\mathbf{y}_{1},\mathbf{y}_{2})=1$.

Now, we know the $\mathcal{S}^{2}$ is a convex function which define $f(\sum_{i=1}^{m}\lambda_{i}e_{i})\leqslant \sum_{i=1}^{m}\lambda_{i}f(e_{i}))$ with $\sum_{i=1}^{m}\lambda_{i}=1$. For specific set $e_{i}=0,1,\, \forall\,i$, we have $\langle\mathcal{\tilde{S}}_{\mathrm{Phy}}^{2}\rangle\leqslant\langle\mathcal{S}^{2}\rangle$.


The physical CHSH parameter is given by applying invariant form of the correlation function. As will point out in Sec. \ref{IVC}, the physical/symmetric form of the correlation function is generated by liner combination of the perfect NL-box correlations, with appropriate coefficients (probabilities), which shall be invariant under the symmetry group transformations $\left\{\mathcal{F}\right\}$. Therefore, the upper bound any physical the CHSH parameters is $\sqrt{\langle\mathcal{S}^{2}\rangle}$.

\subsection{The symmetry group structure determines the upper bound of the fine-grained uncertainty relation}\label{IVB}

The degree of nonlocality of any theory $P^{\mathrm{NL}}$ is characterized by tradeoff between two physical phenomena, the strength of the uncertainty principle $P^{\mathrm{cer}}$ and the strength of the steerability of theory \cite{OW}. The perfect NL-box models have no uncertainty, but unlike in the classical world, we still have perfect steering. The fine-grained uncertainty relation is given by $P^{\mathrm{cer}}(\sigma_{\alpha, \mathbf{x}}, \beta)=\sum_{\mathbf{y}}p(\mathbf{y})p(\beta|\mathbf{y})_{\sigma_{\alpha, \mathbf{x}}}\leqslant\zeta_{\sigma_{\alpha, \mathbf{x}}, \beta}$ and the degree of nonlocality is given by $P^{\mathrm{NL}}=\sum_{\mathbf{x}}p(\mathbf{x})\sum_{\alpha}p(\alpha|\mathbf{x})P^{\mathrm{cer}}(\sigma_{\alpha, \mathbf{x}}, \beta)=1/2+1/(2\sqrt{2})$.

Due to tradeoff relation between the uncertainty and the steerability in characterizing degree of nonlocality, it may seem that there is not one-to-one relation between $\max P^{\mathrm{NL}}$ and $\max P^{\mathrm{cer}}$. To avoiding any ambiguity in derivation of the upper bound $P^{\mathrm{cer}}$ from $\max P^{\mathrm{NL}}$, we directly determine the upper bound of the fine-grained uncertainty relation $\max P^{\mathrm{cer}}(\sigma_{\alpha, \mathbf{x}}, \beta)=\max \sum_{\mathbf{y}}p(\mathbf{y})p(\beta|\mathbf{y})_{\sigma_{\alpha, \mathbf{x}}}=\zeta_{\sigma_{\alpha, \mathbf{x}}, \beta}$ as natural consequence of fundamental symmetry group transformations. For the same probability distribution of inputs $p(\mathbf{y}_{1})=p(\mathbf{y}_{2})=1/2$, the fine-grained uncertainty relation gives
\begin{eqnarray*}
P^{\mathrm{cer}}(\sigma_{\alpha, \mathbf{x}}, \beta)=\sum_{i=1,2}p(\mathbf{y}_{i})p(\beta|\mathbf{y}_{i})_{\sigma_{\alpha, \mathbf{x}}}&=&p(\mathbf{y}_{1})(\frac{1}{2})[1+(-1)^{\mathbf{x}\cdot\mathbf{y}_{1}}]+
p(\mathbf{y}_{2})(\frac{1}{2})[1+(-1)^{\mathbf{x}\cdot\mathbf{y}_{2}}]\\
&=&\frac{1}{2}+p(\mathbf{y}_{1})(-1)^{\mathbf{x}\cdot\mathbf{y}_{1}}/2+p(\mathbf{y}_{2})(-1)^{\mathbf{x}\cdot\mathbf{y}_{2}}/2\\
&=&\frac{1}{2}+\frac{1}{4}\large[(-1)^{\mathbf{x}\cdot\mathbf{y}_{1}}+(-1)^{\mathbf{x}\cdot\mathbf{y}_{2}}\large]\\
&=&\frac{1}{2}+\frac{1}{4}\mathcal{G}(\mathbf{x},\mathbf{y}_{1},\mathbf{y}_{2}).
\end{eqnarray*}
In the last line, we call the summation of the correlation functions as $\mathcal{G}(\mathbf{x},\mathbf{y}_{1},\mathbf{y}_{2})$.

Now, we determine the maximum value $\mathcal{G}$ and know that the parameter $\mathcal{G}(\mathbf{x},\mathbf{y}_{1},\mathbf{y}_{2})$ are not invariant under the symmetry group elements $F\in\left\{\mathcal{F}\right\}$. All of the parameters $\mathcal{G}(\mathbf{R}^{t}\mathbf{T},\mathbf{x},\mathbf{y}_{1},\mathbf{y}_{2}),\, \forall\, \mathbf{R},\, \mathbf{T}$ are equivalent and shall be appeared with the same weight.

Similar to determining the upper bound of the CHSH parameter, we employ variance of
$(\triangle\mathcal{G})^{2}=\langle\mathcal{G}^{2}\rangle-\langle\mathcal{G}\rangle^{2}\geqslant0$ and the determine maximum $\langle\mathcal{G}^{2}\rangle_{\mathrm{max}}=\sum_{\mathbf{R}^{t}\mathbf{T}}
D(\mathbf{R}^{t}\mathbf{T})\mathcal{G}_{\mathrm{max}}^{2}(\mathbf{R}^{t}\mathbf{T},\mathbf{x},\mathbf{y}_{1},\mathbf{y}_{2})$, where the expectation value is taken on all symmetry transformations $\mathbf{R}^{t}\mathbf{T}$ with the same probability distribution $D(\mathbf{R}^{t}\mathbf{T})=1/2^{n-1},\,\forall\,\mathbf{R},\,\mathbf{T}$. Therefore, we have
\begin{eqnarray*}
\langle\mathcal{G}^{2}\rangle_{\mathrm{max}}=(1/2^{n-1})
\sum_{\mathbf{R}^{t}\mathbf{T}}\left[(-1)^{\mathbf{x}\cdot\mathbf{y}_{1}\oplus[\mathbf{x}\oplus\mathbf{y}_{1}]\cdot\mathbf{R}^{t}\mathbf{T}}
+(-1)^{\mathbf{x}\cdot\mathbf{y}_{2}\oplus[\mathbf{x}\oplus\mathbf{y}_{2}]\cdot\mathbf{R}^{t}\mathbf{T}}\right]^{2},
\end{eqnarray*}
by choosing the input settings as $\mathbf{x}\cdot\mathbf{y}_{1}=\mathbf{x}\cdot\mathbf{y}_{2}$ and using $\left[(-1)^{[\mathbf{x}\oplus\mathbf{y}_{1}]\cdot\mathbf{R}^{t}\mathbf{T}}
\right]^{2}=1$, we have
\begin{eqnarray*}
\langle\mathcal{G}^{2}\rangle_{\mathrm{max}}=(1/2^{n-1})
\sum_{\mathbf{R}^{t}\mathbf{T}}\left[1
+(-1)^{[\mathbf{y}_{1}\oplus\mathbf{y}_{2}]\cdot\mathbf{R}^{t}\mathbf{T}}\right]^{2},
\end{eqnarray*}
which is equal to $\langle\mathcal{G}^{2}\rangle_{\mathrm{max}}=2$, or equivalently $|\langle\mathcal{G}\rangle|\leqslant \sqrt{2}$. This gives the upper bound of the fine-grained uncertainty relation $\zeta_{\sigma_{\alpha, \mathbf{x}}, \beta}=1/2+1/(2\sqrt{2})$.

\subsection{The invariant form of the NL-box correlation function}\label{IVC}

In the previous parts, we have derived the maximum value of $\langle\mathcal{S}\rangle_{\mathrm{max}}$, however, derivation of the physical variables are more interesting which shall be invariant under symmetry transformations. Therefore, observable variables, e.g., the correlation function, are generated by employing linear combination of the perfect NL-box correlations with appropriate coefficients (probabilities) which is conditioned to be invariant under the symmetry group transformations. One can see that the NL-box Assumptions $1-3$ that correlation function $E$ depends to input variables $\mathbf{x}$ and $\mathbf{y}$, $E(\mathbf{x},\mathbf{y})$. This outcome may give fresh insight on the intensive efforts to predicting the quantum correlations without referring to operators acting on Hilbert spaces \cite{QC1,QC2,QC3,QC4,QC5,QC6,QC7,QC8}.


Here, we derive an analytic relation for the correlation functions which is invariant under symmetry group elements. The observable correlation functions shall respect the following conditions:
\begin{eqnarray}\label{corp}
-1\leqslant E(\mathbf{x},\mathbf{y})\leqslant1, \hspace {.4cm}E(F(\mathbf{x}),F(\mathbf{y}))=E(\mathbf{x},\mathbf{y}),\hspace {.4cm} F(\mathbf{x})=\mathbf{R}\mathbf{x}\oplus\mathbf{T},\\\nonumber
E(\bar{\mathbf{x}},\mathbf{y})=E(\mathbf{x},\bar{\mathbf{y}})=-E(\mathbf{x},\mathbf{y})=-E{(\bar{\mathbf{x}}},\bar{\mathbf{y}}),\hspace {.1cm} \forall
\mathbf{x}, \mathbf{y}, \mathbf{R}, \mathbf{T}.
\end{eqnarray}
To compare with quantum singlet state (total spin zero), we impose the following constrain on the correlation functions,
\begin{eqnarray}\label{corp1}
E(\mathbf{x},\mathbf{x})=-1, \hspace {.4cm}E(\mathbf{x},\bar{\mathbf{x}})=1,\hspace {.1cm} \forall\mathbf{x}.
\end{eqnarray}
Here, the correlation functions need to be invariant under symmetry group elements:
\begin{eqnarray}\label{inva}
E(\mathbf{x},\mathbf{y})=E(\mathbf{R}\mathbf{x}\oplus\mathbf{T}, \mathbf{R}\mathbf{y}\oplus\mathbf{T}),\hspace{.5cm} \forall \hspace{.2cm}\mathbf{R},\hspace{.1cm} \mathbf{T}.
\end{eqnarray}
Therefore, the same relations (\ref{corp}) and (\ref{corp1}) hold for transferred correlation functions:
\begin{eqnarray}\label{E1p}
-1\leqslant E(\mathbf{x},\mathbf{y},\mathbf{R},\mathbf{T})\leqslant1, \hspace {.4cm}E(\mathbf{x},\mathbf{x},\mathbf{R},\mathbf{T})=-1, \hspace {.4cm} E(\mathbf{x},\bar{\mathbf{x}},\mathbf{R},\mathbf{T})=1,\\\nonumber
E(\bar{\mathbf{x}},\mathbf{y},\mathbf{R},\mathbf{T})=E(\mathbf{x},\bar{\mathbf{y}},\mathbf{R},\mathbf{T})
=-E(\mathbf{x},\mathbf{y},\mathbf{R},\mathbf{T})=-E{(\bar{\mathbf{x}}},\bar{\mathbf{y}},\mathbf{R},\mathbf{T}).
\end{eqnarray}

Moreover, if we know the correlation function $E(\mathbf{R}\mathbf{x}\oplus\mathbf{T}, \mathbf{R}\mathbf{y}\oplus\mathbf{T})$, we have the correlation functions $E(\mathbf{R}\bar{\mathbf{x}}\oplus\mathbf{T}, \mathbf{R}\mathbf{y}\oplus\mathbf{T})$, $E(\mathbf{R}\mathbf{x}\oplus\mathbf{T}, \mathbf{R}\bar{\mathbf{y}}\oplus\mathbf{T})$ and
$E(\mathbf{R}\bar{\mathbf{x}}\oplus\mathbf{T}, \mathbf{R}\bar{\mathbf{y}}\oplus\mathbf{T})$. Therefore, a general expression for correlation functions is given by:
\begin{eqnarray*}\label{}
E(\mathbf{x},\mathbf{y},\mathbf{R},\mathbf{T})=
A(\mathbf{x},\mathbf{y},\mathbf{R},\mathbf{T})(-1)^{(\mathbf{R}\mathbf{x}\oplus\mathbf{T})\cdot(\mathbf{R}\mathbf{y}\oplus\mathbf{T})}
+B(\mathbf{x},\mathbf{y},\mathbf{R},\mathbf{T})(-1)^{(\mathbf{R}\bar{\mathbf{x}}\oplus\mathbf{T})
\cdot(\mathbf{R}\bar{\mathbf{y}}\oplus\mathbf{T})}\\
+C(\mathbf{x},\mathbf{y},\mathbf{R},\mathbf{T})(-1)^{(\mathbf{R}\mathbf{x}\oplus\mathbf{T})
\cdot(\mathbf{R}\bar{\mathbf{y}}\oplus\mathbf{T})}
+D(\mathbf{x},\mathbf{y},\mathbf{R},\mathbf{T})(-1)^{(\mathbf{R}\bar{\mathbf{x}}\oplus\mathbf{T})
\cdot(\mathbf{R}\mathbf{y}\oplus\mathbf{T})}\\
=
\left[A(\mathbf{x},\mathbf{y},\mathbf{R},\mathbf{T})+B(\mathbf{x},\mathbf{y},\mathbf{R},\mathbf{T})\right]
(-1)^{(\mathbf{R}\mathbf{x}\oplus\mathbf{T})\cdot(\mathbf{R}\mathbf{y}\oplus\mathbf{T})}\\
+\left[C(\mathbf{x},\mathbf{y},\mathbf{R},\mathbf{T})+D(\mathbf{x},\mathbf{y},\mathbf{R},\mathbf{T})\right]
(-1)^{(\mathbf{R}\mathbf{x}\oplus\mathbf{T})
\cdot(\mathbf{R}\bar{\mathbf{y}}\oplus\mathbf{T})}\\
=
U(\mathbf{x},\mathbf{y},\mathbf{R},\mathbf{T})
(-1)^{(\mathbf{R}\mathbf{x}\oplus\mathbf{T})\cdot(\mathbf{R}\mathbf{y}\oplus\mathbf{T})}
+V(\mathbf{x},\mathbf{y},\mathbf{R},\mathbf{T})
(-1)^{(\mathbf{R}\mathbf{x}\oplus\mathbf{T})
\cdot(\mathbf{R}\bar{\mathbf{y}}\oplus\mathbf{T})}.
\end{eqnarray*}

The $U$ and $V$ shall respect the conditions (\ref{E1p}). In what follows, we obtain explicit form for them.

If the coefficients be greater than one, $U, V\geqslant1$ so that $U-V\leqslant1$, without loss of generality, they can be scaled to following intervals:
\begin{eqnarray*}\label{}
-1\leqslant U(\mathbf{x},\mathbf{y},\mathbf{R},\mathbf{T}),\hspace{.2cm} V(\mathbf{x},\mathbf{y},\mathbf{R},\mathbf{T})\leqslant1.
\end{eqnarray*}
Moreover, if the coefficients $U$ and $V$ have different signs $UV\leqslant0$, we can re-scale the coefficients so that they have been the same sign $UV\geqslant0$. For example, for $U\geqslant0$, $V\leqslant0$ and $U\geqslant|V|$, the correlation function $U(-1)^{\mathbf{x}\cdot\mathbf{y}}+V(-1)^{\mathbf{x}\cdot\bar{\mathbf{y}}}$ can be re-scaled to same correlation function $U'(-1)^{\mathbf{x}\cdot\mathbf{y}}+V'(-1)^{\mathbf{x}\cdot\bar{\mathbf{y}}}$. Here, $U', V'\geqslant 0$ and $U'=\frac{1}{2}[1+U-V],\hspace{.1cm} V'=\frac{1}{2}[1-(U-V)]$.

Besides, according to Eq. (\ref{inva}), the coefficients $U$ and $V$ save their sign under symmetry operations,
\begin{eqnarray*}\label{}
\mathrm{sgn}[U(\mathbf{x},\mathbf{y})V(\mathbf{x},\mathbf{y})]= \mathrm{sgn}[U(\mathbf{R}\mathbf{x}\oplus\mathbf{T},\mathbf{R}\mathbf{y}\oplus\mathbf{T})
V(\mathbf{R}\mathbf{x}\oplus\mathbf{T},\mathbf{R}\mathbf{y}\oplus\mathbf{T})], \hspace{.2cm} \forall\hspace{.2cm} \mathbf{x},\mathbf{y},\mathbf{R},\mathbf{T}.
\end{eqnarray*}

Here, we select the first interval and the coefficients $U$ and $V$ can be written as
\begin{eqnarray*}\label{}
U(\mathbf{x},\mathbf{y},\mathbf{R},\mathbf{T})=\sin^{2}(\Phi_{Rx\oplus T,Ry\oplus T}),\hspace{.2cm} V(\mathbf{x},\mathbf{y},\mathbf{R},\mathbf{T})=\cos^{2}(\Omega_{Rx\oplus T,Ry\oplus T}).
\end{eqnarray*}
At this step $\Phi$ and $\Omega$ are arbitrary functions and consequently correlation function can be written as
\begin{eqnarray*}\label{E2p}
E(\mathbf{x},\mathbf{y},\mathbf{R},\mathbf{T})=
\sin^{2}(\Phi_{Rx\oplus T,Ry\oplus T})(-1)^{(\mathbf{R}\mathbf{x}\oplus\mathbf{T})\cdot(\mathbf{R}\mathbf{y}\oplus\mathbf{T})}
+\cos^{2}(\Omega_{Rx\oplus T,Ry\oplus T})(-1)^{(\mathbf{R}\mathbf{x}\oplus\mathbf{T})\cdot(\mathbf{R}\bar{\mathbf{y}}\oplus\mathbf{T})}.
\end{eqnarray*}
However, we can rewrite the above equation as
\begin{eqnarray*}\label{}
&&E(\mathbf{x},\mathbf{y},\mathbf{R},\mathbf{T})\\
&&=
\left[1-\cos^{2}(\Phi_{Rx\oplus T,Ry\oplus T})\right](-1)^{(\mathbf{R}\mathbf{x}\oplus\mathbf{T})\cdot(\mathbf{R}\mathbf{y}\oplus\mathbf{T})}
+\left[1-\sin^{2}(\Omega_{Rx\oplus T,Ry\oplus T})\right](-1)^{(\mathbf{R}\mathbf{x}\oplus\mathbf{T})\cdot(\mathbf{R}\bar{\mathbf{y}}\oplus\mathbf{T})}\\
&&=
\cos^{2}(\Phi_{Rx\oplus T,Ry\oplus T})(-1)^{(\mathbf{R}\mathbf{x}\oplus\mathbf{T})\cdot(\mathbf{R}\mathbf{y}\oplus\mathbf{T})}
+\sin^{2}(\Omega_{Rx\oplus T,Ry\oplus T})(-1)^{(\mathbf{R}\mathbf{x}\oplus\mathbf{T})\cdot(\mathbf{R}\bar{\mathbf{y}}\oplus\mathbf{T})}\\
&&=
\sin^{2}(\Omega_{Rx\oplus T,Ry\oplus T})(-1)^{(\mathbf{R}\mathbf{x}\oplus\mathbf{T})\cdot(\mathbf{R}\mathbf{y}\oplus\mathbf{T})}
+\cos^{2}(\Phi_{Rx\oplus T,Ry\oplus T})(-1)^{(\mathbf{R}\mathbf{x}\oplus\mathbf{T})\cdot(\mathbf{R}\bar{\mathbf{y}}\oplus\mathbf{T})}.
\end{eqnarray*}
It shows that we have two different combinations of correlation functions which describe sub-correlation function, therefore, we consider combination of them
\begin{eqnarray}\label{prop}
E(\mathbf{x},\mathbf{y},\mathbf{R},\mathbf{T})=
\frac{1}{2}\left[\sin^{2}(\Phi_{Rx\oplus T,Ry\oplus T})+\sin^{2}(\Omega_{Rx\oplus T,Ry\oplus T})\right](-1)^{(\mathbf{R}\mathbf{x}\oplus\mathbf{T})\cdot(\mathbf{R}\mathbf{y}\oplus\mathbf{T})}\\\nonumber
+\frac{1}{2}\left[\cos^{2}(\Phi_{Rx\oplus T,Ry\oplus T})+\cos^{2}(\Omega_{Rx\oplus T,Ry\oplus T})\right](-1)^{(\mathbf{R}\mathbf{x}\oplus\mathbf{T})\cdot(\mathbf{R}\bar{\mathbf{y}}\oplus\mathbf{T})}\\\nonumber
=\sin^{2}(\Theta_{Rx\oplus T,Ry\oplus T}/2)(-1)^{(\mathbf{R}\mathbf{x}\oplus\mathbf{T})\cdot(\mathbf{R}\mathbf{y}\oplus\mathbf{T})}
+\cos^{2}(\Theta_{Rx\oplus T,Ry\oplus T}/2)(-1)^{(\mathbf{R}\mathbf{x}\oplus\mathbf{T})\cdot(\mathbf{R}\bar{\mathbf{y}}\oplus\mathbf{T})}.
\end{eqnarray}
Here, we have defined new variable $\Theta_{Rx\oplus T,Ry\oplus T}$ as:
\begin{eqnarray*}\label{}
\sin^{2}(\Theta_{Rx\oplus T,Ry\oplus T}/2)=\frac{1}{2}\left[\sin^{2}(\Phi_{Rx\oplus T,Ry\oplus T})+\sin^{2}(\Omega_{Rx\oplus T,Ry\oplus T})\right]\\
\cos^{2}(\Theta_{Rx\oplus T,Ry\oplus T}/2)=\frac{1}{2}\left[\cos^{2}(\Phi_{Rx\oplus T,Ry\oplus T})+\cos^{2}(\Omega_{Rx\oplus T,Ry\oplus T})\right].
\end{eqnarray*}
In Eq. (\ref{prop}), the coefficients of the NL-box machines can be described as probability distribution. In other words, we have derived the probabilistic interpretation of correlation functions by imposing symmetry group structure on the NL-boxes.



Now, we impose other conditions to fix $\Theta$
\begin{eqnarray}\label{condp}
E(\mathbf{x},\bar{\mathbf{y}},\mathbf{R},\mathbf{T})=-E(\mathbf{x},\mathbf{y},\mathbf{R},\mathbf{T})\rightarrow
\Theta_{Rx\oplus T,R\bar{y}\oplus T}-\Theta_{Rx\oplus T,Ry\oplus T}=2k\pi,\\
\hspace{.2cm}k=0,1,...\hspace{.2cm} \forall\hspace{.2cm} \mathbf{x},\mathbf{y},\mathbf{R},\mathbf{T}\nonumber.
\end{eqnarray}
The above relation means that the $\Theta_{Rx\oplus T,R\bar{y}\oplus T}-\Theta_{Rx\oplus T,Ry\oplus T}$ is independent from Alice input, therefore, the $\Theta_{Rx\oplus T,R\bar{y}\oplus T}$ is any combination of inputs which support this condition. There are two combinations which satisfy to this condition. One of them is Eq. (\ref{nl2p}) and the other one is difference of two arbitrary functions of inputs $\theta_{Rx\oplus T}$ and $\theta_{Ry\oplus T}$, we call the rotation angles
\begin{eqnarray}\label{EE1p}
\Theta_{Rx\oplus T,Ry\oplus T}=(\mathbf{R}\mathbf{x}\oplus\mathbf{T})\cdot(\mathbf{R}\mathbf{y}\oplus\mathbf{T})\varphi+ \theta_{Rx\oplus T}-\theta_{Ry\oplus T}.
\end{eqnarray}
Here $\varphi$ is arbitrary angle which will be fixed by constrain
\begin{eqnarray}\label{EEp}
E(\mathbf{x},\mathbf{x},\mathbf{R},\mathbf{T})=-1\rightarrow\Theta_{Rx\oplus T,Rx\oplus T}=\pi, \hspace{.2cm} \forall\mathbf{x},\mathbf{R},\mathbf{T},
\end{eqnarray}
which means the $\varphi=\pi$. Moreover, by looking to Eq. (\ref{condp}), we have $\theta_{R\bar{x}+T}=\theta_{Rx+T}+\pi$.
Hence, the rotation angles, before and after operation, are given by
\begin{eqnarray*}\label{}
\theta_{Rx+T}=\theta_{x}+\delta_{x,Rx+T},\hspace{1cm}\theta_{Ry+T}=\theta_{y}+\epsilon_{y,Ry+T}.
\end{eqnarray*}
It is clear from Eqs. (\ref{EE1p}) and (\ref{EEp}) which $\delta_{x,Rx+T}=\epsilon_{x,Rx+T}\hspace{.2cm} \forall\mathbf{x},\mathbf{R},\mathbf{T}$. Besides, the $\delta$ function for two arbitrary measurement settings $\mathbf{R}'\mathbf{x}\oplus\mathbf{T}'$ and $\mathbf{R}\mathbf{x}\oplus\mathbf{T}$ are given by difference of $\delta$ functions which calculate to a reference input.
\begin{eqnarray*}\label{}
\theta_{R'x+T'}-\theta_{Rx+T}=\delta_{Rx+T,R'x+T'}=\delta_{x,R'x+T'}-\delta_{x,Rx+T},
\end{eqnarray*}
without any ambiguity, we can remove reference measurement settings $\mathbf{x}$ from $\delta$ functions $\delta_{x,Rx+T}\equiv\delta_{R,T}$, the same relation holds for $\mathbf{y}$. Therefore, we have
\begin{eqnarray*}\label{}
\theta_{Rx+T}=\theta_{x}+\delta_{R,T},\hspace{1cm}\theta_{Ry+T}=\theta_{y}+\delta_{R,T}.
\end{eqnarray*}
The observed/physical correlation function is given by averaging on all of correlations which are given by symmetry group transformations $E(\mathbf{x},\mathbf{y})=\sum_{R,T}P(\mathbf{R}^{t}\mathbf{T})E(\theta_{Rx+T},\theta_{Ry+T})$ which is equal to $-\cos(\theta_{x}-\theta_{y})$, and similarly the corresponding probability is give by averaging on all of equivalent probabilities $P(\alpha,\beta|\mathbf{x},\mathbf{y})$ $=\frac{1}{2}\left[1+(-1)^{\alpha\oplus\beta}E(\mathbf{x},\mathbf{y})\right]$. The maximum value of violation is given by suitable selection of $(\theta_{x_{i}},\theta_{y_{j}}),\hspace {.1cm} i,j=1,2$ which is equal to the Tsirelson bound $2\sqrt{2}\nleqslant2$. These results show that the Tsirelson bound is a consequence of the symmetry of the correlation functions and quantum mechanics is one of the optimal nonlocal models.

\subsection{Multipartite and high-dimension postquantum models}\label{IVD}

Recently, efforts have been devoted to study the multipartite Bell type inequalities \cite{MP1,MP2,MP3,MP4,MP5,MP6,MP7,MP8,MP9} and the high-dimension input/output Bell type inequalities \cite{HD1,HD2,HD3,HD4,HD5,HD6,HD7,HD8,HD9} which have crucial applications in wide range from fundamental physics to information theory \cite{QKD1,NT4,NT5,NT6,NT7,NT8,NT9,GM1}. The fundamental problem in this scenario is determination of the upper bound of the various type of Bell parameters \cite{MP1,MP2,MP3,MP4,MP5,MP6,MP7,MP8,MP9,HD1,HD2,HD3,HD4,HD5,HD6,HD7,HD8,HD9} as criterion to confidence of information process, in theory and experiment. We usually are interested to characterize the maximum value of the Bell parameters from a theory-independent viewpoint by introducing mathematical abstract nonlocal models (postquantum) and proposing a fundamental principle(s) which gives the upper bound of the Bell parameters (maximum nonlocality).


The several information-theoretic principles \cite{Dam,Bra2,NC,ML,ML1,IC,IC2,IC3,MBL,LO,Alm1,Alm3} are mainly proposed to rule out \emph{specific} postquantum models \cite{PR1,BCP,JM,BP1,BP2,Mas,Alm,QC1}. These principles are designed to detect bipartite postquantum models \cite{Dam,Bra2,NC,ML,ML1,IC,IC2,IC3,MBL,LO,Alm1,Alm3}, however, they fail to distinguish the multipartite postquantum models from multipartite quantum correlations \cite{NIP,NIP1}. The almost quantum correlations highlights this problem by proposing tripartite version of almost quantum correlations which meets all bipartite information-theoretic principles \cite{Alm}, meanwhile, violate the upper bounds of quantum predictions. Although there already exist tripartite, binary-input and binary-output Bell type inequalities that witness almost-quantum correlations beyond quantum theory \cite{MuBell}, an exact characterization of the set of multipartite quantum correlations has remained inaccessible.

In another work \cite{Almsy}, we have applied our scenario to the almost quantum correlations models \cite{Alm} and have imposed the standard representations of the isotropy/homogeneity symmetry group transformations on them. We have requested that the probability distributions (the Born rule) remain invariant. This condition is \textit{sufficient} to reduce the almost quantum correlations to standard quantum mechanics. Interestingly, unlike to the information theoretic principles \cite{Dam,Bra2,NC,ML,ML1,IC,IC2,IC3,MBL,LO,Alm1,Alm3} which fail to distinguish the multipartite postquantum models from multipartite quantum correlations \cite{NIP,NIP1}, the number of parties and dimension of input/output spaces do not play any role in this framework which makes the isotropy and homogeneity principles as criteria to rule out the multipartite/high-dimension postquantum models \cite{Alm}. We anticipate that impact of the isotropy/homogeneity principle goes beyond merely ruling out multipartite the almost quantum correlations and it is applicable to the multipartite or the high-dimension postquantum models \cite{PR1,Sv,Non1,Non5,Non3,Non2,Non7,Non4,Non6,PR1,BCP,JM,BP1,BP2,Mas,Alm,QC1}. In another work at progress, we will discus these models.

As indicated in main text, we have imposed the flat space symmetry-group structure on the NL-box models \cite{BCP,JM,BP1,BP2,Mas} and have derived the upper bound of the CHSH parameter. In addition, as we have shown in Sec. \ref{IVC}, the symmetric version of NL-box correlation function is exactly equal to the singlet state quantum correlation function. This outcome can give upper bound of the bipartite, arbitrary-input and binary-output Bell type inequalities, e.g., the generalized CHSH inequality \cite{We} or other types of three-input and binary-output inequalities \cite{QC1}, as predicted by quantum mechanics. We would like to emphasis that the fundamental variable to determine the upper bound of any Bell type parameters is the corresponding \textit{symmetric NL-box correlation function}. Other approaches, e.g., derivation of the upper bound of the CHSH parameter in the main text, only confide our scenario.

We stress that our scenario does not \textit{necessarily} characterize the upper bound of any multipartite/high-dimension type of the Bell parameters. However, we \textit{expect} that our approach rule out \textit{all} of postquantum models. The exclusion of multipartite/high-dimension postquantum models is out of scope of present article, meanwhile, to show ability of our approach, we determine the upper bound of a tripartite Bell parameter \cite{QC1} without invoking the explicit form of the symmetric tripartite NL-box correlation. Here, we employ similar approach as Sec. \ref{IVA} to determine upper bound, but, this is not a general approach to characterize the upper bound of every Bell parameter.


The wide range of Bell-type parameters were proposed to characterizing the facets of the correlation polytope \cite{WW}. Here, we consider three observables, Alice, Bob, and Charlie, with the binary-inputs $(\mathbf{x}_{i},\mathbf{y}_{j},\mathbf{z}_{k})\, i,j,k=0,1$ and binary-outputs $\alpha,\beta,\gamma=\{0,1\}$ which the corresponding Bell parameter is defined as
\begin{eqnarray*}\label{},
\mathcal{I}=E(\mathbf{x}_{0},\mathbf{y}_{0},\mathbf{y}_{0})+E(\mathbf{x}_{0},\mathbf{y}_{1},\mathbf{z}_{1})
+E(\mathbf{x}_{1},\mathbf{y}_{0},\mathbf{z}_{0})-E(\mathbf{x}_{1},\mathbf{y}_{1},\mathbf{z}_{1}).
\end{eqnarray*}
The Bell parameter takes the upper bound $\mathcal{I}_{\mathrm{L}}=2$ for local models, and the upper bound  $\mathcal{I}_{\mathrm{NLB}}=4$ for the NL-box models which have been found by enumerating the vertices of the respective polytopes \cite{QC1}. The upper bound of $\mathcal{I}$ in quantum framework is equal to $\mathcal{I}_{\mathrm{QM}}=2\sqrt{2}$.


The proposed tripartite NL-box model is $\alpha\oplus\beta\oplus\gamma=(x+1)y+z,\,(\mathrm{mod}\hspace{.1 cm}2)$ (\cite{QC1}) which the parties inputs are binary numbers $x, y, z\in\{0,1\}$. Similar to earlier version of the bipartite NL-box models, this model is not consistence with the NL-box Assumption $1$. Here, we firstly suggest a tripartite NL-box model which is consistence with the NL-box Assumption $1$ and gives the upper bound of $\mathcal{I}_{\mathrm{NLB}}=4$. Afterwards, we impose the symmetry transformations $F\in\{\mathcal{F}\}$ on the tripartite NL-box model and require the $\max\mathcal{I}$ remains invariant. This condition is sufficient to characterize maximum value of $\mathcal{I}$.

We suggest the following tripartite NL-box correlation function:
\begin{eqnarray}\label{3NLB}
\alpha\oplus\beta\oplus\gamma=\mathbf{x}\cdot\mathbf{y}\oplus\mathbf{x}\cdot\mathbf{z}\oplus\mathbf{y}\cdot\mathbf{z}
\oplus\mathbf{x}\cdot\mathbf{c}\oplus\mathbf{y}\cdot\mathbf{c}\oplus\mathbf{z}\cdot\mathbf{c}, \hspace{1 cm}(\mathrm{mod}\hspace{.1 cm}2)
\end{eqnarray}
where, the $\mathbf{c}$ is a constant $n$-tuple with the number of characters 1 (0) is odd and transformed under symmetry group elements $F$ as inputs $F(\mathbf{c})=\mathbf{R}\mathbf{c}\oplus\mathbf{T}$.

The same as the main text, we define the reverse of inputs by $\bar{\mathbf{x}}=\mathbf{1}\oplus\mathbf{x}$ and similar relations hold for $\bar{\mathbf{y}}$ and $\mathbf{z}$, where $\mathbf{1}$ is the string of $n$ $1$'s, $\mathbf{1}=(1,1,\cdots,1)$. These nonlocal models satisfy the NL-box Assumption $1$, for example,
\begin{eqnarray*}\label{}
(\alpha\oplus1)\oplus\beta\oplus\gamma&=&\bar{\mathbf{x}}\cdot\mathbf{y}\oplus\bar{\mathbf{x}}\cdot\mathbf{z}\oplus\mathbf{y}\cdot\mathbf{z}
\oplus\bar{\mathbf{x}}\cdot\mathbf{c}\oplus\mathbf{y}\cdot\mathbf{c}\oplus\mathbf{z}\cdot\mathbf{c}\\
&=&(\mathbf{x}\cdot\mathbf{y}\oplus1)\oplus(\mathbf{x}\cdot\mathbf{z}\oplus1)\oplus\mathbf{y}\cdot\mathbf{z}
\oplus(\mathbf{x}\cdot\mathbf{c}\oplus1)\oplus\mathbf{y}\cdot\mathbf{c}\oplus\mathbf{z}\cdot\mathbf{c}\\
&=&\mathbf{x}\cdot\mathbf{y}\oplus\mathbf{x}\cdot\mathbf{z}\oplus\mathbf{y}\cdot\mathbf{z}
\oplus\mathbf{x}\cdot\mathbf{c}\oplus\mathbf{y}\cdot\mathbf{c}\oplus\mathbf{z}\cdot\mathbf{c}\oplus1.
\end{eqnarray*}
Moreover, there are various sets of measurement inputs which gives the upper bound of $\mathcal{I}$, for example, the set
\begin{eqnarray}\label{Set}
\mathbf{x}_{0}=[1,0,0,0,0,0,0,0,\cdots,0],\hspace{.7cm}\mathbf{x}_{1}=[1,0,0,1,1,0,0,0,\cdots,0],\nonumber\\
\mathbf{y}_{0}=[0,1,0,0,0,0,0,0,\cdots,0],\hspace{.7cm}\mathbf{y}_{1}=[0,0,0,0,1,0,0,0,\cdots,0],\nonumber\\
\mathbf{z}_{0}=[0,0,1,0,0,0,0,0,\cdots,0],\hspace{.7cm}\mathbf{z}_{1}=[0,0,0,0,0,1,0,0,\cdots,0],\nonumber\\
\mathbf{c}=[0,0,0,0,0,0,1,0,\cdots,0],
\end{eqnarray}
gives the correlation functions
\begin{eqnarray}\label{SC}
E(\mathbf{x}_{0},\mathbf{y}_{0},\mathbf{z}_{0})=1,
\hspace{.2cm} E(\mathbf{x}_{0},\mathbf{y}_{1},\mathbf{z}_{1})=1,
\hspace{.2cm}E(\mathbf{x}_{1},\mathbf{y}_{0},\mathbf{z}_{0})=1,
\hspace{.2cm}E(\mathbf{x}_{1},\mathbf{y}_{1},\mathbf{z}_{1})=-1,
\end{eqnarray}
with algebraic maximum value $\mathcal{I}_{\mathrm{NLB}}=4$.

We now impose the symmetry transformations on the parameter $\mathcal{I}$ to characterize the upper bound of it. We firstly take absolute value of correlation functions as
\begin{eqnarray*}\label{I32}
\mathcal{I}\leqslant |E(\mathbf{x}_{0},\mathbf{y}_{0},\mathbf{z}_{0})+E(\mathbf{x}_{0},\mathbf{y}_{1},\mathbf{z}_{1})|
+|E(\mathbf{x}_{1},\mathbf{y}_{0},\mathbf{z}_{0})-E(\mathbf{x}_{1},\mathbf{y}_{1},\mathbf{z}_{1})|,
\end{eqnarray*}
and similar to determination of the upper bound of CHSH parameter, we employ variance of the Bell parameter $(\triangle\mathcal{I})^{2}=\langle\mathcal{I}^{2}\rangle-\langle\mathcal{I}\rangle^{2}\geqslant0$ and the determine maximum $\langle\mathcal{I}^{2}\rangle_{\mathrm{max}}$ which is
\begin{eqnarray}\label{I33}
\langle\mathcal{I}^{2}\rangle\leqslant\frac{1}{2^{n-1}}\sum_{\mathbf{R}\mathbf{T}}
\left[|E(\mathbf{x}_{0},\mathbf{y}_{0},\mathbf{z}_{0})+E(\mathbf{x}_{0},\mathbf{y}_{1},\mathbf{z}_{1})|
+|E(\mathbf{x}_{1},\mathbf{y}_{0},\mathbf{z}_{0})-E(\mathbf{x}_{1},\mathbf{y}_{1},\mathbf{z}_{1})|
\right]^{2}.
\end{eqnarray}
Due to the measurement inputs (\ref{Set}) and the correlation functions (\ref{SC}), Eq. (\ref{I33}) reduces to
\begin{eqnarray}\label{I34}
\mathcal{I}\leqslant |(-1)^{
(\mathbf{x}_{0}\oplus\mathbf{y}_{0}\oplus\mathbf{z}_{0}\oplus\mathbf{c})\cdot\mathbf{R}\mathbf{T}}
+(-1)^{(\mathbf{x}_{0}\oplus\mathbf{y}_{1}\oplus\mathbf{z}_{1}\oplus\mathbf{c})\cdot\mathbf{R}\mathbf{T}}|\nonumber\\
+|(-1)^{(\mathbf{x}_{1}\oplus\mathbf{y}_{0}\oplus\mathbf{z}_{0}\oplus\mathbf{c})\cdot\mathbf{R}\mathbf{T}}
+(-1)^{(\mathbf{x}_{1}\oplus\mathbf{y}_{1}\oplus\mathbf{z}_{1}\oplus\mathbf{c})\cdot\mathbf{R}\mathbf{T}}|\nonumber\\
=2|1+(-1)^{(\mathbf{y}_{0}\oplus\mathbf{z}_{0}\oplus\mathbf{y}_{1}\oplus\mathbf{z}_{1})\cdot\mathbf{R}\mathbf{T}}|,
\end{eqnarray}
where we used $|(-1)^{(\mathbf{x}_{0}\oplus\mathbf{x}_{1}\oplus\mathbf{c})\cdot\mathbf{R}\mathbf{T}}|=1$.

Eq. (\ref{I34}) is similar to variance of the parameter $\mathcal{S}$. After simple algebra, we obtain $\langle\mathcal{I}^{2}\rangle_{\mathrm{max}}=8$, which is the very quantum bound prediction $|\langle\mathcal{I}\rangle|\leqslant 2\sqrt{2}$.

In similar manner, we can determine the upper bound of another tripartite Bell type inequality with the binary-input and binary-output \cite{Sve},
\begin{eqnarray*}\label{}
\mathcal{J}=|E(\mathbf{x}_{0},\mathbf{y}_{0},\mathbf{z}_{0})+E(\mathbf{x}_{0},\mathbf{y}_{0},\mathbf{z}_{1})
+E(\mathbf{x}_{1},\mathbf{y}_{0},\mathbf{z}_{0})-E(\mathbf{x}_{1},\mathbf{y}_{0},\mathbf{z}_{1})\\
+E(\mathbf{x}_{0},\mathbf{y}_{1},\mathbf{z}_{0})-E(\mathbf{x}_{0},\mathbf{y}_{1},\mathbf{z}_{1})
-E(\mathbf{x}_{1},\mathbf{y}_{1},\mathbf{z}_{0})-E(\mathbf{x}_{1},\mathbf{y}_{1},\mathbf{z}_{1})|.
\end{eqnarray*}
The upper bound of the $\mathcal{J}$ for local models gives $\mathcal{J}_{\mathrm{L}}=4$ and for the Nl-box models is $\mathcal{J}_{\mathrm{NLB}}=8$ which have been found by enumerating the vertices of the respective polytopes. The upper bound of quantum mechanics has been given $\mathcal{J}_{\mathrm{QM}}=5.66$ \cite{Sve}. We again use the tripartite NL-box models (\ref{3NLB}) and determine the upper bound which gives $\mathcal{J}_{\mathrm{max}}=4\sqrt{2}\backsimeq 5.65685$.

These results highlight ability of the flat space symmetry-group structure to exactly distinguish \textit{physical} correlations from \textit{nonphysical} ones in the multipartite systems. This ability does not mean that we can characterize the upper bound of multipartite or high-dimension any type of Bell parameters. We anticipate the flat space symmetry-group structure is sufficient to rule out all of postquantum models.

\section{An explicit representation of measurement input spaces and transformation matrices $\mathcal{R}$ and $\mathcal{T}$ in $n=6$}\label{V}

Here, we represent an explicit example of inputs space $\mathcal{X}$ and $\mathcal{Y}$; and transformation matrix spaces $\mathcal{R}$ and $\mathcal{T}$ in the $n=6$. As we mentioned in Eq. (\ref{nup}), the $n$-tuple inputs have odd number of $1$ ($0$), which are:
\begin{eqnarray*}\label{st1}
&&[1,0,0,0,0,0], \hspace{.1cm}[0,1,0,0,0,0], \hspace{.1cm}[0,0,1,0,0,0], \hspace{.1cm} [0,0,0,1,0,0], \hspace{.1cm}[0,0,0,0,1,0], \hspace{.1cm}[0,0,0,0,0,1],\nonumber\\\nonumber
&&[0,1,1,1,1,1], \hspace{.1cm}[1,0,1,1,1,1], \hspace{.1cm}[1,1,0,1,1,1], \hspace{.1cm}[1,1,1,0,1,1], \hspace{.1cm}[1,1,1,1,0,1],
\hspace{.1cm}[1,1,1,1,1,0], \\\nonumber
&&[1,1,1,0,0,0], \hspace{.1cm}[0,0,0,1,1,1],  \hspace{.1cm}[1,1,0,1,0,0], \hspace{.1cm}[0,0,1,0,1,1],  \hspace{.1cm} [1,1,0,0,1,0],
\hspace{.1cm}[0,0,1,1,0,1],  \\\nonumber
&&[1,1,0,0,0,1], \hspace{.1cm}[0,0,1,1,1,0],  \hspace{.1cm}[1,0,1,1,0,0], \hspace{.1cm}[0,1,0,0,1,1],  \hspace{.1cm}[1,0,1,0,1,0],
\hspace{.1cm}[0,1,0,1,0,1], \\\nonumber
&&[1,0,1,0,0,1], \hspace{.1cm}[0,1,0,1,1,0],  \hspace{.1cm}[1,0,0,1,1,0], \hspace{.1cm}[0,1,1,0,0,1],  \hspace{.1cm}[1,0,0,1,0,1],
\hspace{.1cm}[0,1,1,0,1,0], \\
&&[1,0,0,0,1,1], \hspace{.1cm}[0,1,1,1,0,0],  \hspace{.1cm}
\end{eqnarray*}

$\vspace{.1cm}$

The elements of Boolean translation space $\mathbf{T}\in\{\mathcal{T}\}$ are $n$-tuple of even $1$ ($0$) numbers $\mathbf{T}=(\underbrace{T_{1,1},T_{1,2},\hdots,T_{1,n}}_{\textrm{number of 1 (0) is even}})$,

\begin{eqnarray*}\label{st2}
[0,0,0,0,0,0], \hspace{.1cm}[1,1,1,1,1,1],  \hspace{.1cm}[1,1,0,0,0,0], \hspace{.1cm}[0,0,1,1,1,1],  \hspace{.1cm}\\\nonumber
[1,0,1,0,0,0], \hspace{.1cm}[0,1,0,1,1,1],  \hspace{.1cm}[1,0,0,1,0,0], \hspace{.1cm}[0,1,1,0,1,1],  \hspace{.1cm}\\\nonumber
[1,0,0,0,1,0], \hspace{.1cm}[0,1,1,1,0,1],  \hspace{.1cm}[1,0,0,0,0,1], \hspace{.1cm}[0,1,1,1,1,0],  \hspace{.1cm}\\\nonumber
[0,1,1,0,0,0], \hspace{.1cm}[1,0,0,1,1,1],  \hspace{.1cm}[0,1,0,1,0,0], \hspace{.1cm}[1,0,1,0,1,1],  \hspace{.1cm}\\\nonumber
[0,1,0,0,1,0], \hspace{.1cm}[1,0,1,1,0,1],  \hspace{.1cm}[0,1,0,0,0,1], \hspace{.1cm}[1,0,1,1,1,0],  \hspace{.1cm}\\\nonumber
[0,0,1,1,0,0], \hspace{.1cm}[1,1,0,0,1,1],  \hspace{.1cm}[0,0,1,0,1,0], \hspace{.1cm}[1,1,0,1,1,1],  \hspace{.1cm}\\\nonumber
[0,0,1,0,0,1], \hspace{.1cm}[1,1,0,1,1,0],  \hspace{.1cm}[0,0,0,1,1,0], \hspace{.1cm}[1,1,1,0,0,1],  \hspace{.1cm}\\\nonumber
[0,0,0,1,0,1], \hspace{.1cm}[1,1,1,0,1,0],  \hspace{.1cm}[0,0,0,0,1,1], \hspace{.1cm}[1,1,1,1,0,0],  \hspace{.1cm}\nonumber
\end{eqnarray*}

$\vspace{.1cm}$

The rotation matrices $\mathbf{R}$ are constructed from inputs space. The rows (columns) of each Boolean rotation matrices $\mathbf{R}_{i,(n)}$ are $n$-tuples of inputs space $\mathbf{R}_{i,(n)}\in\{\mathcal{X}\}$, so that $\mathbf{R}_{i,(n)}=(\underbrace{R_{i,1},R_{i,2},\hdots,R_{i,n}}_{\textrm{number of 1 (0) is odd}})$ and $\mathbf{R}_{i,(n)}\cdot\mathbf{R}_{l,(n)}=\delta_{il}, \hspace{.2 cm}(\text{mod}\hspace{.1 cm}2)$, in each rotation matrices $\mathbf{R}$. Here we represent some of them:

$\vspace{.1cm}$

$\left(
  \begin{array}{cccccc}
    1 & 0 & 0 & 0 & 0 & 0 \\
    0 & 1 & 0 & 0 & 0 & 0 \\
    0 & 0 & 0 & 1 & 1 & 1 \\
    0 & 0 & 1 & 0 & 1 & 1 \\
    0 & 0 & 1 & 1 & 0 & 1 \\
    0 & 0 & 1 & 1 & 1 & 0 \\
\end{array}
\right), \hspace{.2cm}\left(
  \begin{array}{cccccc}
    0 & 1 & 1 & 1 & 0 & 0 \\
    1 & 0 & 1 & 1 & 0 & 0 \\
    1 & 1 & 0 & 1 & 0 & 0 \\
    1 & 1 & 1 & 0 & 0 & 0 \\
    0 & 0 & 0 & 0 & 1 & 0 \\
    0 & 0 & 0 & 0 & 0 & 1 \\
  \end{array}
\right), \hspace{.2cm} \left(
  \begin{array}{cccccc}
    0 & 1 & 1 & 1 & 0 & 0 \\
    1 & 0 & 1 & 1 & 0 & 0 \\
    1 & 1 & 1 & 0 & 1 & 1 \\
    1 & 1 & 0 & 1 & 1 & 1 \\
    0 & 0 & 1 & 1 & 0 & 1 \\
    0 & 0 & 1 & 1 & 1 & 0 \\
  \end{array}
\right), \hspace{.2cm}\left(
  \begin{array}{cccccc}
    1 & 0 & 0 & 0 & 0 & 0 \\
    0 & 0 & 0 & 1 & 1 & 1 \\
    0 & 0 & 1 & 0 & 0 & 0 \\
    0 & 1 & 0 & 0 & 1 & 1 \\
    0 & 1 & 0 & 1 & 0 & 1 \\
    0 & 1 & 0 & 1 & 1 & 0 \\
  \end{array}
\right), \hdots$

$\vspace{.3cm}$

We impose homogeneity and isotropy conditions (\ref{nl2p}) and (\ref{conp}) on the inputs and transformation matrices spaces to obtain subsets which respect symmetries criteria. The conditions (\ref{conp}) imply that the inputs and transformation matrices spaces are divided to five subsets for $n=6$; $\{\mathcal{X}_{k}\}$, $\{\mathcal{Y}_{k}\}$, $\{\mathcal{R}_{k}\}$ and $\{\mathcal{T}_{k}\}$, $k=1,\ldots,5$.

$\vspace{.1cm}$

The elements of the first sub-space are:

$\vspace{0cm}$

\begin{eqnarray*}\label{st21}
\{\mathcal{X}_{1}\},\{\mathcal{Y}_{1}\}=\left\{[1,0,0,0,0,0],  \hspace{.1cm}[0,1,0,0,0,0], \hspace{.1cm}[1,0,1,1,0,0],\hspace{.1cm}[1,0,0,0,1,1], \hspace{.1cm}\right.\\\nonumber
\left.[0,1,1,1,1,1],  \hspace{.1cm}[1,0,1,1,1,1], \hspace{.1cm}[0,1,0,0,1,1],\hspace{.1cm}[0,1,1,1,0,0]\right\} \hspace{.1cm}
\end{eqnarray*}
\begin{eqnarray*}\label{st21}
\{\mathcal{T}_{1}\}=\left\{[0,0,0,0,0,0],  \hspace{.1cm}[1,1,0,0,0,0], \hspace{.1cm}[0,0,1,1,0,0],\hspace{.1cm}[0,0,0,0,1,1], \right.\hspace{.1cm}\\\nonumber
\left.[1,1,1,1,0,0],  \hspace{.1cm}[1,1,0,0,1,1], \hspace{.1cm}[0,0,1,1,1,1],\hspace{.1cm}[1,1,1,1,1,1]\right\} \hspace{.1cm}
\end{eqnarray*}
\begin{eqnarray*}
\{\mathcal{R}_{1}\}=\left(
  \begin{array}{cccccc}
    1 & 0 & 0 & 0 & 0 & 0 \\
    0 & 1 & 0 & 0 & 0 & 0 \\
    0 & 0 & 0 & 1 & 1 & 1 \\
    0 & 0 & 1 & 0 & 1 & 1 \\
    0 & 0 & 1 & 1 & 0 & 1 \\
    0 & 0 & 1 & 1 & 1 & 0 \\
\end{array}
\right), \hspace{.2cm}\left(
  \begin{array}{cccccc}
    0 & 1 & 0 & 0 & 1 & 1 \\
    1 & 0 & 0 & 0 & 1 & 1 \\
    0 & 0 & 1 & 0 & 0 & 0 \\
    0 & 0 & 0 & 1 & 0 & 0 \\
    1 & 1 & 0 & 0 & 0 & 1 \\
    1 & 1 & 0 & 0 & 1 & 0 \\
  \end{array}
\right), \hspace{.2cm} \left(
  \begin{array}{cccccc}
    0 & 1 & 1 & 1 & 0 & 0 \\
    1 & 0 & 1 & 1 & 0 & 0 \\
    1 & 1 & 0 & 1 & 0 & 0 \\
    1 & 1 & 1 & 0 & 0 & 0 \\
    0 & 0 & 0 & 0 & 1 & 0 \\
    0 & 0 & 0 & 0 & 0 & 1 \\
  \end{array}
\right), \hspace{.2cm}\left(
  \begin{array}{cccccc}
     0 & 0 & 1 & 1 & 0 & 1 \\
    0 & 0 & 1 & 1 & 1 & 0 \\
    0 & 0 & 0 & 1 & 1 & 1 \\
    0 & 0 & 1 & 0 & 1 & 1 \\
    1 & 0 & 0 & 0 & 0 & 0 \\
    0 & 1 & 0 & 0 & 0 & 0 \\
  \end{array}
\right), \hdots\nonumber\\
\end{eqnarray*}

$\vspace{.1cm}$

The elements of the second sub-space are:

\begin{eqnarray*}\label{st22}
\{\mathcal{X}_{2}\},\{\mathcal{Y}_{2}\}=\left\{[1,0,0,0,0,0], \hspace{.1cm}[0,0,1,0,0,0], \hspace{.1cm}[1,0,0,1,0,1],  \hspace{.1cm}[1,1,0,0,1,0],  \hspace{.1cm}\right.\\\nonumber
\left\{[0,0,1,1,0,1], \hspace{.1cm}[0,1,1,0,1,0],  \hspace{.1cm}[1,1,0,1,1,1], \hspace{.1cm}[0,1,1,1,1,1]\right\}  \hspace{.1cm}
\end{eqnarray*}
\begin{eqnarray*}\label{st22}
\{\mathcal{T}_{2}\}=\left\{[0,0,0,0,0,0], \hspace{.1cm}[1,0,1,0,0,0], \hspace{.1cm}[0,0,0,1,0,1],  \hspace{.1cm}[0,1,0,0,1,0],\right.  \hspace{.1cm}\\\nonumber
\left.[1,0,1,1,0,1], \hspace{.1cm}[1,1,1,0,1,0],  \hspace{.1cm}[0,1,0,1,1,1], \hspace{.1cm}[1,1,1,1,1,1]\right\}  \hspace{.1cm}
\end{eqnarray*}

\begin{eqnarray*}\label{st22}
\{\mathcal{R}_{2}\}=\left(
  \begin{array}{cccccc}
    1 & 0 & 0 & 0 & 0 & 0 \\
    0 & 1 & 0 & 1 & 1 & 0 \\
    0 & 0 & 1 & 0 & 0 & 0 \\
    0 & 1 & 0 & 1 & 0 & 1 \\
    0 & 1 & 0 & 0 & 1 & 1 \\
    0 & 0 & 0 & 1 & 1 & 1 \\
\end{array}
\right), \hspace{.2cm}\left(
  \begin{array}{cccccc}
    1 & 1 & 1 & 0 & 0 & 0 \\
    1 & 1 & 0 & 0 & 1 & 0 \\
    1 & 0 & 0 & 0 & 1 & 0 \\
    0 & 0 & 0 & 1 & 0 & 0 \\
    0 & 1 & 1 & 0 & 1 & 0 \\
    0 & 0 & 0 & 0 & 0 & 1 \\
  \end{array}
\right), \hspace{.2cm} \left(
  \begin{array}{cccccc}
    1 & 0 & 1 & 1 & 0 & 0 \\
    0 & 1 & 0 & 0 & 0 & 0 \\
    1 & 0 & 1 & 0 & 0 & 1 \\
    1 & 0 & 0 & 1 & 0 & 1 \\
    0 & 0 & 0 & 0 & 1 & 0 \\
    0 & 0 & 1 & 1 & 0 & 1 \\
  \end{array}
\right), \hspace{.2cm}\left(
  \begin{array}{cccccc}
    0 & 1 & 0 & 1 & 0 & 1 \\
    0 & 1 & 0 & 1 & 1 & 0 \\
    0 & 0 & 0 & 1 & 1 & 1 \\
    1 & 0 & 0 & 0 & 0 & 0 \\
    0 & 1 & 0 & 0 & 1 & 1 \\
    0 & 0 & 1 & 0 & 0 & 0 \\
  \end{array}
\right), \hdots\nonumber\\
\end{eqnarray*}\label{st22}

$\vspace{.1cm}$

The elements of the third sub-space are:

\begin{eqnarray*}\label{st23}
\{\mathcal{X}_{3}\},\{\mathcal{Y}_{3}\}=\left\{[1,0,0,0,0,0], \hspace{.1cm}[0,0,0,1,0,0],  \hspace{.1cm}[1,1,0,0,0,1],\right. \hspace{.1cm}[1,0,1,0,1,0],  \hspace{.1cm}\\\nonumber
\left.[0,1,0,1,0,1], \hspace{.1cm}[0,0,1,1,1,0],  \hspace{.1cm}[1,1,1,0,1,1], \hspace{.1cm}[0,1,1,1,1,1]\right\}  \hspace{.1cm}
\end{eqnarray*}
\begin{eqnarray*}\label{st23}
\{\mathcal{T}_{3}\}=\left\{[0,0,0,0,0,0], \hspace{.1cm}[1,0,0,1,0,0],  \hspace{.1cm}[0,1,0,0,0,1], \hspace{.1cm}[0,0,1,0,1,0], \right. \hspace{.1cm}\\\nonumber
\left.[1,1,0,1,0,1], \hspace{.1cm}[1,0,1,1,1,0],  \hspace{.1cm}[0,1,1,0,1,1], \hspace{.1cm}[1,1,1,1,1,1]\right\} \hspace{.1cm}
\end{eqnarray*}

\begin{eqnarray*}\label{st23}
\{\mathcal{R}_{3}\}=\left(
  \begin{array}{cccccc}
    1 & 0 & 0 & 0 & 0 & 0 \\
    0 & 1 & 1 & 0 & 1 & 0 \\
    0 & 1 & 1 & 0 & 0 & 1 \\
    0 & 0 & 0 & 1 & 0 & 0 \\
    0 & 1 & 0 & 0 & 1 & 1 \\
    0 & 0 & 1 & 0 & 1 & 1 \\
\end{array}
\right), \hspace{.2cm}\left(
  \begin{array}{cccccc}
    1 & 0 & 1 & 1 & 0 & 0 \\
    0 & 1 & 0 & 0 & 0 & 0 \\
    1 & 0 & 1 & 0 & 1 & 0 \\
    1 & 0 & 0 & 1 & 1 & 0 \\
    0 & 0 & 1 & 1 & 1 & 0 \\
    0 & 0 & 0 & 0 & 0 & 1 \\
  \end{array}
\right), \hspace{.2cm} \left(
  \begin{array}{cccccc}
    1 & 1 & 0 & 1 & 0 & 0 \\
    1 & 1 & 0 & 0 & 0 & 1 \\
    0 & 0 & 1 & 0 & 0 & 0 \\
    1 & 0 & 0 & 1 & 0 & 1 \\
    0 & 0 & 0 & 0 & 1 & 0 \\
    0 & 1 & 0 & 1 & 0 & 1 \\
  \end{array}
\right), \hspace{.2cm}\left(
  \begin{array}{cccccc}
    0 & 1 & 1 & 0 & 1 & 0 \\
    1 & 0 & 0 & 0 & 0 & 0 \\
    0 & 1 & 1 & 0 & 0 & 1 \\
    0 & 0 & 1 & 0 & 1 & 1 \\
    0 & 1 & 0 & 0 & 1 & 1 \\
    0 & 0 & 0 & 1 & 0 & 0 \\
  \end{array}
\right), \hdots\nonumber\\
\end{eqnarray*}\label{st23}

$\vspace{.1cm}$

The elements of the fourth sub-space are:

\begin{eqnarray*}\label{st24}
\{\mathcal{X}_{4}\},\{\mathcal{Y}_{4}\}=\left\{[1,0,0,0,0,0], \hspace{.1cm}[0,0,0,0,1,0],  \hspace{.1cm}[1,1,0,1,0,0], \hspace{.1cm}[1,0,1,0,0,1], \right. \hspace{.1cm}\\\nonumber
\left.[0,1,0,1,1,0], \hspace{.1cm}[0,0,1,0,1,1],  \hspace{.1cm}[0,1,0,1,1,0], \hspace{.1cm}[0,1,1,1,1,1]\right\}  \hspace{.1cm}
\end{eqnarray*}
\begin{eqnarray*}\label{st24}
\{\mathcal{T}_{4}\}=\left\{[0,0,0,0,0,0], \hspace{.1cm}[1,0,0,0,1,0],  \hspace{.1cm}[0,1,0,1,0,0], \hspace{.1cm}[0,0,1,0,0,1],\right.  \hspace{.1cm}\\\nonumber
\left.[1,1,0,1,1,0], \hspace{.1cm}[1,0,1,0,1,1],  \hspace{.1cm}[1,1,0,1,1,0], \hspace{.1cm}[1,1,1,1,1,1]\right\}  \hspace{.1cm}
\end{eqnarray*}

\begin{eqnarray*}\label{st24}
\{\mathcal{R}_{4}\}=\left(
  \begin{array}{cccccc}
    1 & 0 & 0 & 0 & 0 & 0 \\
    0 & 1 & 1 & 1 & 0 & 0 \\
    0 & 1 & 1 & 0 & 0 & 1 \\
    0 & 1 & 0 & 1 & 0 & 1 \\
    0 & 0 & 0 & 0 & 1 & 0 \\
    0 & 0 & 1 & 1 & 0 & 1 \\
\end{array}
\right), \hspace{.2cm}\left(
  \begin{array}{cccccc}
    1 & 0 & 1 & 0 & 1 & 0 \\
    0 & 1 & 0 & 0 & 0 & 0 \\
    1 & 0 & 1 & 0 & 1 & 1 \\
    0 & 0 & 0 & 1 & 0 & 0 \\
    1 & 0 & 0 & 0 & 1 & 1 \\
    0 & 0 & 1 & 0 & 1 & 1 \\
  \end{array}
\right), \hspace{.2cm} \left(
  \begin{array}{cccccc}
    1 & 1 & 0 & 1 & 0 & 0 \\
    1 & 1 & 0 & 0 & 1 & 0 \\
    0 & 0 & 1 & 0 & 0 & 0 \\
    1 & 0 & 0 & 1 & 1 & 0 \\
    0 & 1 & 0 & 1 & 1 & 0 \\
    0 & 0 & 0 & 0 & 0 & 1 \\
  \end{array}
\right), \hspace{.2cm}\left(
  \begin{array}{cccccc}
    0 & 0 & 0 & 1 & 1 & 1 \\
    0 & 1 & 0 & 0 & 0 & 0 \\
    0 & 0 & 1 & 0 & 0 & 0 \\
    0 & 1 & 0 & 1 & 0 & 1 \\
    0 & 1 & 0 & 0 & 1 & 1 \\
    0 & 1 & 0 & 1 & 1 & 0 \\
  \end{array}
\right), \hdots\nonumber\\
\end{eqnarray*}\label{st24}

$\vspace{.1cm}$

The elements of the fifth sub-space are:

\begin{eqnarray*}\label{st25}
\{\mathcal{X}_{5}\},\{\mathcal{Y}_{5}\}=\left\{[1,0,0,0,0,0], \hspace{.1cm}[0,0,0,0,0,1],  \hspace{.1cm}[1,1,1,0,0,0], \hspace{.1cm}[1,0,0,1,1,0], \right. \hspace{.1cm}\\\nonumber
\left.[0,1,1,0,0,1], \hspace{.1cm}[0,0,0,1,1,1],  \hspace{.1cm}[0,1,1,0,0,1], \hspace{.1cm}[0,1,1,1,1,1]\right\}  \hspace{.1cm}
\end{eqnarray*}
\begin{eqnarray*}\label{st25}
\{\mathcal{T}_{5}\}=\left\{[0,0,0,0,0,0], \hspace{.1cm}[1,0,0,0,0,1],  \hspace{.1cm}[0,1,1,0,0,0], \hspace{.1cm}[0,0,0,1,1,0],\right.  \hspace{.1cm}\\\nonumber
\left.[1,1,1,0,0,1], \hspace{.1cm}[1,0,0,1,1,1],  \hspace{.1cm}[1,1,1,0,0,1], \hspace{.1cm}[1,1,1,1,1,1]\right\}  \hspace{.1cm}
\end{eqnarray*}

\begin{eqnarray*}\label{st25}
\{\mathcal{R}_{5}\}=\left(
  \begin{array}{cccccc}
    1 & 0 & 0 & 0 & 0 & 0 \\
    0 & 1 & 1 & 1 & 0 & 0 \\
    0 & 1 & 1 & 0 & 1 & 0 \\
    0 & 1 & 0 & 1 & 1 & 0 \\
    0 & 0 & 1 & 1 & 1 & 0 \\
    0 & 0 & 0 & 0 & 0 & 1 \\
\end{array}
\right), \hspace{.2cm} \left(
  \begin{array}{cccccc}
    0 & 1 & 1 & 1 & 0 & 0 \\
    1 & 0 & 1 & 1 & 0 & 0 \\
    1 & 1 & 0 & 1 & 0 & 0 \\
    1 & 1 & 1 & 0 & 0 & 0 \\
    0 & 0 & 0 & 0 & 1 & 0 \\
    0 & 0 & 0 & 0 & 0 & 1 \\
  \end{array}
\right), \hspace{.2cm}\left(
  \begin{array}{cccccc}
    1 & 1 & 1 & 0 & 0 & 0 \\
    1 & 1 & 0 & 0 & 0 & 1 \\
    1 & 0 & 1 & 0 & 0 & 1 \\
    0 & 0 & 0 & 1 & 0 & 0 \\
    0 & 0 & 0 & 0 & 1 & 0 \\
    0 & 1 & 1 & 0 & 0 & 1 \\
  \end{array}
\right), \hspace{.2cm} \left(
  \begin{array}{cccccc}
    0 & 1 & 1 & 1 & 0 & 0 \\
    1 & 0 & 0 & 0 & 0 & 0 \\
    0 & 0 & 0 & 0 & 0 & 1 \\
    0 & 1 & 0 & 1 & 1 & 0 \\
    0 & 0 & 1 & 1 & 1 & 0 \\
    0 & 1 & 1 & 0 & 1 & 0 \\
\end{array}
\right),\hdots\nonumber\\
\end{eqnarray*}



\end{document}